\documentclass{article}

\usepackage{arxiv}
\usepackage{graphicx,times} 
\usepackage[utf8]{inputenc} % allow utf-8 input
\usepackage[T1]{fontenc}    % use 8-bit T1 fonts
\usepackage{hyperref}       % hyperlinks
\usepackage{url}
\usepackage{longtable} % 加载longtable宏包
\usepackage{rotating} % 导入rotating宏包 % 提供更专业的表格线条% simple URL typesetting
\usepackage{booktabs}       % professional-quality tables
\usepackage{amsfonts}       % blackboard math symbols
\usepackage{wrapfig}
\usepackage{nicefrac}       % compact symbols for 1/2, etc.
\usepackage{microtype}      % microtypography
\usepackage{lipsum}		% Can be removed after putting your text content
\usepackage{amsmath} % 用于数学环境
\usepackage{array} % 用于表格
\usepackage{graphicx}
\usepackage{float}
\usepackage{natbib}
\usepackage{doi}

\title{Molecular Gas Content in Dust-Rich Virgo Galaxies. I. First Results from PMO Telescope Observations}

\author{ ZhiYong Hu  \\
	School of Physics and Electronic Information Engineering, Qinghai Normal University, Xining  810000, China
	\And
	JingFu Hu \\
	School of Physics and Electronic Information Engineering, Qinghai Normal University, Xining  810000, China
	\AND
     YiPing Ao\\
	Purple Mountain Observatory\\
	Chinese Academy of sciences, 10 Yuanhua Road, Nanjing, Jiangsu 210023, People's Republic of china\\
	School of Astronomy and Space Sciences\\
	University of cience and Technology of China, Hefei, Anhui 230026, People's Republic of china\\
}

\renewcommand{\shorttitle}{\textit{arXiv} Template}

\hypersetup{
pdftitle={A template for the arxiv style},
pdfsubject={q-bio.NC, q-bio.QM},
pdfauthor={David S.~Hippocampus, Elias D.~Striatum},
pdfkeywords={First keyword, Second keyword, More},
}

\begin{document}
\maketitle

\begin{abstract}
In this study, we used the 13.7m telescope at Qinghai Station to observe CO data in 48 galaxies in Virgo clusters, of which 41 sources observed CO signals. The properties of molecular gas are deduced by co-to-$H_2$ factor. We also collected and investigated the relationship between $M_{H_2}$ and other galactic properties ($M_B$, $L_K$, sfr, and $def_{HI}$).
We found correlations between $M_{H_2}$ and $M_B$, $L_K$, sfr, and $def_{HI}$.  It has the strongest correlation with Lk. There is a certain correlation between MB and MB, but the scattering is larger.  For sfr, the more abundant the atomic and molecular gases, the stronger the sfr activity.
\end{abstract}

% keywords can be removed
\keywords{galaxies: clusters: individual(Virgo) - galaxies:star formation rate - molecular data(CO, H$_2$)}

\section{Introduction}           %% first-level sections will be auto-capitalized
\label{sect:intro}
In galaxy clusters, the distribution of molecular gas is influenced by the hot gas within the cluster. Galaxies in clusters can lose gas due to interactions with hot gas.In addition, galactic interactions in galaxy clusters also affect the distribution of molecular gas and star formation(\citealt{Braine+etal+1993};\citealt{Combes+etal+1994};\citealt{Casasola+etal+2004}).(\citealt{Kenney+etal+1989};\citealt{Boselli+etal+1997}).The distribution of molecular gas in the interior of a galaxy is usually closely related to star formation activity.The dissipation time scale of molecular gas (that is, the time it takes for molecular gas to transform into stars) varies from galaxy to galaxy, but is generally related to the star formation rate (SFR) of the galaxy.

The proximity of the Virgo Cluster makes it possible to observe its member galaxies with excellent spatial resolution.The Virgo cluster has been the object of recent multi-band observations, for example HI (\citealt{Chung+etal+2009}),radio(\citealt{Yun+etal+2001}),IR (Spitzer(\citealt{Kenney+etal+2012})),FIR(\citealt{Planck+2014} \& Herschel Virgo Cluster Survey (HeViCS)(\citealt{Edvige Corbelli+etal+2012})), submm(\citealt{Angus Mok+etal+2016} use JCMT), optical(\citealt{Côté+etal+2004} HST/ACS Virgo Cluster Survey \& \citealt{Suk Kim+etal+2014} SDSS), H$\alpha$ (\citealt{Koopmann+etal+2001}), and UV (\citealt{Brosch+etal+1997} and \citealt{Voyer+etal+2014} use GALEX).Therefore, it is possible to complete the analysis and processing of the first batch of spectral line data of the member galaxies of the Virgo cluster through the joint analysis of a large number of multi-band data and the CO data currently observed by us, and combine the multi-band data to study the evolution process of galactic gas in the cluster environment and its relationship with the star formation process.

The section 2 introduces sample selection and observation.The section 3 introduces the collection of multi-band data.The section 4 describes the data processing methods. In section 5 we analyze our results and discussion.Section 6 summarizes the main conclusions of this work.

%% Authors can give a citation as 'Michel et al. 1992'.
%% You may also use \cite, \citep and \citet for citation, and use Table~1 or Figure~1
%% and so forth. Using \ref and \label for cross-references of Tables/Figures
%% is a good way in adjusting/adding/removing text, tables or figures.

\section{Sample selection and observations}
\subsection{Sample selection}
\label{sect:Obs}

The sample observed at the 13.7m telescope is defined by the following criteria: i)Selected from the extend virgo clusters catalog (EVCC) directory(\citealt{Suk Kim+etal+2014}),ii)The flux of 857Ghz via the PLanck(\citealt{Planck+2014}) satellite is selected from high to low  of Sources for which no proceed co observations have been made in the EVCC ( because Far-infrared radiation is closely related to that of interstellar dust.  Dust particles absorb shorter wavelengths of light and re-radiate longer wavelengths of infrared light, including the far infrared band.  This helps us understand the properties of the interstellar medium, including the distribution and dynamics of gas and dust,Therefore, it can be considered that the higher the flow rate of the fir band, the higher the probability of observing CO).The basic information of these sources is given in Table~\ref{Tab1}, such as: name, ra, dec, v, distance.

%
%               one-column-spanning table
%________________________________________ Table 1: Use_of_the routines

\begin{longtable}{l l l l l l}
\caption{EVCC Ra Dec Type T} \label{Tab1} \\
% 表头
\toprule
EVCC & RA & DEC & Type & T & Dist \\
     & h:m:s & d:m:s & & & Mpc \\
\midrule
\endfirsthead
% 每页重复的表头
\toprule
EVCC & RA & DEC & Type & T & Dist \\
     & h:m:s & d:m:s & & & Mpc \\
\midrule
\endhead
% 每页重复的表尾
\midrule
\multicolumn{5}{r}{\textit{}} \\
\endfoot
% 最后一页的表尾
\bottomrule
\endlastfoot
EVCC429	 &  12:21:55&	+04:28:26.04&	Sbc	&   4.00  & 24.2 \\
EVCC2159 &	12:32:45&	+00:06:50.04&	Scd	&   6.00  & 8.3  \\
EVCC2184 &	12:36:50&	+13:09:55.08&	Sab	&   2.40  & 16.7 \\
EVCC1099 &	12:43:33&	+11:34:51.24&	SABc&	5.10  & 16.7 \\
EVCC171	 &  12:12:46&	+10:51:55.8	&   Scd	&   6.90  & 16.7 \\
EVCC2174 &	12:35:27&	+14:29:43.8	&   Sb	&   3.10  & 15.0 \\
EVCC233	 &  12:15:39&	+13:54:07.92&	Sc	&   4.90  & 16.7 \\
EVCC808	 &  12:31:39&	+03:56:23.64&	Scd	&   7.40  & 14.5 \\
EVCC1314 &	13:11:37&	+22:54:55.44&	Sc	&   5.10  & 42.5 \\
EVCC2209 &	12:42:32&	-00:04:50.52&	Sc	&   5.00  & 14.4 \\
EVCC952	 &  12:36:56&	+14:13:04.08&	Scd	&   6.40  & 16.7 \\
EVCC84	 &  12:02:42&	+01:58:37.56&	Sa	&   1.30  & 33.7 \\
EVCC884	 &  12:34:03&	+07:41:57.48&	S0	&   -1.90 & 17.3  \\
EVCC1281 &	13:00:39&	+02:30:04.68&	Sc	&   5.10  & 16.8 \\
EVCC1074 &	12:42:41&	+14:17:44.16&	Sbc	&   5.80  & 16.7 \\
EVCC107	 &  12:08:11&	+02:52:42.96&	Sc	&   5.00  & 22.9 \\
EVCC1202 &	12:51:55&	+12:04:58.8	&   Sb	&   3.20  & 33.7 \\
EVCC1182 &	12:49:39&	+15:09:56.52&	S0-a&	-0.90 & 16.7  \\
EVCC673	 &  12:27:45&	+13:00:39.6	&   Sa	&   0.60  & 16.7 \\
EVCC59	 &  11:55:57&	+06:44:55.32&	Sb	&   3.20  & 36.8 \\
EVCC104	 &  12:07:37&	+02:41:28.32&	SBcd&	7.30  & 22.8 \\
EVCC574	 &  12:25:42&	+07:13:05.16&	SBc	&   6.00  & 22.9 \\
EVCC631	 &  12:26:55&	-00:52:39.36&	SABa&	1.10  &      \\
EVCC461	 &  12:22:42&	+09:19:57.72&	Sc	&   5.50  & 22.9 \\
EVCC497	 &  12:23:54&	-03:26:34.8	&   Sbc	&   3.90  &      \\
EVCC153	 &  12:11:53&	+24:07:23.88&	Sbc	&   4.00  & 43.9 \\
EVCC488	 &  12:23:39&	+06:57:15.12&	Sb	&   3.00  & 22.9 \\
EVCC439	 &  12:22:06&	+09:02:37.32&	SBb	&   3.20  & 16.4 \\
EVCC1080 &	12:42:52&	+13:15:25.56&	Sbc	&   3.60  & 21.0 \\
EVCC126	 &  12:10:38&	+16:01:59.16&	SABc&	4.90  & 16.7 \\
EVCC1178 &	12:49:12&	+03:23:19.68&	Sc	&   5.90  & 16.6 \\
EVCC1282 &	13:00:59&	-00:01:39.72&	Sc	&   5.80  & 17.8 \\
EVCC1217 &	12:53:21&	+01:16:09.12&	Sc	&   6.00  & 20.2 \\
EVCC996	 &  12:39:19&	-00:31:53.76&	Sd	&   7.90  &      \\
EVCC587	 &  12:25:58&	+03:25:47.64&	Sc	&   5.70  & 26.1 \\
EVCC660	 &  12:27:26&	+06:15:50.04&	Sb	&   3.50  & 16.4 \\
EVCC1223 &	12:53:51&	+09:42:36.36&	SBc	&   4.60  & 38.9 \\
EVCC1069 &	12:42:31&	+03:57:31.32&	Ir	&   9.60  & 16.6 \\
EVCC1104 &	12:43:51&	-00:33:39.6	&   SABc&	5.90  &      \\
EVCC635	 &  12:26:58&	+02:29:39.84&	SBc	&   4.70  & 15.1 \\
EVCC176	 &  12:13:03&	+07:02:20.04&	SABa&	2.00  & 24.2 \\
EVCC208	 &  12:14:39&	+05:48:23.76&	SABc&	6.70  & 16.5 \\
EVCC231	 &  12:15:30&	+09:35:06.36&	Sd	&   7.70  & 16.7 \\
EVCC1243 &	12:55:12&	+00:06:57.96&	SBcd&	6.90  & 14.4 \\
EVCC192	 &  12:13:54&	+13:10:21.72&	SABb&	4.10  & 30.1 \\
EVCC776	 &  12:30:27&	+04:14:46.68&	SABc&	5.10  & 43.5 \\
EVCC1209 &	12:52:44&	+15:50:46.68&	SBm	&   9.00  & 13.0 \\
EVCC451	&   12:22:31&	+15:32:16.44&	Sab &	2.10  & 16.7 \\

\end{longtable}

\subsection{13.7m telescope observations}
\subsubsection{13.7m telescope}
The CO data used in this article was observed using the 13.7m telescope at the Qinghai station,The 13.7-m millimeter-wave radio telescope of Purple Mountain Observatory operates at 3200-m above the sea level near Delingha, Qinghai Province.China. Equipped with a superconducting SIS receiver, the telescope is used inthe millimeter-wave band ranging from 85 to 115GHz. The superconducting imaging spectrometer has 3×3 beams, the single beam has a half-power beam width of 50 "(115.2 GHz), the spacing distance between adjacent beams is 60 mm, and the corresponding spacing Angle is 175".

\subsubsection{Method of observation}
Starting from April 30, 2024, the observations will be made in an ON-OFF mode, with an integration time of 10 seconds each time, and a pointing calibration of the telescope will be performed before each observation.One source per day is observed for a total observation time of about 3 hours, and the observation time for each source is about 60 minutes.  And our detection sensitivity can detect the molecular line emission of a significant number of dusty galaxies.  The CO signal was observed in 46 galaxies, with an upper limit of 3$\sigma$ given for other sources.

\subsection{Optical HI and FIR data}
\subsubsection{Optical data}

The collected optical data are listed in Table~\ref{Tab2} and will be compared with the CO data,The type of pattern shown in col(2) is:contains the machine-readable version of the Third Reference Catalogue of Bright Galaxies (RC2) by \citealt{De Vaucouleurs+etal+1976},B-band in col(3) and BT-band in col(4),from \citealt{De Vaucouleurs+etal+1976} ( RC2 p33 rel.25) BT is the absolute magnitude.

\begin{longtable}{l l l l l l l } 
\caption{EVCC and optical data} \label{Tab2} \\
% 表头
\toprule
EVCC & Bband & BT & Kband & \( L_{K} \) & cz & d25 \\
     & mag & mag & mag & \( L_{\odot} \) & km\ s\(^{-1}\) & KPC \\
\midrule
\endfirsthead
% 每页重复的表头
\toprule
EVCC & Bband & BT & Kband & \( L_{K} \) & cz & d25 \\
     & mag & mag & mag & \( L_{\odot} \) & km\ s\(^{-1}\) & KPC \\
\midrule
\endhead
% 每页重复的表尾
\midrule
\multicolumn{5}{r}{\textit{}} \\
\endfoot
% 最后一页的表尾
\bottomrule
\endlastfoot
EVCC429 &	10.16 ± 0.06  &	-21.76 ± 0.118 &	6.87 ± 0.09 &	8.31E+11 &	1613.5  &	28.30 \\
EVCC2159 &	11.19 ± 0.08  & -18.41 ± 0.133 &	7.38 ± 0.11	&   6.11E+10 &	1126.0  &	34.85 \\ 
EVCC2184 &	10.18 ± 0.14  & -20.93 ± 0.379 &	6.60 ± 0.08 &	5.07E+11 &	-235.0  &	40.55 \\
EVCC1099 &	11.60 ± 0.48  &	-19.51 ± 0.572 &	8.07 ± 0.08 &	1.31E+11 &	1417.2  &	23.93 \\
EVCC171 &	11.92 ± 0.07  &	-19.10 ± 0.211 &	9.64 ± 0.15 &	2.83E+10 &	441.9  &	35.96 \\
EVCC2174 &	10.94 ± 0.06  &	-19.94 ± 0.073 &	7.18 ± 0.14 &	2.40E+11 &	486.0  &	32.36 \\ 
EVCC233	 &  11.78 ± 0.09  &	-19.33 ± 0.185 &	8.37 ± 0.01 &	9.94E+10 &	-156.8  &	16.96 \\
EVCC808	 &  12.11 ± 0.09  & -18.70 ± 0.093 &	9.73 ± 0.25 &	2.14E+10 &	1746.4  &	17.58 \\
EVCC1314 &	12.40 ± 0.07  &	-20.74 ± 0.159 &	8.78 ± 0.04 &	4.41E+11 &	2622.5  &	32.26 \\ 
EVCC2209 &	12.27 ± 0.04  &	-18.52 ± 0.757 &	9.30 ± 0.07 &	3.14E+10 &	1721.0  &	15.45 \\
EVCC952  &	11.92 ± 0.22  &	-19.19 ± 0.281 &	8.64 ± 0.15 &	7.75E+10 &	340.8  &	21.96 \\
EVCC84   &	12.61 ± 0.10  &	-20.03 ± 0.307 &	10.77 ± 0.07 &	4.44E+10 &	1996.9  &	26.82 \\
EVCC884  &	10.55 ± 0.09  &	-20.64 ± 0.094 &	6.48 ± 0.05 &	6.08E+11 &	427.3  &	35.07 \\
EVCC1281 &	11.89 ± 0.09  &	-19.24 ± 1.046 &	8.68 ± 0.07 &	7.56E+10 &	959.7  &	20.26 \\
EVCC1074 &	13.04 ± 0.07  &	-18.07 ± 0.144 &	9.27 ± 0.04 &	4.34E+10 &	232.6  &	19.84 \\
EVCC107	 &  11.96 ± 0.11  &	-19.84 ± 0.222 &	8.86 ± 0.14 &	1.19E+11 &	1287.1  &	22.67 \\
EVCC1202 &	13.33 ± 0.02  &	-19.31 ± 0.24  &    9.51 ± 0.06 &	1.42E+11 &	1754.8  &	47.60 \\
EVCC1182 &	11.60 ± 0.15  &	-19.51 ± 0.893 &	7.58 ± 0.05 &	2.06E+11 &	1190.6  &	31.03 \\
EVCC673  &	10.97 ± 0.08  &	-20.14 ± 0.36  &    7.28 ± 0.06 &	2.71E+11 &	9.5  &	32.77 \\
EVCC59   &	12.45 ± 0.03  &	-20.38 ± 0.235 &	8.98 ± 0.36 &	2.75E+11 &	2695.4  &	35.81 \\
EVCC104	 &  12.47 ± 0.08  &	-19.32 ± 0.33  &    10.39 ± 0.19 &	2.88E+10 &	1331.3  &	20.01 \\
EVCC574  &  13.56 ± 0.07  &	-18.24 ± 0.239 &	9.79 ± 0.02 &	5.05E+10 &	1002.3  &	27.10 \\
EVCC631	 &  13.83 ± 0.03  &	-17.30 ± 0.501 &	10.17 ± 0.08 &	1.92E+10 &	2123.1  &	11.06 \\
EVCC461	 &  13.54 ± 0.06  &	-18.26 ± 0.26  &    9.26 ± 0.03 &	8.24E+10 &	1248.6  &	25.49 \\
EVCC497	 &  13.11 ± 0.07  &	-18.02 ± 0.248 &	8.96 ± 0.16 &	5.84E+10 &	2028.0  &	31.31 \\
EVCC153	 &  12.68 ± 0.19  &	-20.53 ± 0.496 &	9.38 ± 0.03 &	2.71E+11 &	2688.1  &	26.31 \\
EVCC488	 &  13.31 ± 0.38  &	-18.49 ± 0.503 &	9.00 ± 0.05 &	1.05E+11 &	1000.0  &	20.69 \\
EVCC439	 &  12.71 ± 0.06  &	-18.36 ± 0.366 &	8.73 ± 0.01 &	6.88E+10 &	983.4  &	28.79 \\
EVCC1080 &	12.08 ± 0.10  &	-19.53 ± 0.100 &	8.81 ± 0.05 &	1.05E+11 &	1091.6  &	21.04 \\
EVCC126  &	12.84 ± 0.19  &	-18.27 ± 0.331 &	9.65 ± 0.03 &	3.06E+10 &	2173.1  &	21.92 \\
EVCC1178 &	12.79 ± 0.04  &	-18.31 ± 0.351 &	9.80 ± 0.07 &	2.63E+10 &	753.0  &	23.22 \\
EVCC1282 &	12.59 ± 0.09  &	-18.66 ± 0.285 &	9.52 ± 0.05 &	3.92E+10 &	1188.6  &	14.65 \\
EVCC1217 &	12.73 ± 0.07  &	-18.80 ± 0.343 &	9.01 ± 0.02 &	8.07E+10 &	1125.4  &	25.63 \\
EVCC996	 &  12.32 ± 0.04  &	-18.81 ± 0.328 &	10.21 ± 0.08 &	1.85E+10 &	1082.4  &	19.85 \\
EVCC587	 &  14.23 ± 0.12  &	-17.86 ± 0.541 &	10.17 ± 0.03 &	9.55E+10 &	1409.6  &	46.34 \\
EVCC660	 &  12.65 ± 0.09  &	-18.42 ± 0.510 &	9.56 ± 0.23 &	3.20E+10 &	1472.5  &	20.63 \\
EVCC1223 &	13.07 ± 0.05  &	-19.88 ± 0.262 &	9.91 ± 0.07 &	1.31E+11 &	2860.9  &	28.62 \\
EVCC1069 &	13.04 ± 0.06  &	-18.06 ± 0.272 &	9.95 ± 0.12 &	2.29E+10 &	732.4  &	12.72 \\
EVCC1104 &	12.77 ± 0.09  &	-18.36 ± 0.415 &	10.11 ± 0.20 &	2.03E+10 &	2630.5  &	36.13 \\
EVCC635	 &  12.77 ± 0.06  &	-18.12 ± 0.532 &	9.67 ± 0.05 &	2.45E+10 &	1722.0  &	10.88 \\
EVCC176	 &  13.22 ± 0.04  & -18.70 ± 0.287 &	9.12 ± 0.03 &	1.05E+11 &	2263.0  &	23.54 \\
EVCC208	 &  13.36 ± 0.05  & -17.73 ± 0.198 &	10.25 ± 0.08 &	1.72E+10 &	2052.8  &	30.29 \\
EVCC231  &	13.40 ± 0.05  &	-17.71 ± 0.330 &	9.46 ± 0.04 &	3.64E+10 &	637.9  &	11.09 \\
EVCC1243 &	12.94 ± 0.10  &	-17.85 ± 0.334 &	11.59 ± 0.46 &	3.81E+09 &	1309.4  &	16.46 \\
EVCC192  &	13.12 ± 0.06  &	-19.27 ± 0.110 &	9.35 ± 0.05 &	1.31E+11 &	2464.1  &	30.82 \\
EVCC776	 &  13.10 ± 0.03  &	-20.09 ± 0.255 &	9.77 ± 0.09 &	1.86E+11 &	2442.8  &	30.03 \\
EVCC1209 &	13.41 ± 0.07  & -17.16 ± 0.437 &	10.12 ± 0.22 &	1.20E+10 &	1306.2  &	15.14 \\
EVCC451  &	12.48 ± 0.06  &	-18.63 ± 0.277 &	8.85 ± 0.11 &	6.39E+10 &	178.9  &	18.02 \\

\end{longtable}

\subsubsection{HI data}
We use of H I sources extracted from the Arecibo Legacy Fast ALFA (ALFALFA) extragalactic survey(\citealt{Riccardo Giovanelli+etal+2007}, \citealt{Brian R. Kent+etal+2008}),initiated in 2005. Sources have been extracted from three-dimensional spectral data cubes exploiting a matched ﬁltering technique and then examined interactively to yield global H I parameters.(\citealt{Riccardo Giovanelli+etal+2007})The H I flux of the galaxy is given.

The H I mass is determined using the following relation:
$$
M_{HI}(M_\odot)=2.356\times10^5d^2S_{HI}
$$
Where d is the distance to the source in MPC,S$_{HI}$ is the flux of HI, in $Jy km s^{-1}$.

To measure the abundance of atomic gas in Virgo galaxies, we adopt the H I deficiency parameter as an indicator of atomic gas.The good correlation of H I mass and optical diameter in isolated galaxies allows a useful parameterization of the normalcy of any galaxy’s H I content to be defined as (\citealt{Giovanelli+etal+1983} GH83);In a study of the H i emission from 324 isolated galaxies, \citealt{Haynes+etal+1984} (1984, hereafter HG84) demonstrated that the H i mass of an isolated galaxy depends primarily on its optical diameter, and only slightly on its morphological type. Since the type dependence for Sa-Sm is slight, we follow GH83 in adopting an H i deficiency parameter which is independent of morphological type. The formula we use for the “expected” H i mass is (GH83):
$$
log{M_{HI}}(isolated) = 7.01 + 0.88log{D^{2}_{H_{0}}}
$$
where M$_{HI}$  is measured in solar units, and the optical diameter $D_{H_{0}}$ is the Holmberg diameter, measured in kiloparsecs to a surface brightness of 26.5 mag arcsec$^{-2}$. The good correlation of H I mass and optical diameter in isolated galaxies allows a useful parameterization of the normalcy of any galaxy’s H I content to be defined as (GH83):
$$
H I def = log M_{HI}(isolated)- log M_{HI}(actual)
$$

\subsubsection{FIR data}

The FIR data used in this paper are two sky survey data, respectively Planck Catalog of Compact Sources (PCCS)(\citealt{Planck+2014}) 545 and 857 GHz flux in two band,And Herschel SPIRE(\citealt{Edvige Corbelli+etal+2012}) Data 250,300,500$\mu m$ flux in three bands. The Planck and Herschel data are listed in Table 2. These data will be compared with CO data.

The photutils\footnote{https://photutils.readthedocs.io/en/stable/index.html} package in python is used to conduct photometry for Herschel, and the 2D Background and Noise Estimation is used to analyze the background noise.  In order to shield the influence of sources on the background noise, It is considered that the pixel mask in the image is larger than the first median noise 2$\sigma$, and the background noise of the image after the mask is analyzed once, and its root mean square is considered as the metering error.

\section{Data reduction}
\label{sect:data}
\subsection{CO data processing}
The tool used in this article to process co data is GILDAS-CLASS(\citealt{pety+etal+2005})\footnote{https://www.iram.fr/IRAMFR/GILDAS/},GILDAS-CLASS is a software package for reducing spectroscopic data obtained on a single-dish telescope. It also has basic functionalities to reduce continuum drifts like pointing or focus.

First, find all the spectral lines where the observation file position error is 0,These lines are averaged and smoothed to V=4.1km s$^{-1}$, after which baseline fitting is performed,The rms noise level of the ﬁnal spectra ranges between 7 mK and 13 mK in CO(1-0) (Table~\ref{Tab3}).Integrated intensities, I$_{CO}$, were calculated way is : we ﬁtted a Gauss function (single-peak proﬁles) or a linear combination ofthe ﬁrst two Hermite functions (double-peak proﬁles;\citealt{Saintonge+etal+2007}) to the spectra.We detected CO (1-0) emission lines in 42 of the first 48 galaxies, and gave an upper limit of 3$\sigma$ for undetected galaxies.In Figure
~\ref{Fig1}, we give the spectral line intensity diagram after processing with class.

In Table 2 we list the following items:

1.rms:root mean square noise in mK for a velocity resolution of 4.71 km s$^{-1}$ .

2.I$_{CO}$:velocity integrated CO line temperature Tmb dv,in Kkm s$^{-1}$, and its error;

3.V$_{CO}$:mean velocity of the CO line, in km s$^{-1}$

4.W$_{CO}$:Velocity linewidth of CO spectrum,in km s$^{-1}$

Conversion between main beam temperature in Kelvin and flux in Jansky (S$\nu$) for a point-like source were made using the relation
$$
G = S_v / T_{mb} = 2k/(\pi r^2/4.0)\times1.0\times10^{26}\times0.51/0.38[Jy/K]
$$

Where k is the Boltzmann constant, r is the radius of the telescope 13.7m, and 0.51 and 0.38 are the main beam efficiency and lunar efficiency, respectively.

The CO luminosity is calculated as follows:

$$L_{CO} =0.119S_{CO}d^{2}$$

where d is the distance to the source in Mpc and S$_{CO}$ is the total CO flux in Jy km s$^{-1}$ (\citealt{Kenney+etal+1989}).The molecular gas mass (M$_{H2}$) is calculated using a Galactic conversion factor of N(H$_2$)/I$_{CO}$ = $2.8 \times 10^{20} cm^{-2}(K km s^{-1})^{-1}$(\citeauthor{Bloemen+etal+1986}\citeyearpar{Bloemen+etal+1986}) yielding:

$$ M_{H_{2}}[M_\odot] = 1.1 \times 10^4D^{2}S_{co}$$

where S$_{CO}$ is the velocity integrated CO line intensity in Jy km s$^{-1}$, D is the distance in Mpc.

\begin{longtable}{l l l l l l  } 

\caption{EVCC and CO data} \label{Tab3} \\
% 表头
\toprule
EVCC & \( V_{CO} \) & \( W_{CO} \) & \( T_{peak} \) & \( I_{CO} \)  & \( L_{CO} \) \\
     & km\ s\(^{-1}\) & km\ s\(^{-1}\) & K  & K\ km\ s\(^{-1}\)  & L\(_{\odot}\)\\
\midrule
\endfirsthead
% 每页重复的表头
\toprule
EVCC & \( V_{CO} \) & \( W_{CO} \) & \( T_{peak} \) & \( I_{CO} \)  & \( L_{CO} \) \\
     & km\ s\(^{-1}\) & km\ s\(^{-1}\) & K  & K\ km\ s\(^{-1}\)  & L\(_{\odot}\)\\
\midrule
\endhead
% 每页重复的表尾
\midrule
\multicolumn{5}{r}{\textit{}} \\
\endfoot
% 最后一页的表尾
\bottomrule
\endlastfoot
EVCC429	 & 1592.00 ± 1.00    &97.18  ± 4.21 	 &0.170 &  20.6 ±	0.40  &	3.61E+04   \\
EVCC2159 & 1126.20 ± 2.60    &94.52  ± 6.75	 &0.058 &	5.90 ±  0.30  &	1.21E+03   \\
EVCC2184 & -211.60 ± 2.30    &214.95 ± 4.80	 &0.112 &	25.9 ±	0.50  &	2.16E+04   \\
EVCC1099 & 1426.10 ± 2.80    &93.46  ± 5.09	 &0.074 &	7.30 ±  0.40  &	6.09E+03   \\
EVCC171	 & 381.85  ± 8.16    &101.47 ± 16.53    &0.022 &  2.33 ±	0.36  &	1.78E+03   \\
EVCC2174 & 451.72  ± 15.69   &188.39 ± 28.91    &0.014 &	2.84 ±	0.43  &	1.91E+03   \\
EVCC233  & -39.61  ± 5.49    &155.00 ± 10.90    &0.040 &	6.63 ±	0.45  &	5.53E+03   \\
EVCC808  & 1716.11 ± 3.64    &56.41  ± 9.05	 &0.035 &	2.08 ±	0.27  &	1.31E+03   \\
EVCC1314 & 2727.01 ± 6.72    &113.37 ± 12.18    &0.020 &	2.44 ±	0.27  &	1.32E+04   \\
EVCC2209 & 1724.36 ± 7.76    &131.00 ± 13.91    &0.021 &	2.87 ±	0.32  &	1.78E+03   \\
EVCC952  & 322.25  ± 3.93    &84.79  ± 8.09	 &0.027 &	2.43 ±	0.21  &	2.03E+03   \\
EVCC84   & 1971.71 ± 6.78    &261.07 ± 13.59    &0.050 &	13.96±	0.70  &	4.74E+04   \\
EVCC884  & 366.12  ± 8.18    &189.07 ± 15.12    &0.042 &	8.51 ±	0.70  &	7.61E+03   \\
EVCC1281 & 962.00  ± 1.90    &55.25  ± 4.19	 &0.079 &	4.60 ±	0.30  &	3.88E+03   \\
EVCC1074 & 105.41  ± 6.22    &289.94 ± 14.37    &0.043 &	13.21±	0.58 &	1.10E+04   \\
EVCC107	 & 1282.46 ± 4.15    &180.60 ± 9.12	 &0.059 &  11.32±	0.49 &	1.77E+04   \\
EVCC1202 & 1841.64 ± 10.51   &293.06 ± 25.49    &0.026 &	8.06 ±	0.59 &	2.74E+04   \\
EVCC1182 & 1154.07 ± 8.88    &237.70 ± 18.31    &0.035 &	8.82 ±	0.64 &	7.35E+03   \\
EVCC673  & 219.98  ± 6.87	  &95.01  ± 14.61	 &0.034 &	3.45 ±	0.46 &	2.88E+03   \\
EVCC59	 & 2639.91 ± 10.53   &191.95 ± 21.68    &0.020 &  4.12 ±	0.44 &	1.67E+04   \\
EVCC104	 & 1292.76 ± 3.74    &33.65  ± 6.53	 &0.029 &  1.05 ±	0.21 &	1.63E+03   \\
EVCC574	 & 943.03  ± 8.75	  &179.74 ± 21.79    &0.024 &  4.56 ±	0.46 &	7.15E+03   \\
EVCC631	 & 2086.67 ± 2.87    &44.40  ± 5.66	 &0.037 &  1.77 ±	0.22 &	1.49E+03   \\
EVCC461	 & 1209.85 ± 9.04    &157.54 ± 18.12    &0.020 &  3.38 ±	0.38 &	5.30E+03   \\
EVCC497  & 2170.83 ± 10.76   &227.75 ± 26.71    &0.023 &	5.61 ±	0.54 &	4.73E+03   \\
EVCC153	 & 2436.25 ± 4.19    &61.94  ± 9.76	 &0.035 &  2.32 ±	0.31 &	1.34E+04   \\
EVCC439	 & 1090.70 ± 8.91    &150.74 ± 19.39    &0.015 &  2.39 ±	0.27 &	1.92E+03   \\
EVCC1080 & 982.60  ± 15.35   &179.60 ± 34.75    &0.008 &	1.46 ±	0.24 &	1.92E+03   \\
EVCC126	 & 2113.32 ± 7.61    &93.76  ± 25.36	 &0.017 &  1.69 ±	0.31 &	1.41E+03   \\
EVCC1178 & 725.28  ± 3.73	  &16.43  ± 7.49	 &0.195 &	3.42 ±	1.50 &  2.82E+03   \\
EVCC1282 & 1238.42 ± 6.13    &40.19  ± 10.29	 &0.020 &	0.87 ±	0.23 &	8.24E+02   \\
EVCC1217 & 1120.03 ± 14.19   &161.88 ± 28.71    &0.013 &	2.31 ±	0.39 &	2.82E+03   \\
EVCC587	 & 1474.44 ± 4.15    &30.99  ± 10.25	 &0.002 &  0.61 ±	0.16 &	1.25E+03   \\
EVCC660	 & 1413.37 ± 5.14    &35.53  ± 7.87	 &0.028 &  1.07 ±	0.27 &	8.60E+02   \\
EVCC1223 & 3030.06 ± 4.62    &28.73  ± 7.14	 &0.020 &	0.61 ±	0.17 &	2.76E+03   \\
EVCC176	 & 2208.29 ± 9.02    &122.29 ± 21.35    &0.038 &  4.95 ±	0.73 &	8.67E+03 	 \\
EVCC231	 & 583.56  ± 12.32   &111.28 ± 18.65    &0.016 &  1.86 ±	0.37 &	1.55E+03 	 \\
EVCC192	 & 2449.16 ± 34.84   &241.73 ± 60.10    &0.005 &  1.47 ±	0.4	 &  3.98E+03 	 \\
EVCC776	 & 2452.99 ± 14.92   &371.71 ± 32.15    &0.015 &  6.14 ±	0.48 &	3.47E+04 	 \\
EVCC1209 & 1236.47 ± 14.35   &98.69  ± 29.70	 &0.007 &	0.76 ±	0.21 &	3.84E+02 	 \\
EVCC451	 & 129.54  ± 4.16    &149.10 ± 9.33	 &0.03 &  4.77 ±	0.26 &	3.98E+03 	 \\

\end{longtable}

\begin{figure}
     \centering
     \includegraphics[width=0.8\textwidth, angle=0]{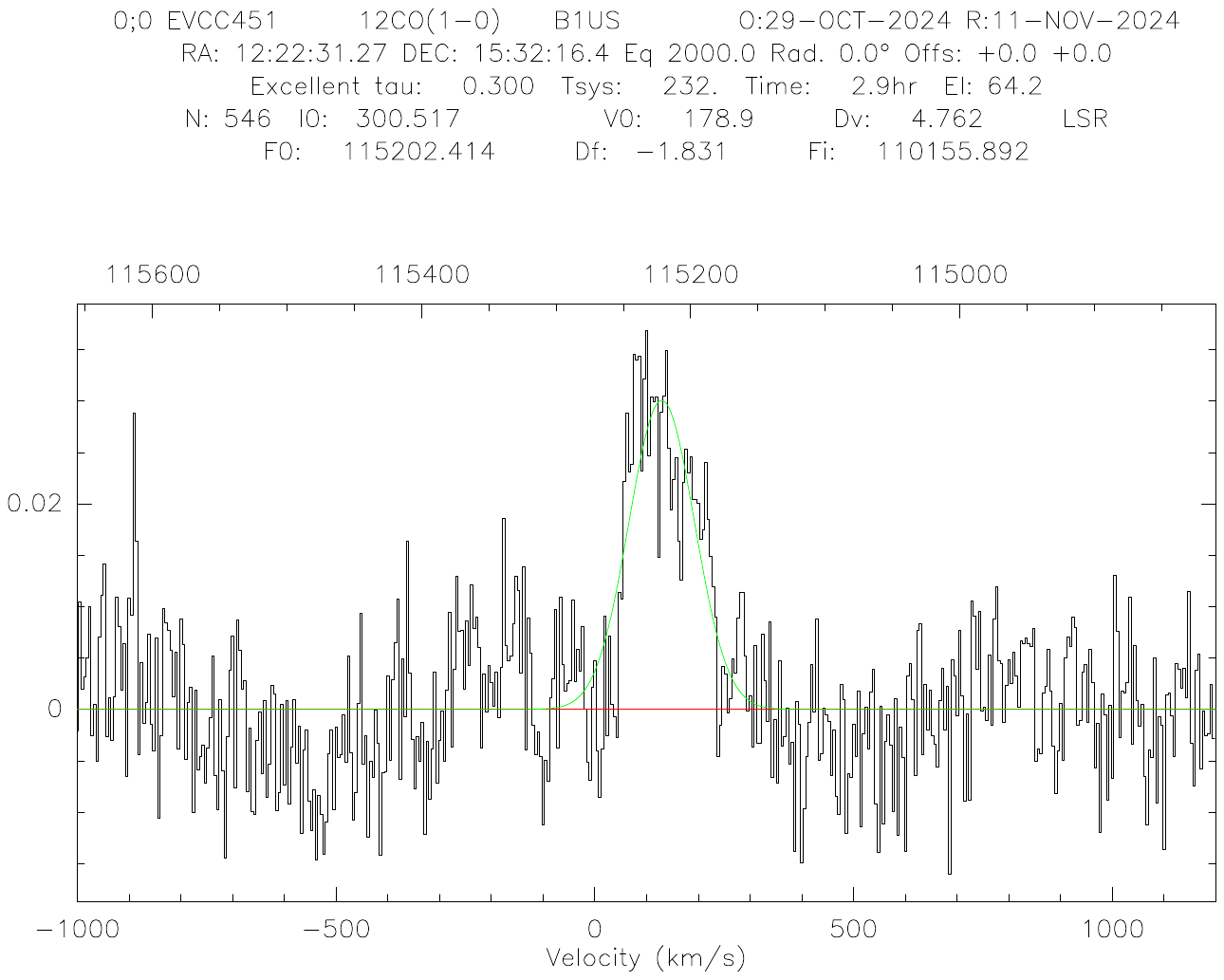}
     \includegraphics[width=0.8\textwidth, angle=0]{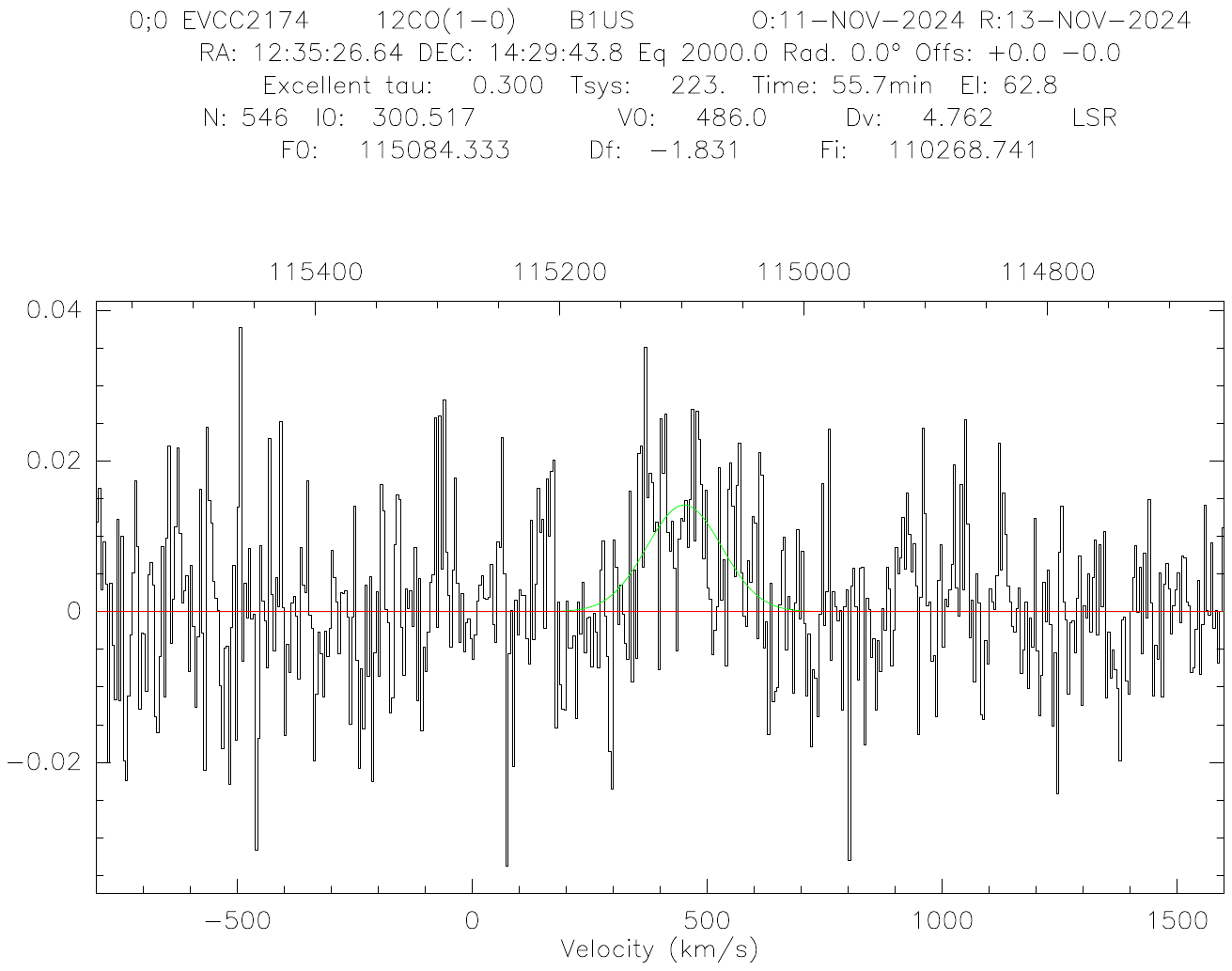}
     \caption{CO spectra  of  Virgo cluster galaxies.}
     \label{Fig1}
\end{figure}

\begin{figure}
     \centering
     \includegraphics[width=0.8\textwidth, angle=0]{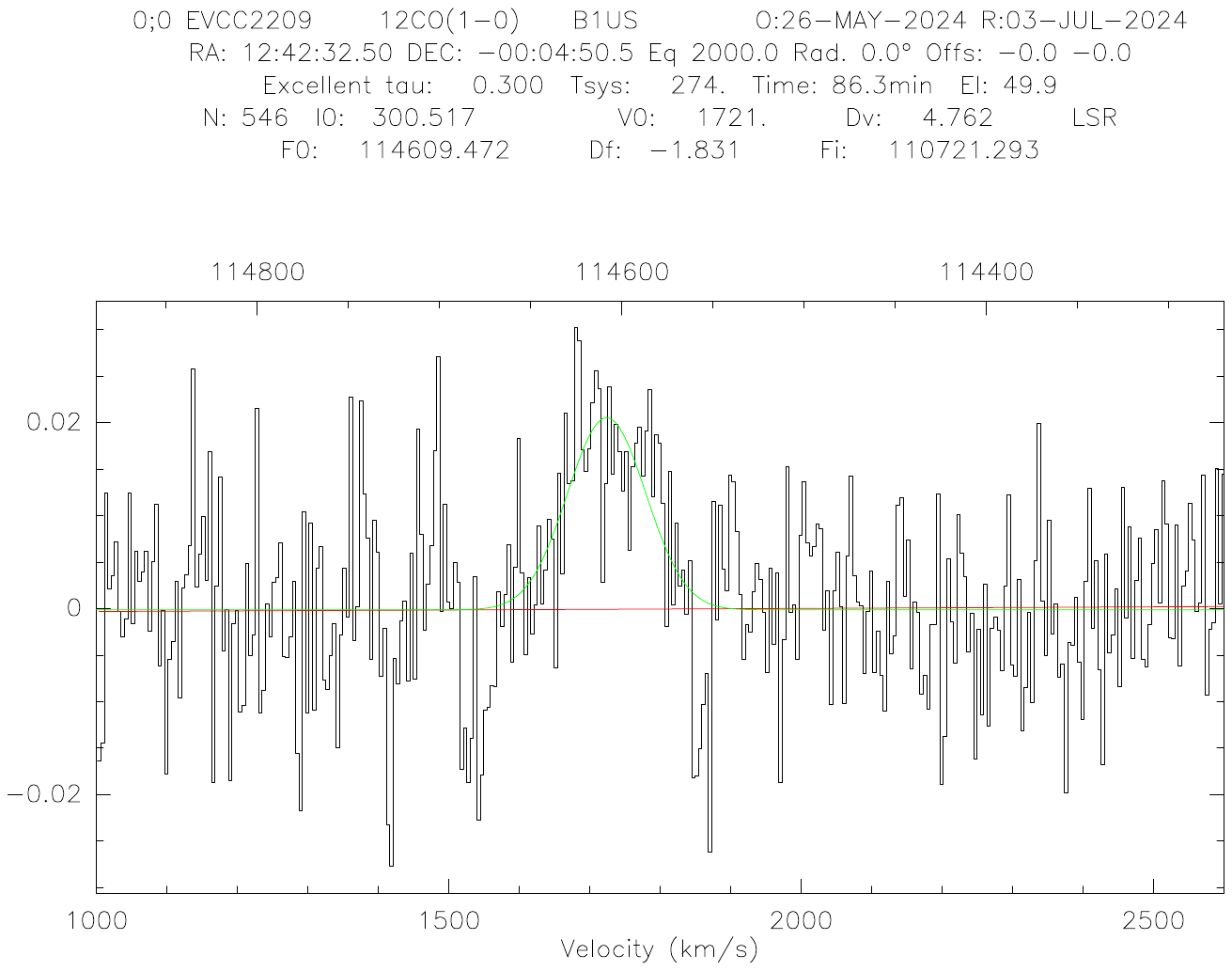}
     \includegraphics[width=0.8\textwidth, angle=0]{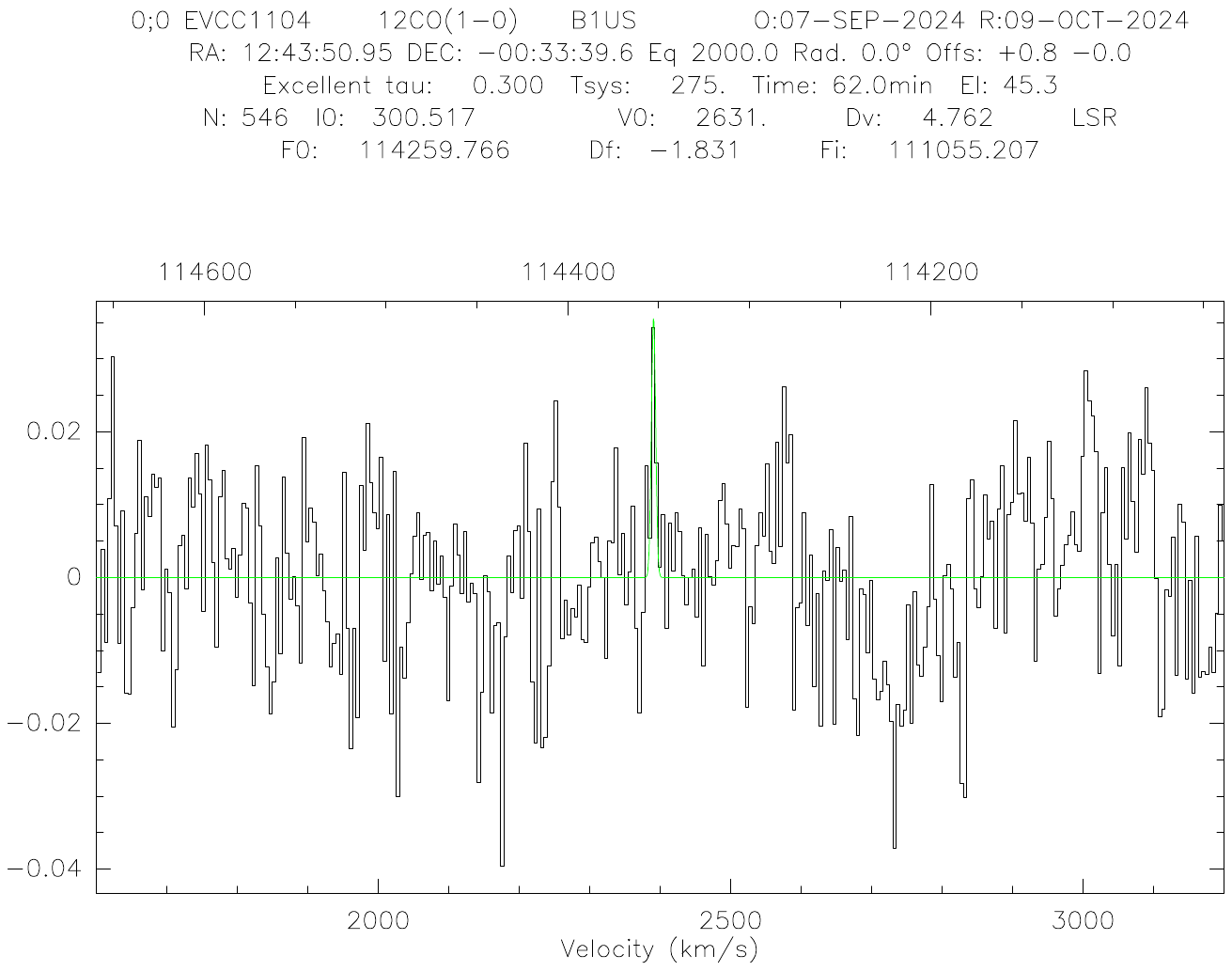}
     \addtocounter{figure}{-1} 
     \caption{(continued)} 
\end{figure}

\begin{figure}
     \centering
     \includegraphics[width=0.8\textwidth, angle=0]{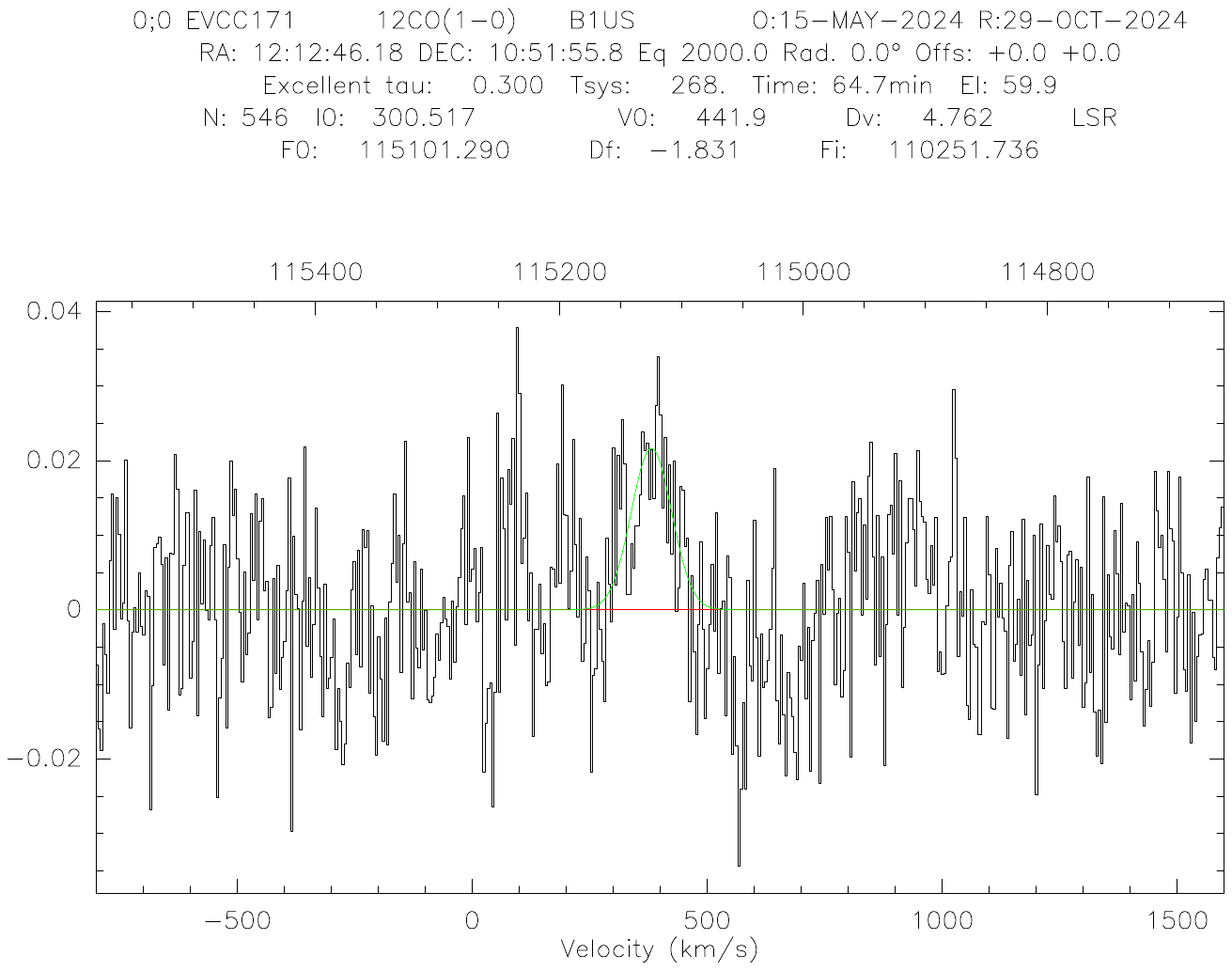}
     \includegraphics[width=0.8\textwidth, angle=0]{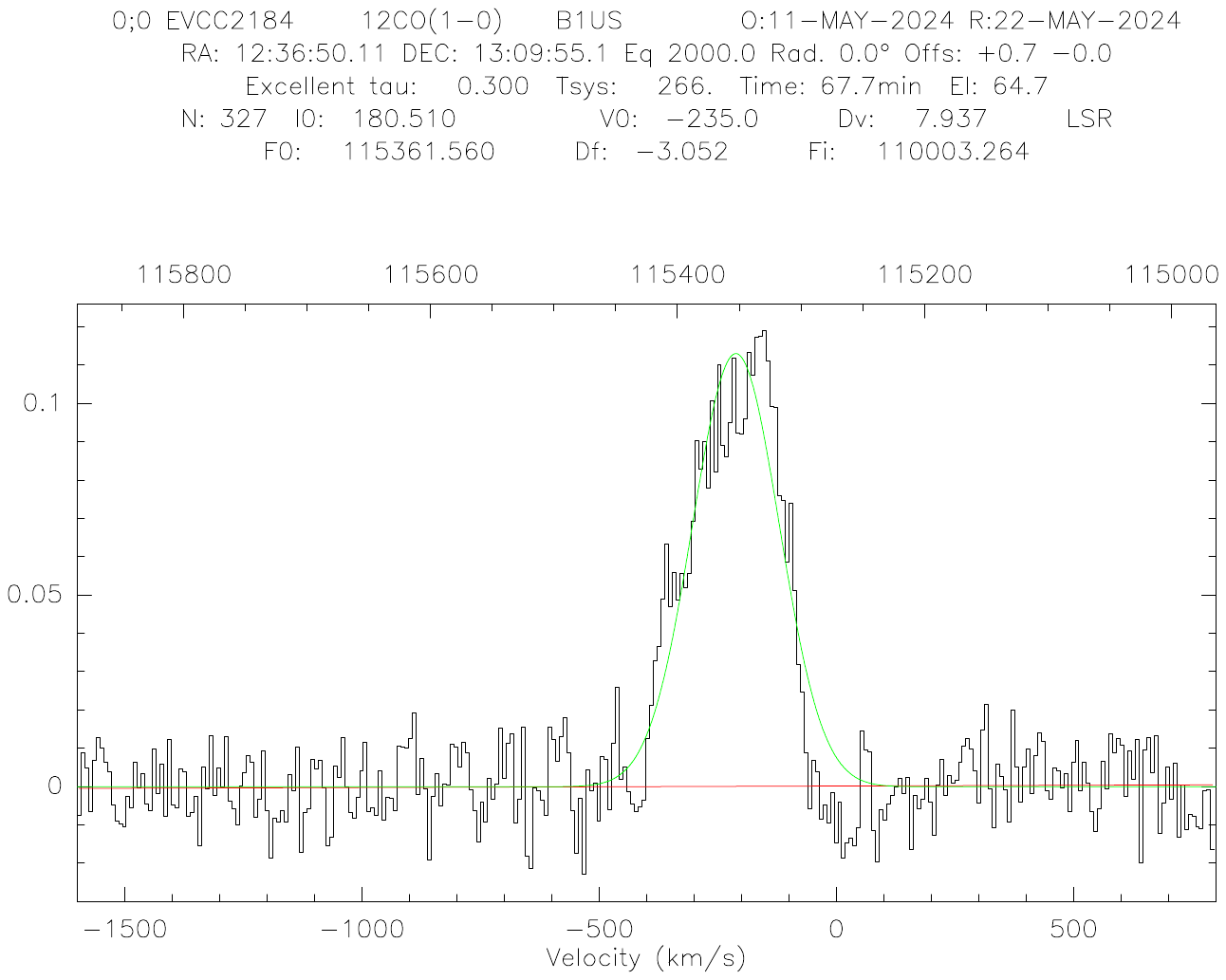}
     \addtocounter{figure}{-1} 
     \caption{(continued)} 
\end{figure}

\begin{figure}
     \centering
     \includegraphics[width=0.8\textwidth, angle=0]{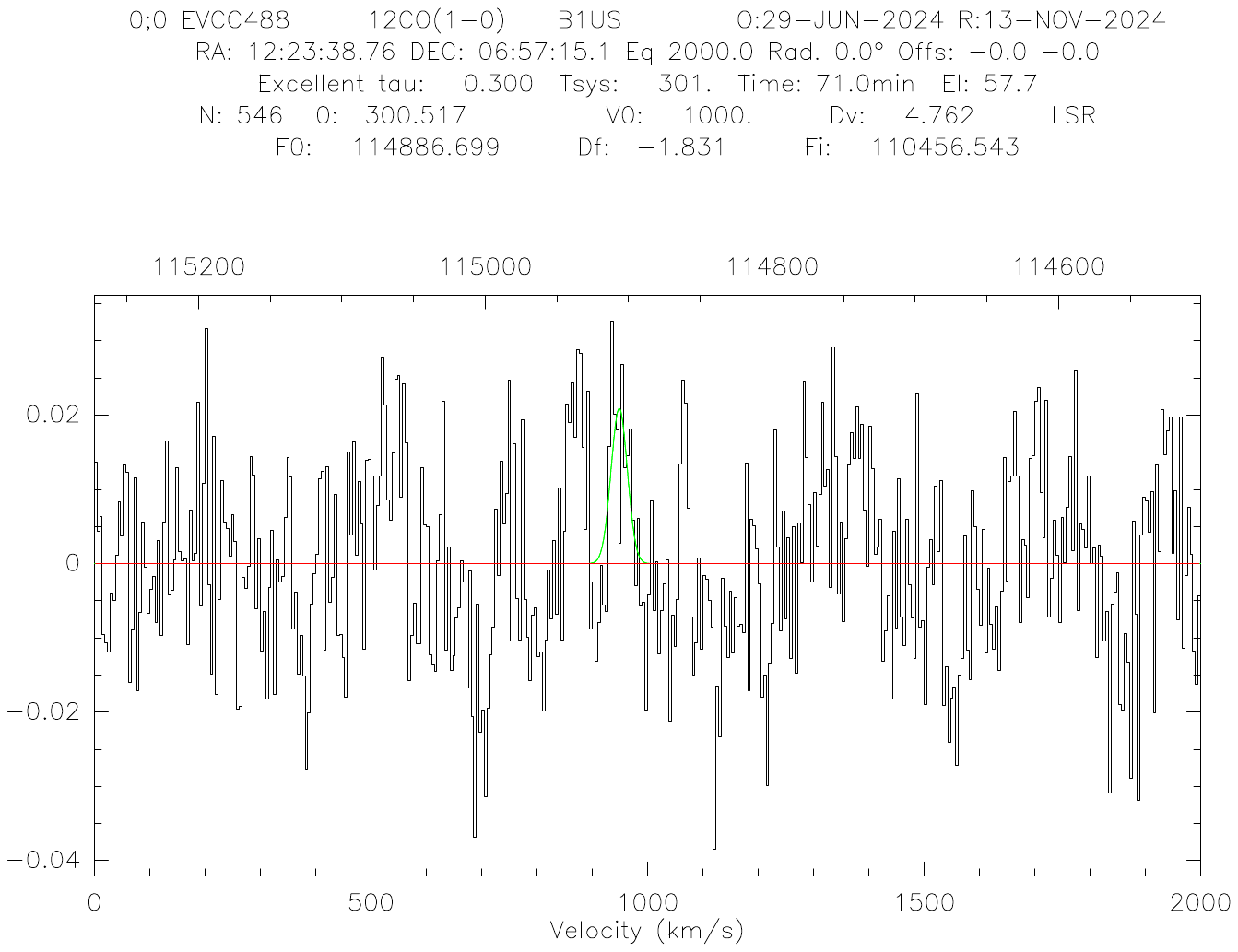}
     \includegraphics[width=0.8\textwidth, angle=0]{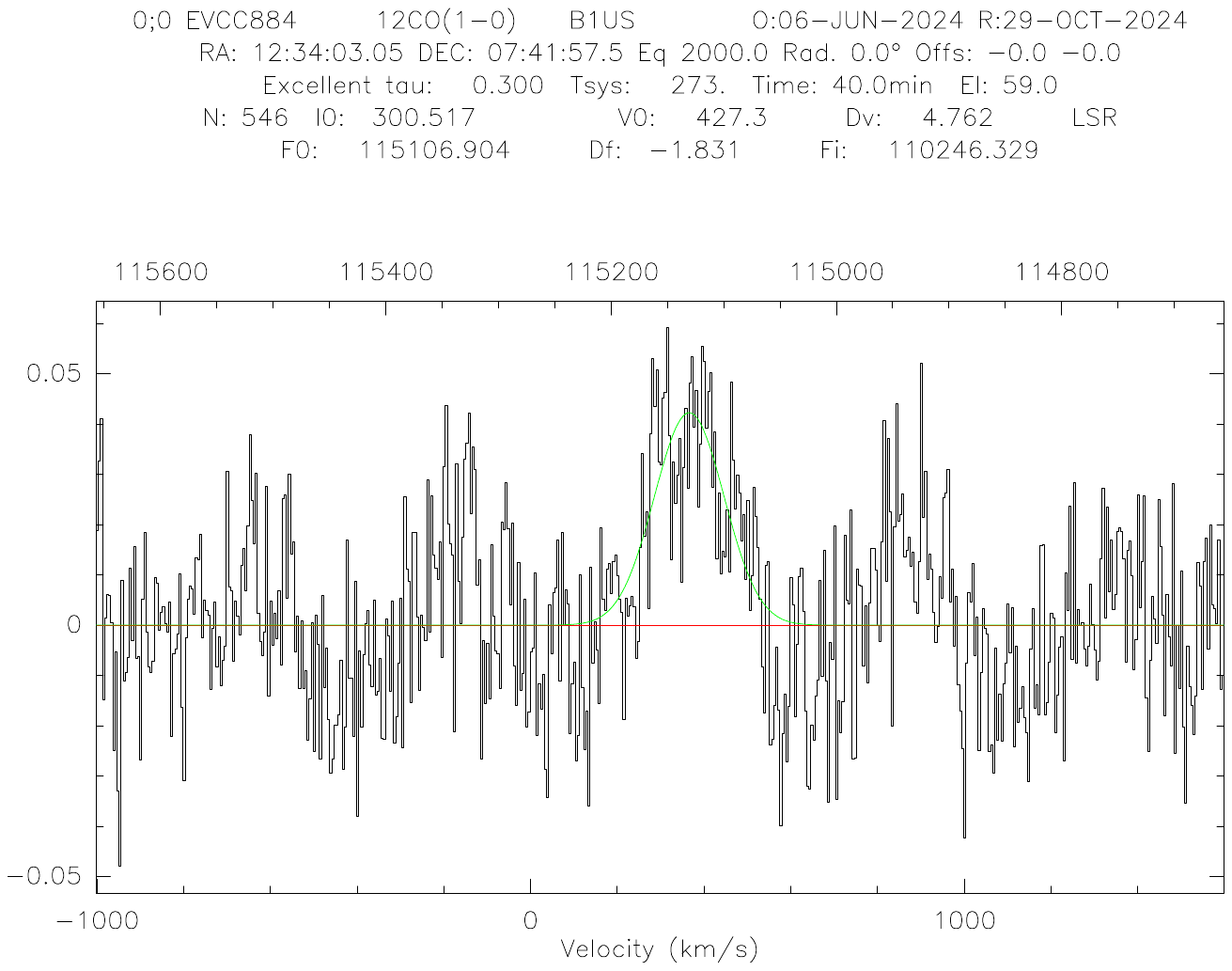}
     \addtocounter{figure}{-1} 
     \caption{(continued)} 
\end{figure}

\begin{figure}
     \centering
     \includegraphics[width=0.8\textwidth, angle=0]{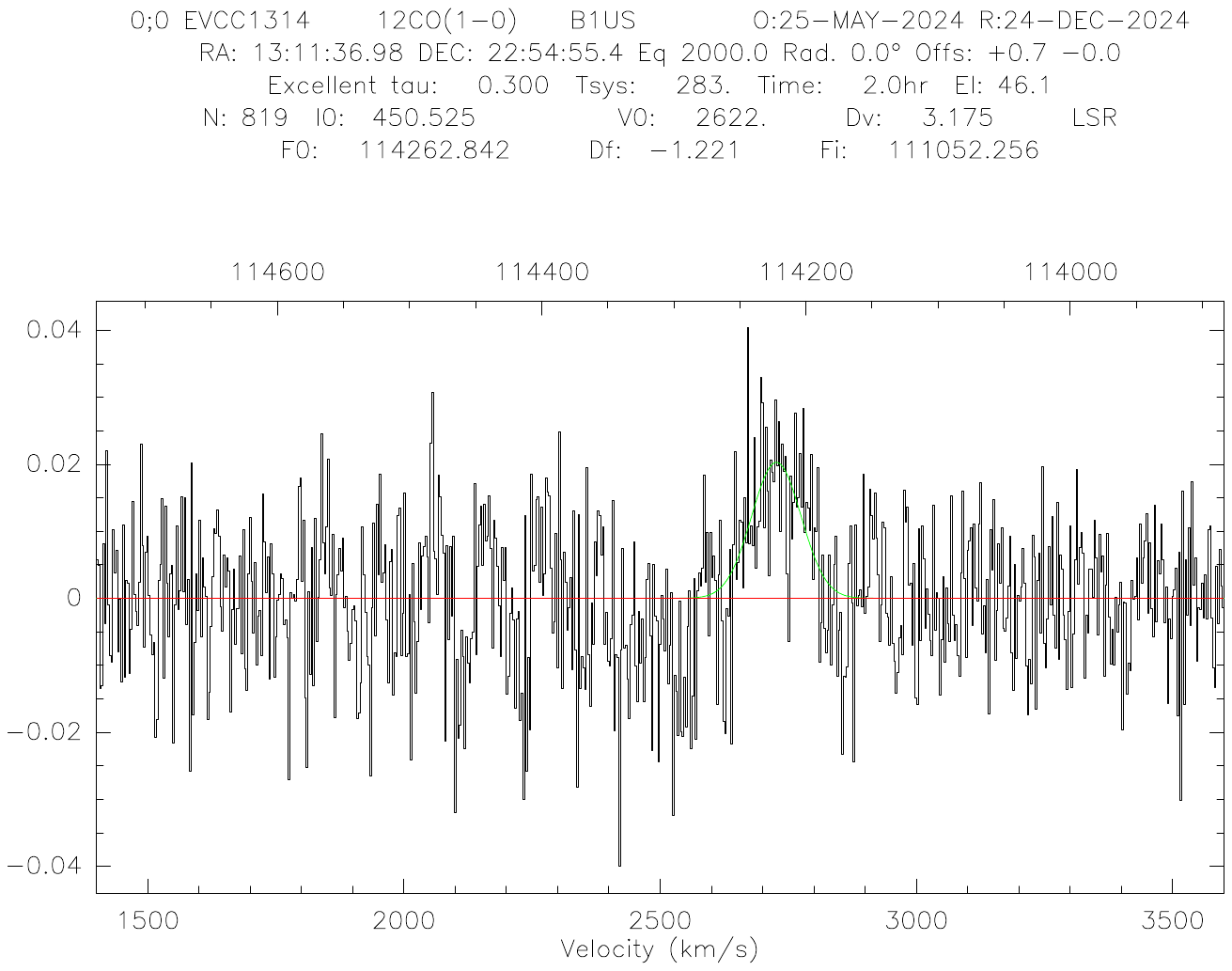}
     \includegraphics[width=0.8\textwidth, angle=0]{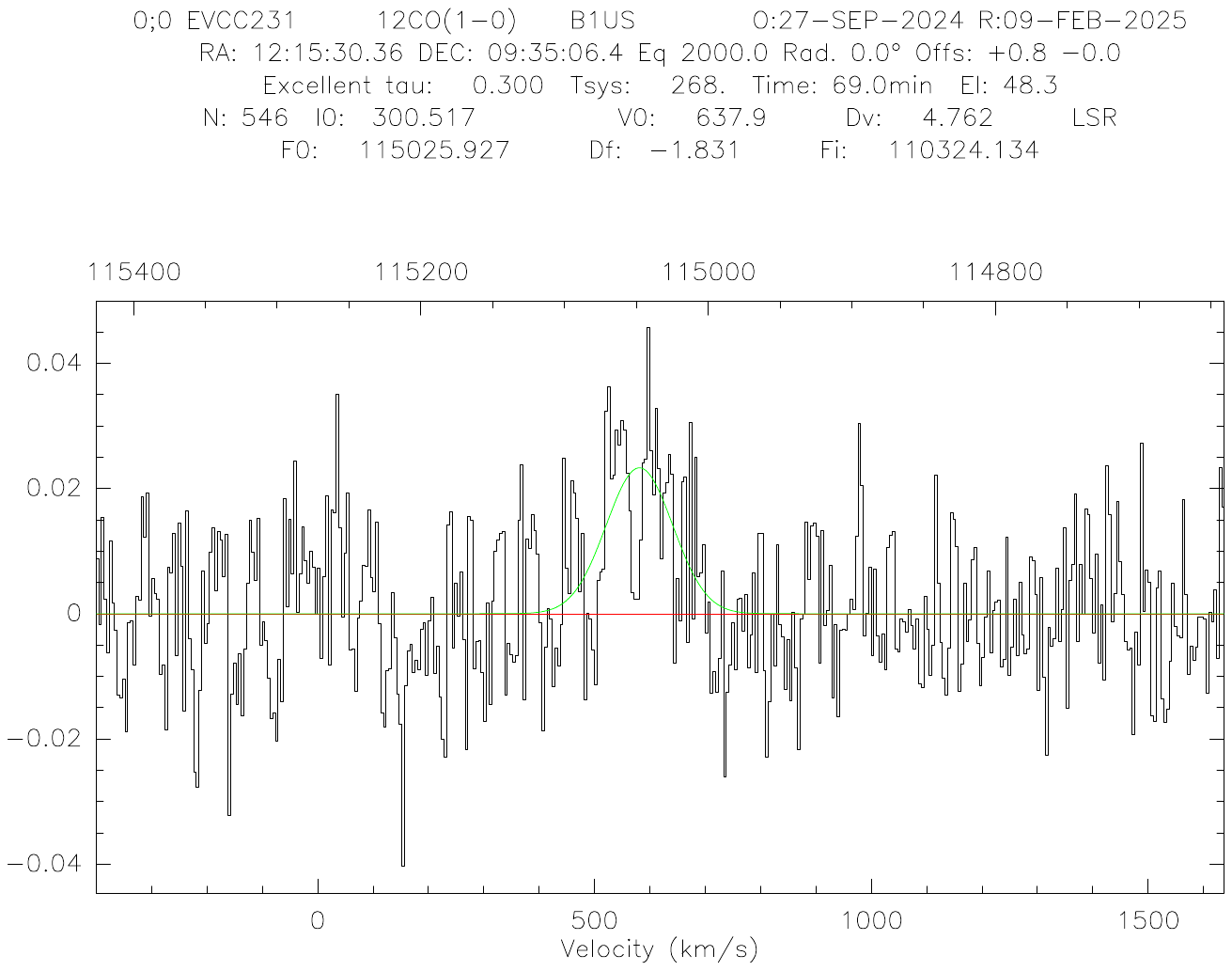}
     \addtocounter{figure}{-1} 
     \caption{(continued)} 
\end{figure}

\begin{figure}
     \centering
     \includegraphics[width=0.8\textwidth, angle=0]{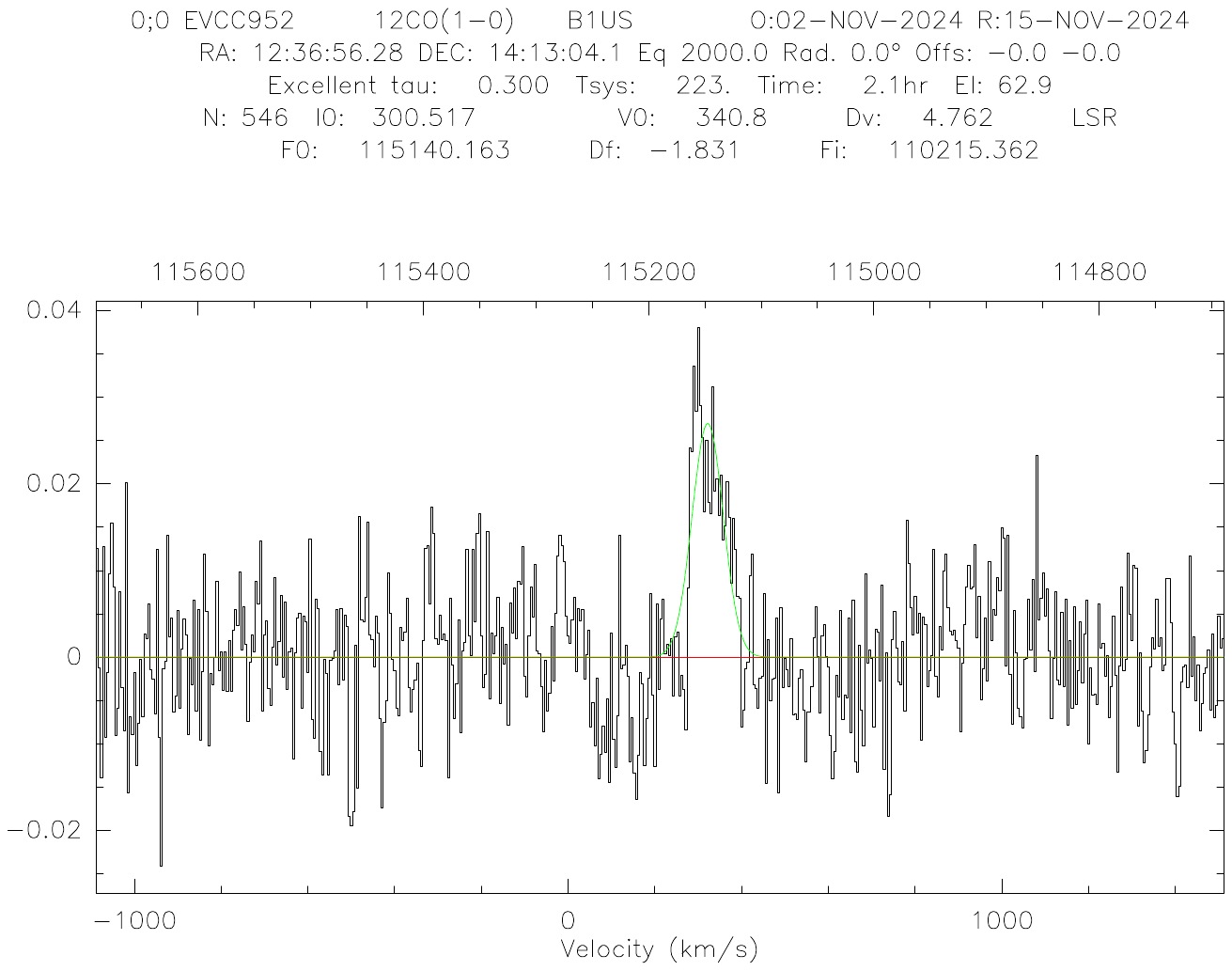}
     \includegraphics[width=0.8\textwidth, angle=0]{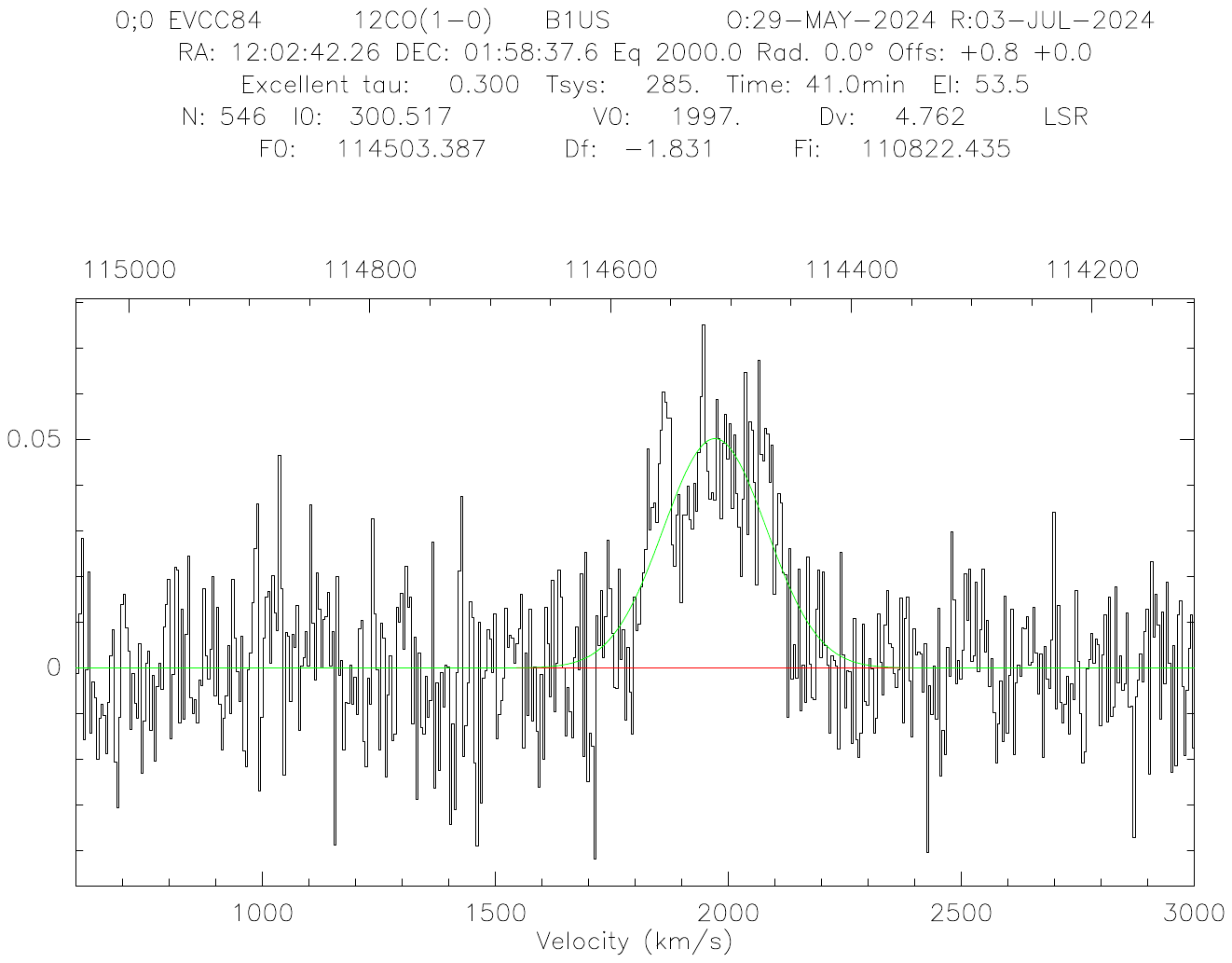}
     \addtocounter{figure}{-1} 
     \caption{(continued)} 
\end{figure}

\begin{figure}
     \centering
     \includegraphics[width=0.8\textwidth, angle=0]{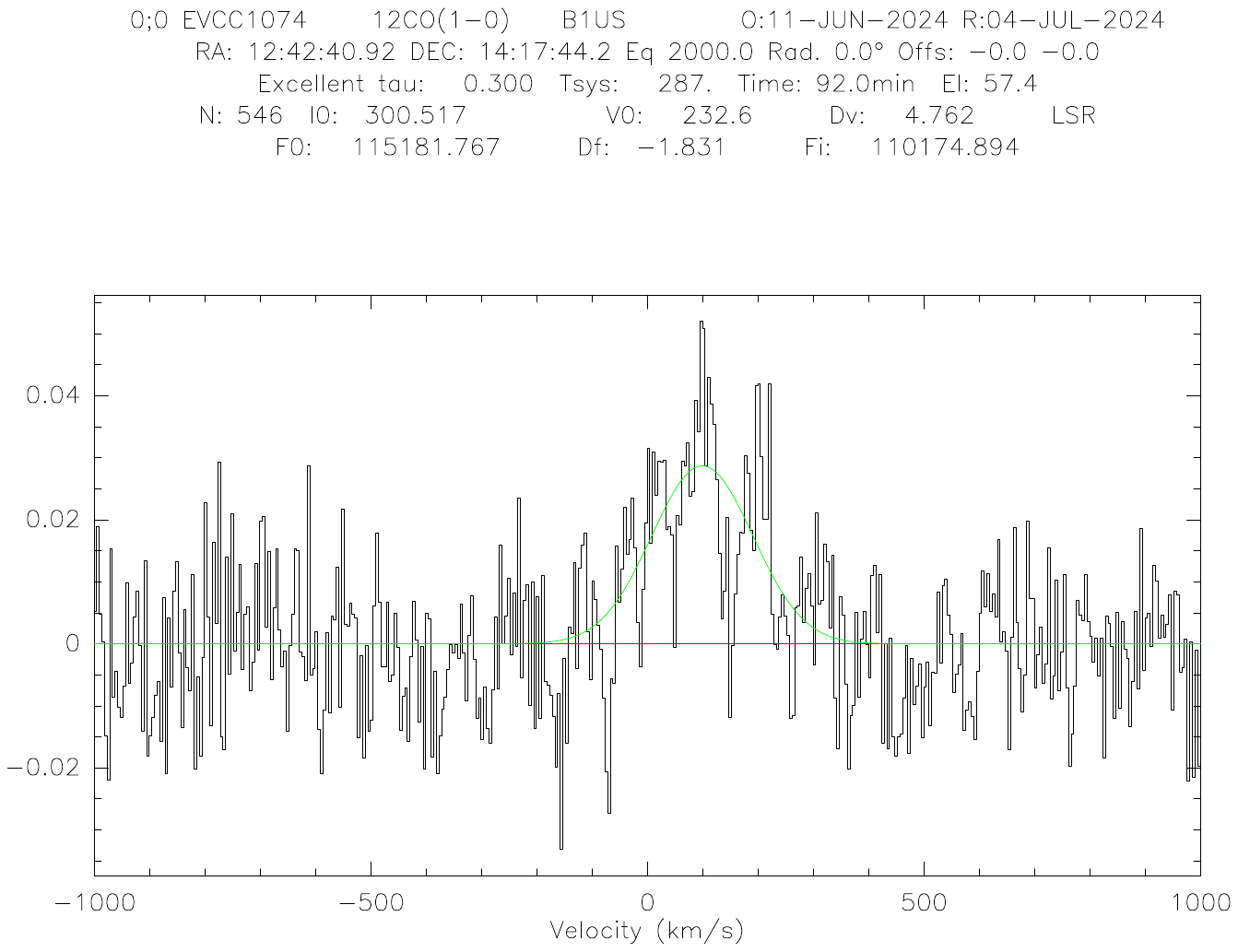}
     \includegraphics[width=0.8\textwidth, angle=0]{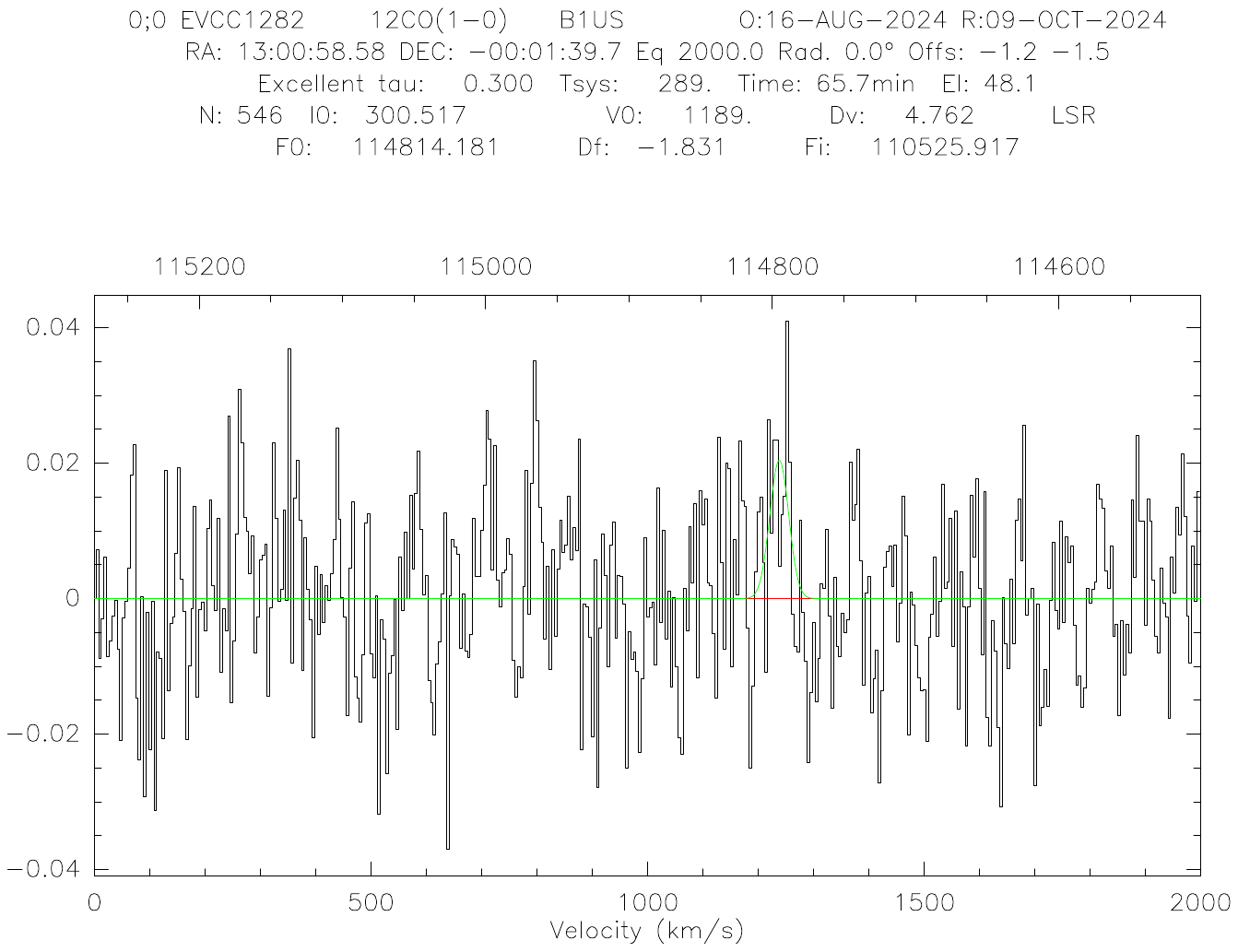}
     \addtocounter{figure}{-1} 
     \caption{(continued)} 
\end{figure}

\begin{figure}
     \centering
     \includegraphics[width=0.8\textwidth, angle=0]{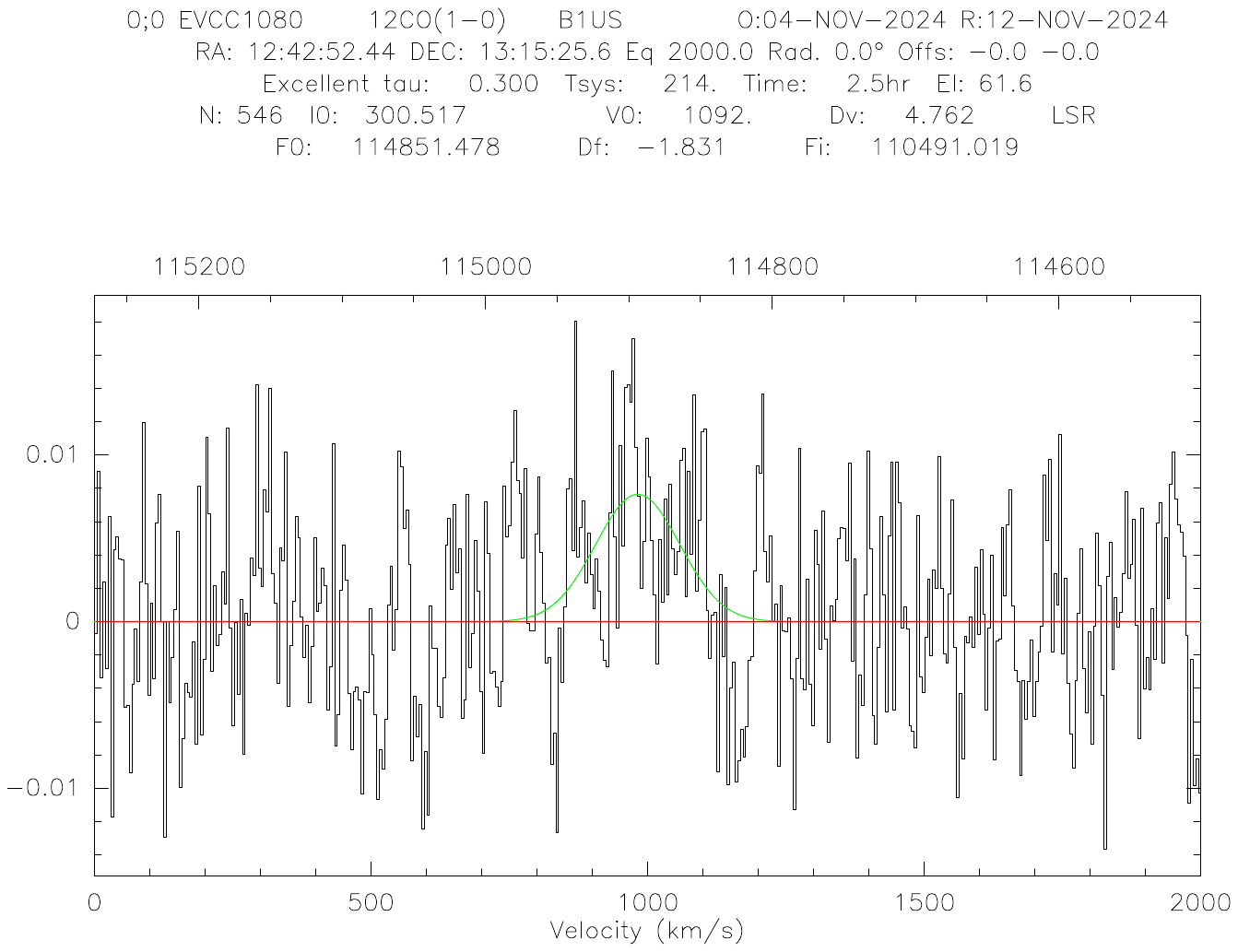}
     \includegraphics[width=0.8\textwidth, angle=0]{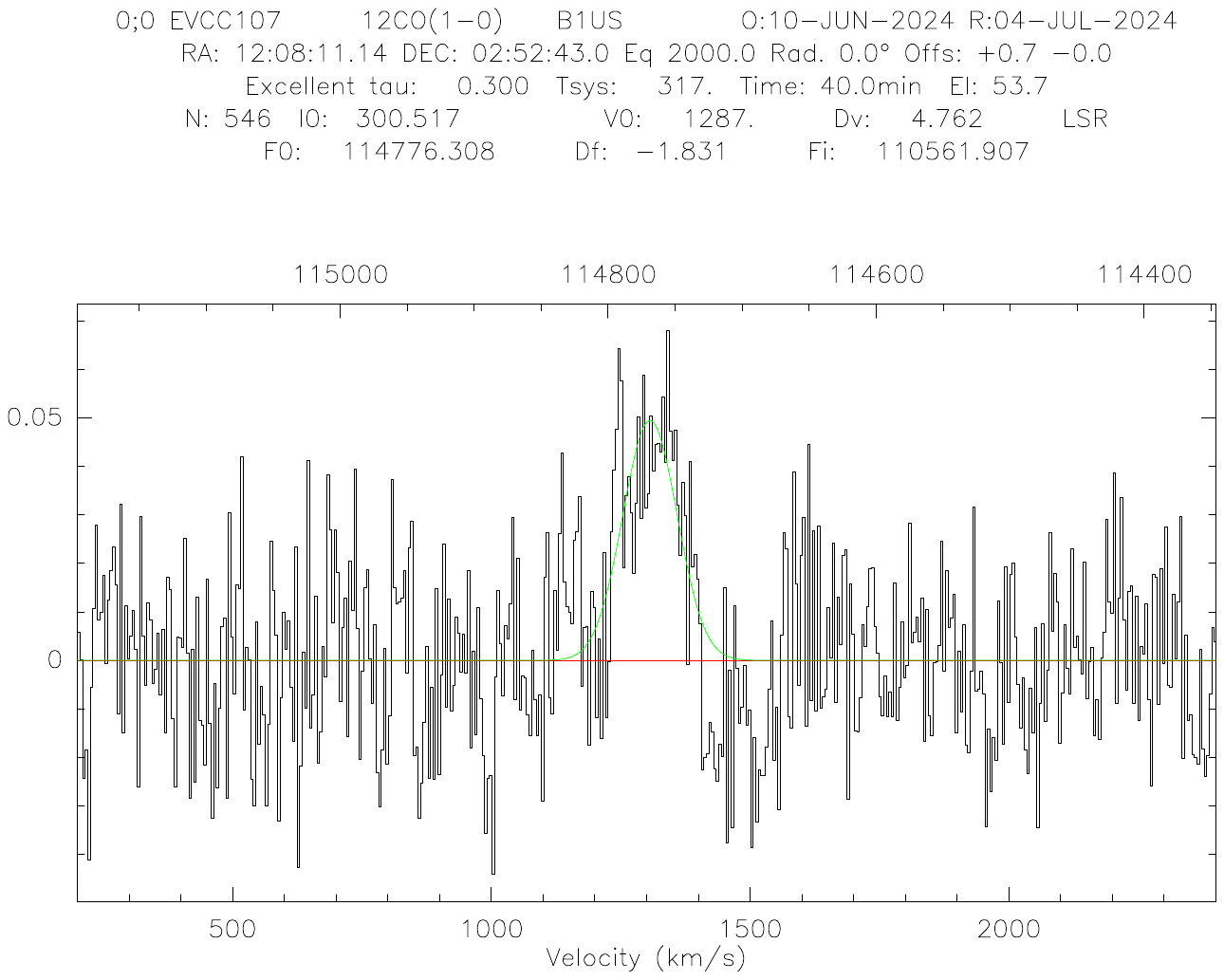}
     \addtocounter{figure}{-1} 
     \caption{(continued)} 
\end{figure}

\begin{figure}
     \centering
     \includegraphics[width=0.8\textwidth, angle=0]{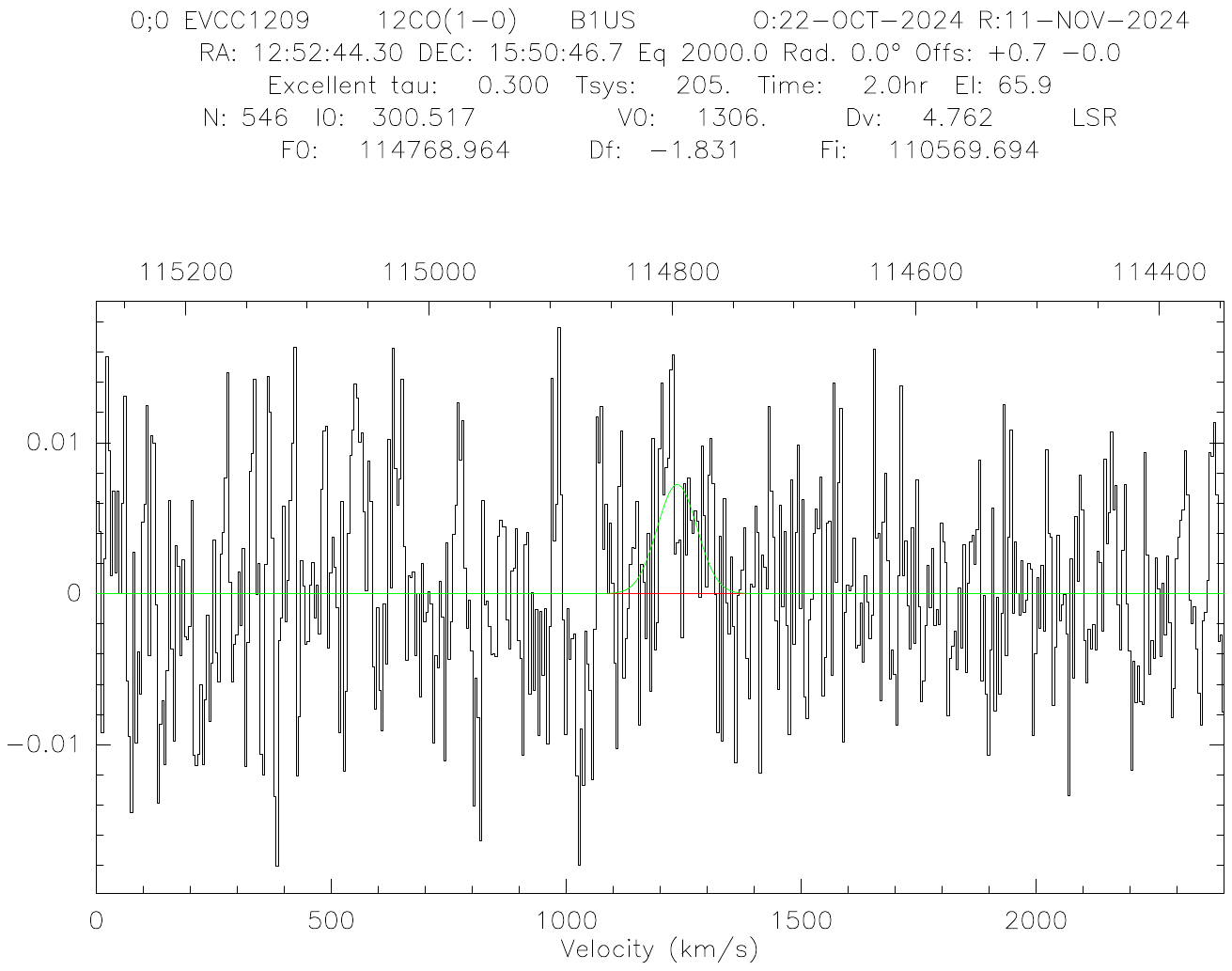}
     \includegraphics[width=0.8\textwidth, angle=0]{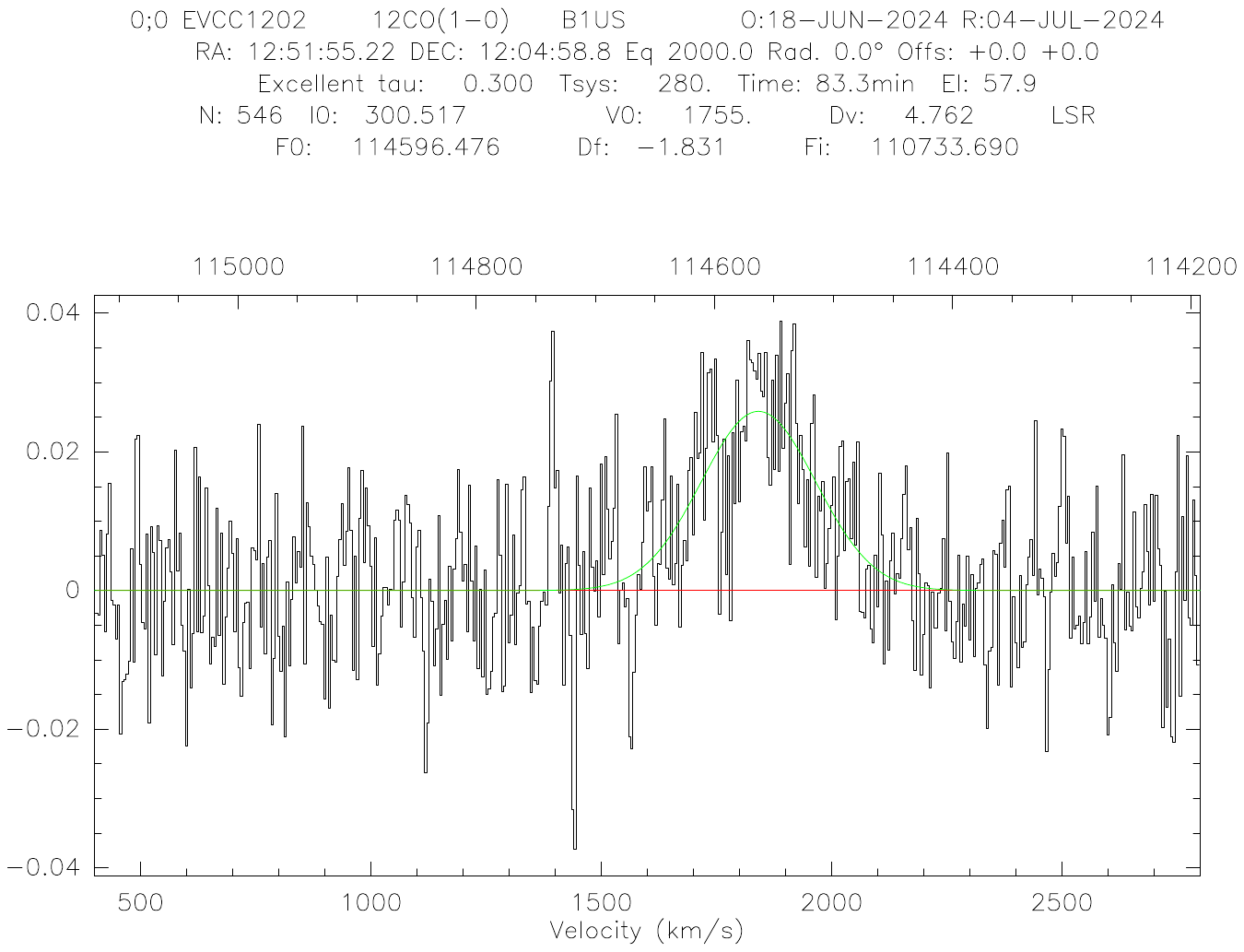}
     \addtocounter{figure}{-1} 
     \caption{(continued)} 
\end{figure}

\begin{figure}
     \centering
     \includegraphics[width=0.8\textwidth, angle=0]{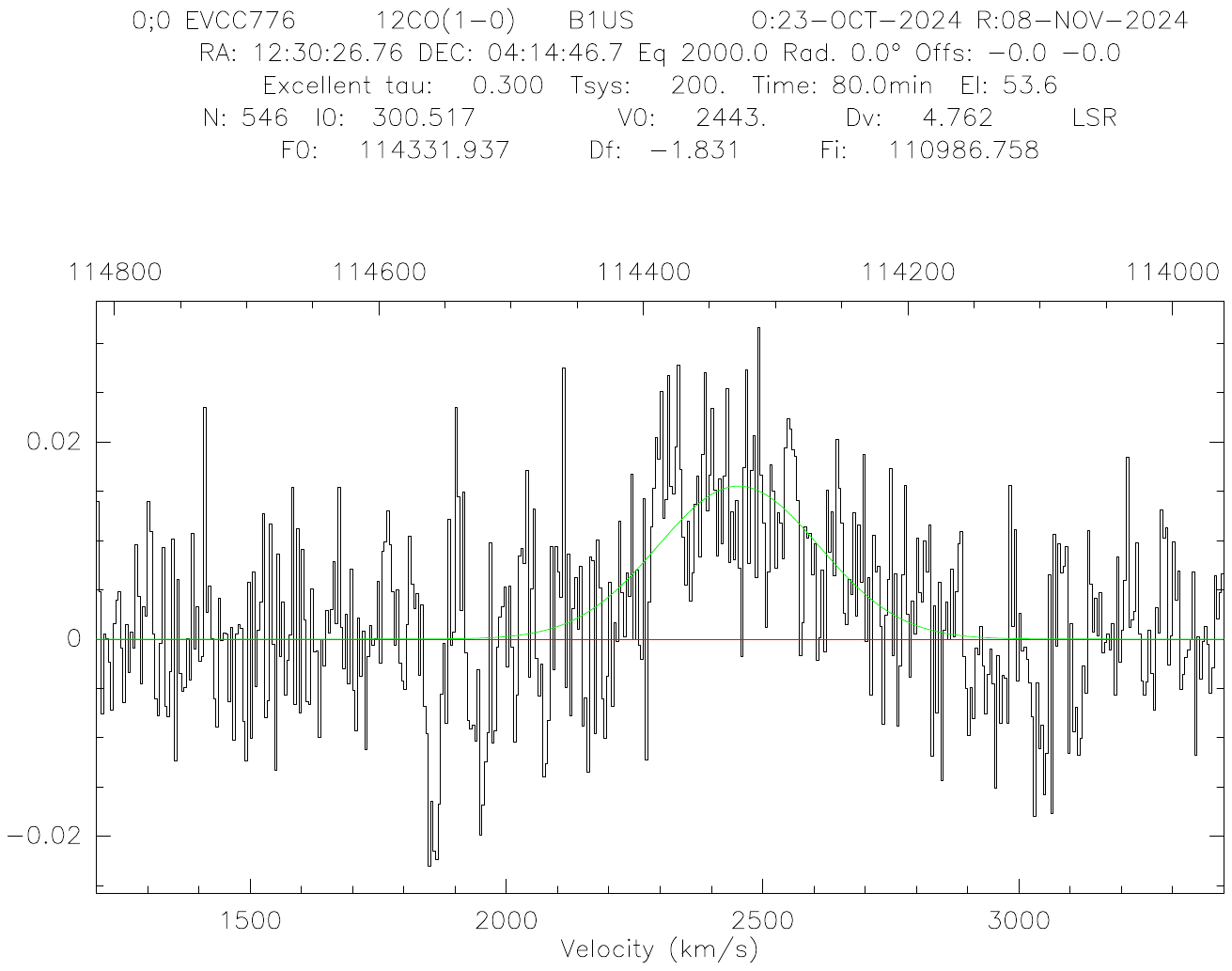}
     \includegraphics[width=0.8\textwidth, angle=0]{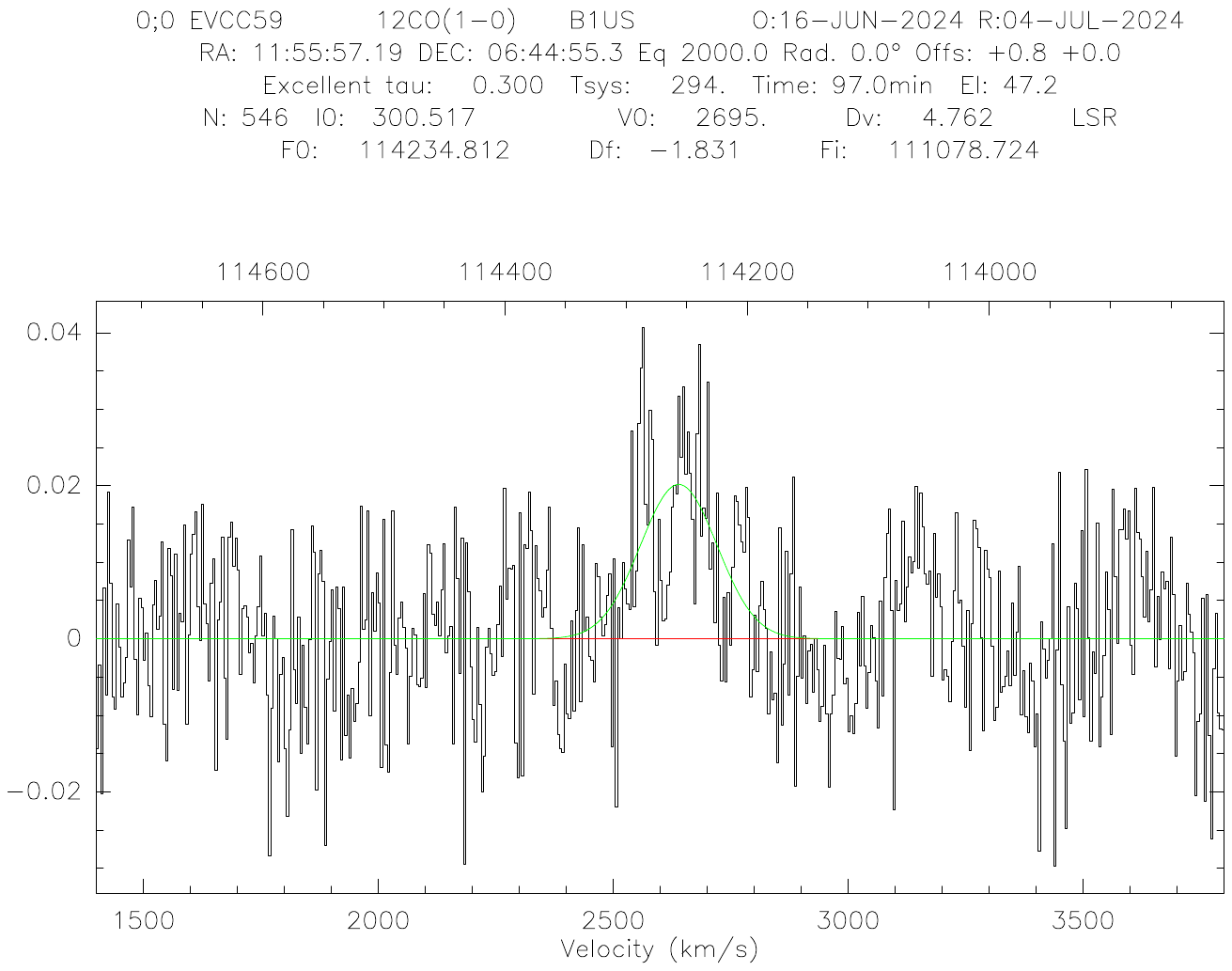}
     \addtocounter{figure}{-1} 
     \caption{(continued)} 
\end{figure}

\begin{figure}
     \centering
     \includegraphics[width=0.8\textwidth, angle=0]{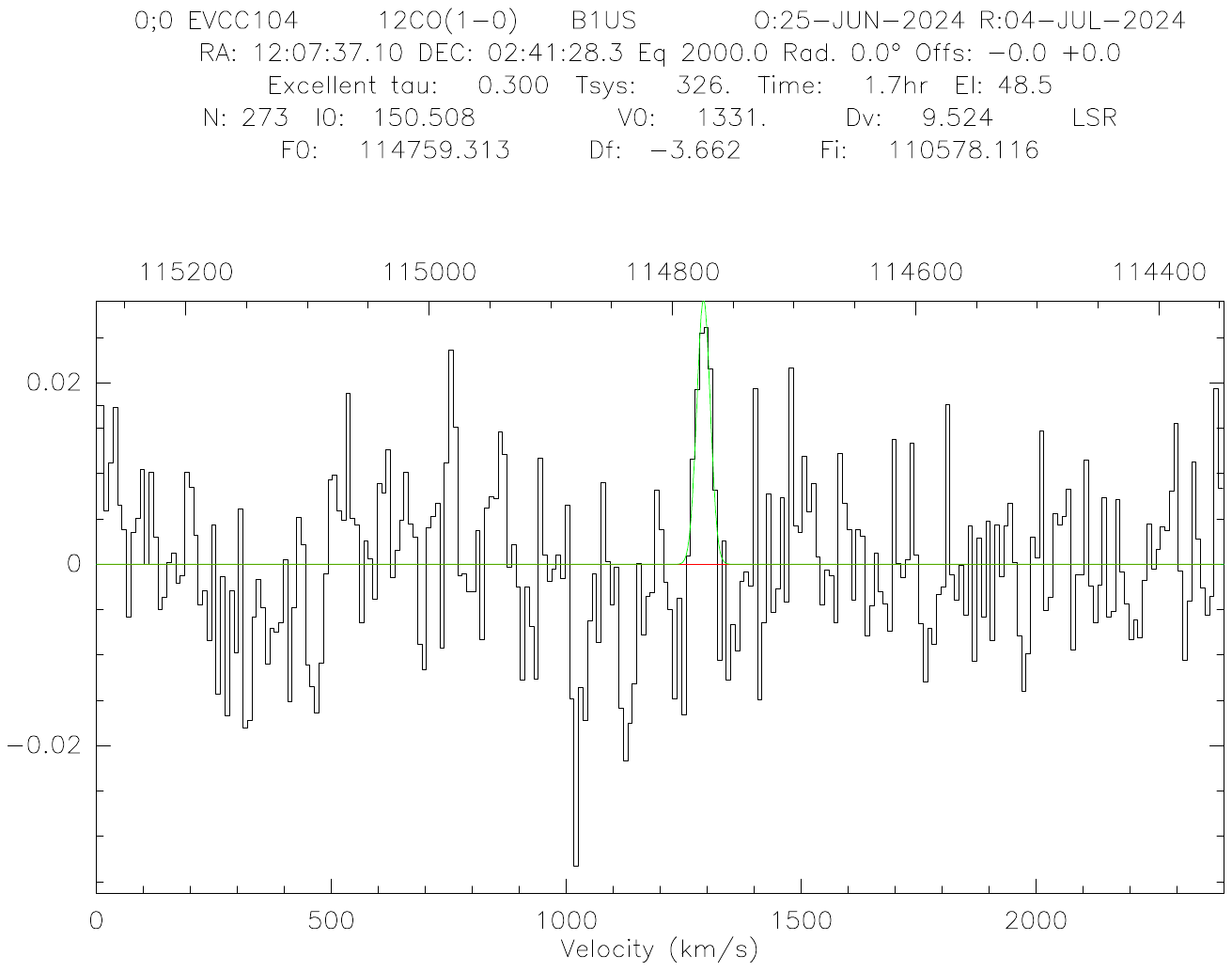}
     \includegraphics[width=0.8\textwidth, angle=0]{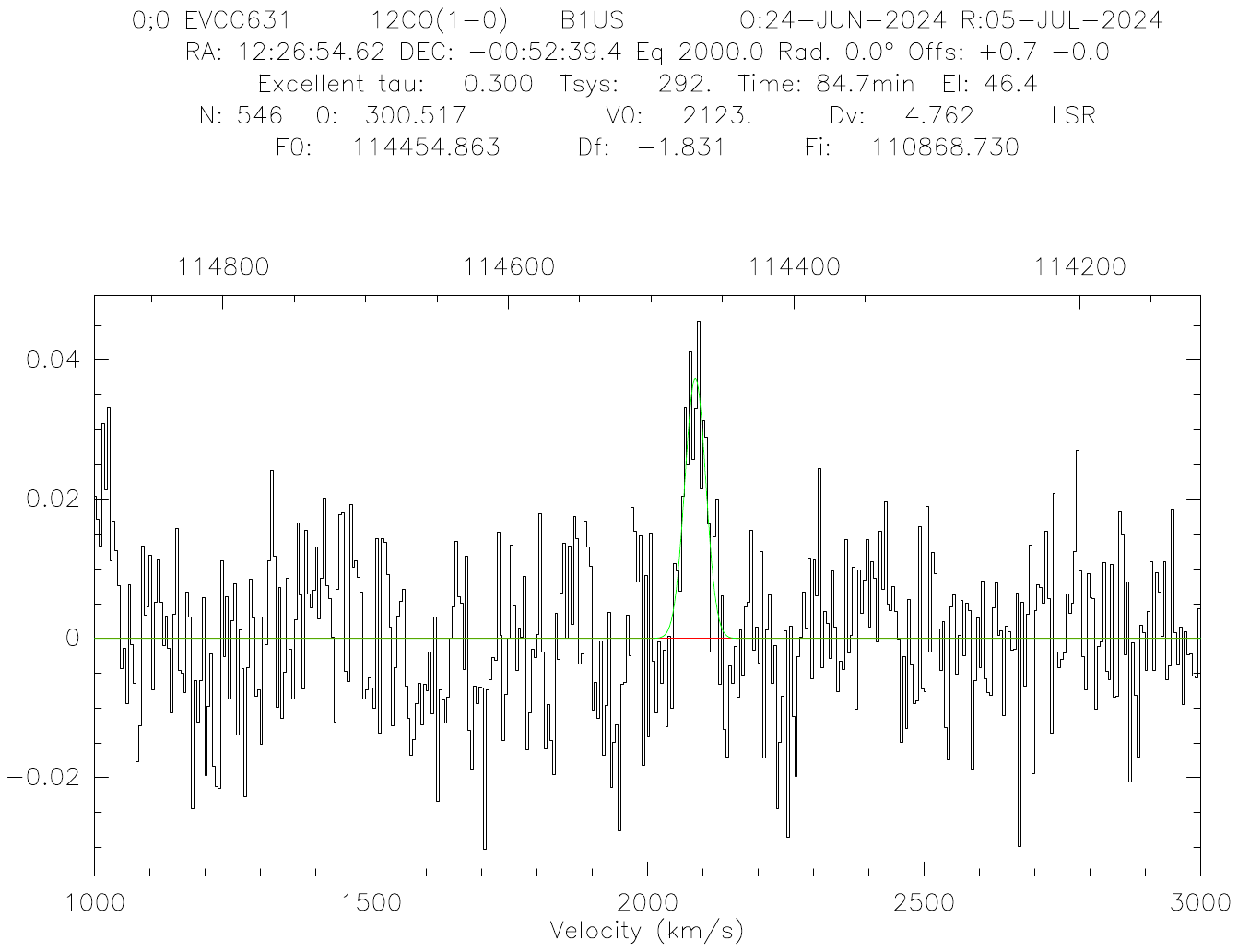}
     \addtocounter{figure}{-1} 
     \caption{(continued)} 
\end{figure}

\begin{figure}
     \centering
     \includegraphics[width=0.8\textwidth, angle=0]{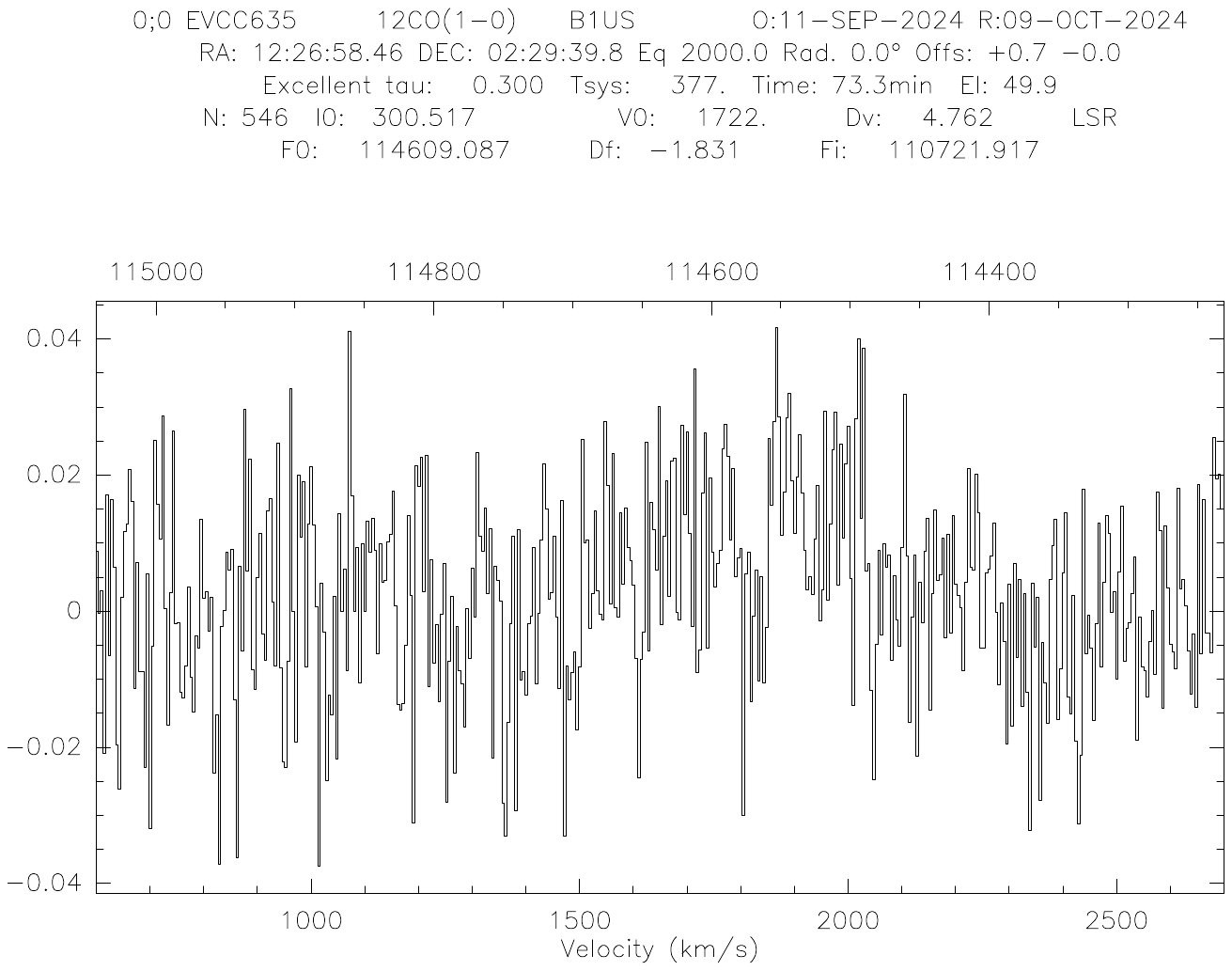}
     \includegraphics[width=0.8\textwidth, angle=0]{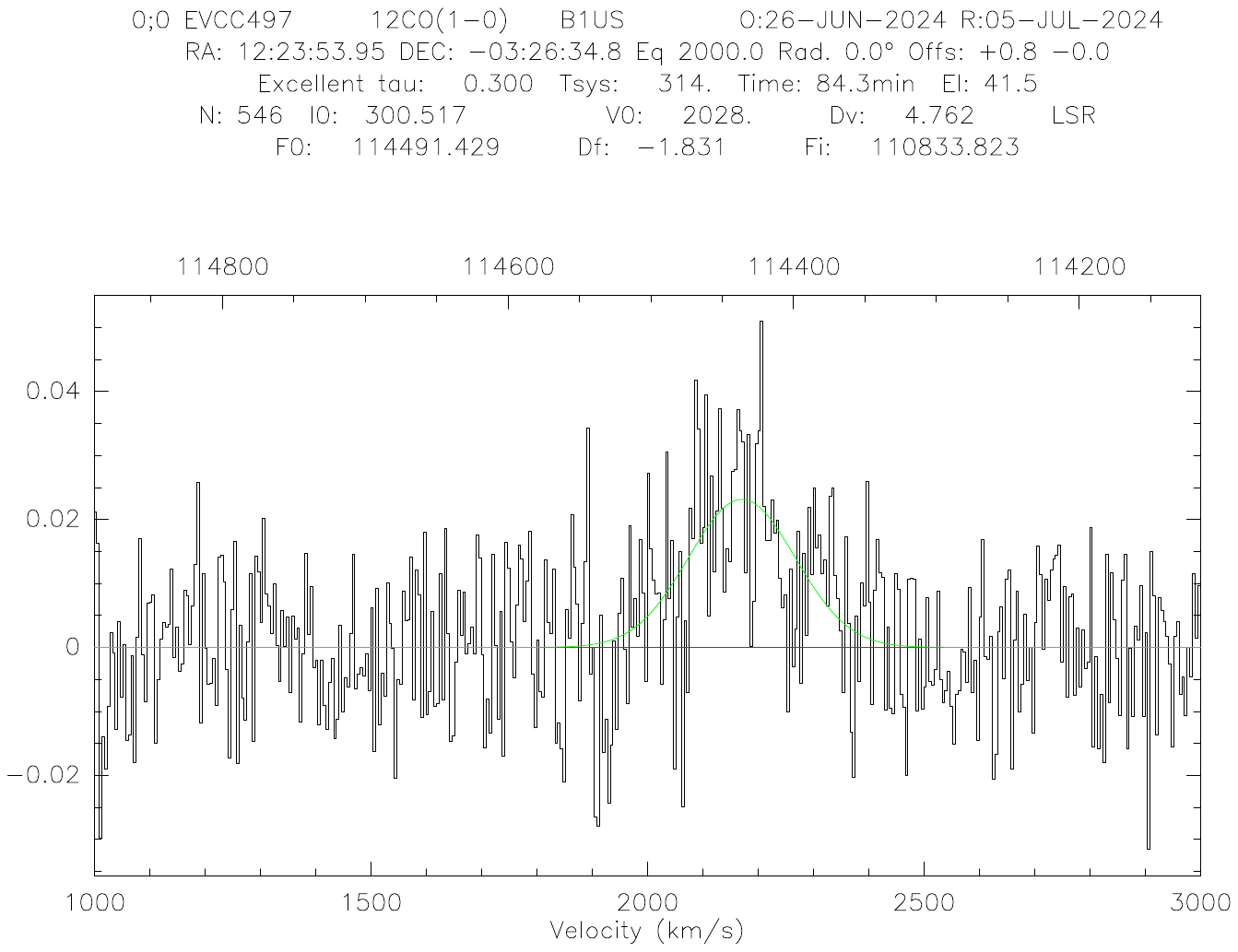}
     \addtocounter{figure}{-1} 
     \caption{(continued)} 
\end{figure}

\begin{figure}
     \centering
     \includegraphics[width=0.8\textwidth, angle=0]{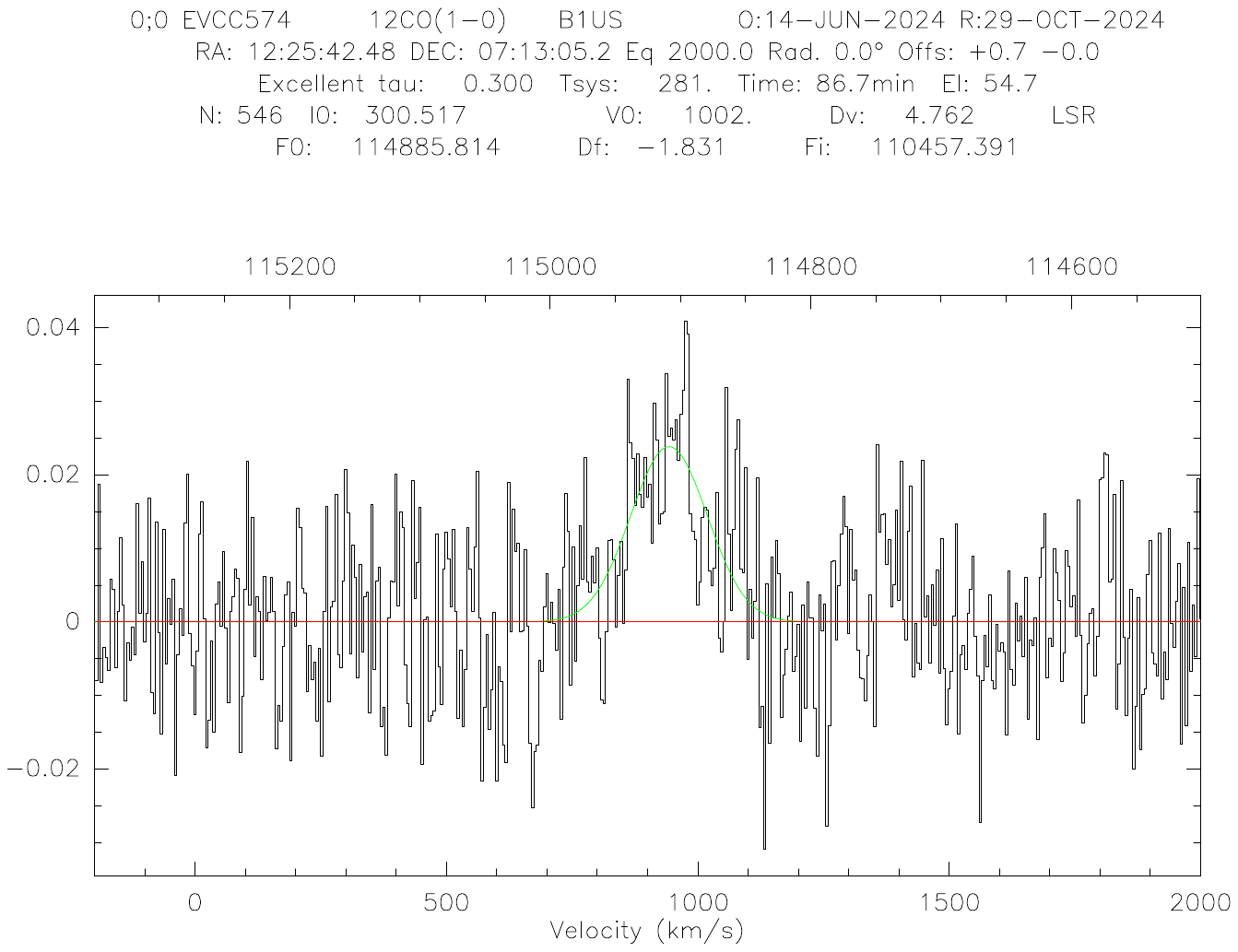}
     \includegraphics[width=0.8\textwidth, angle=0]{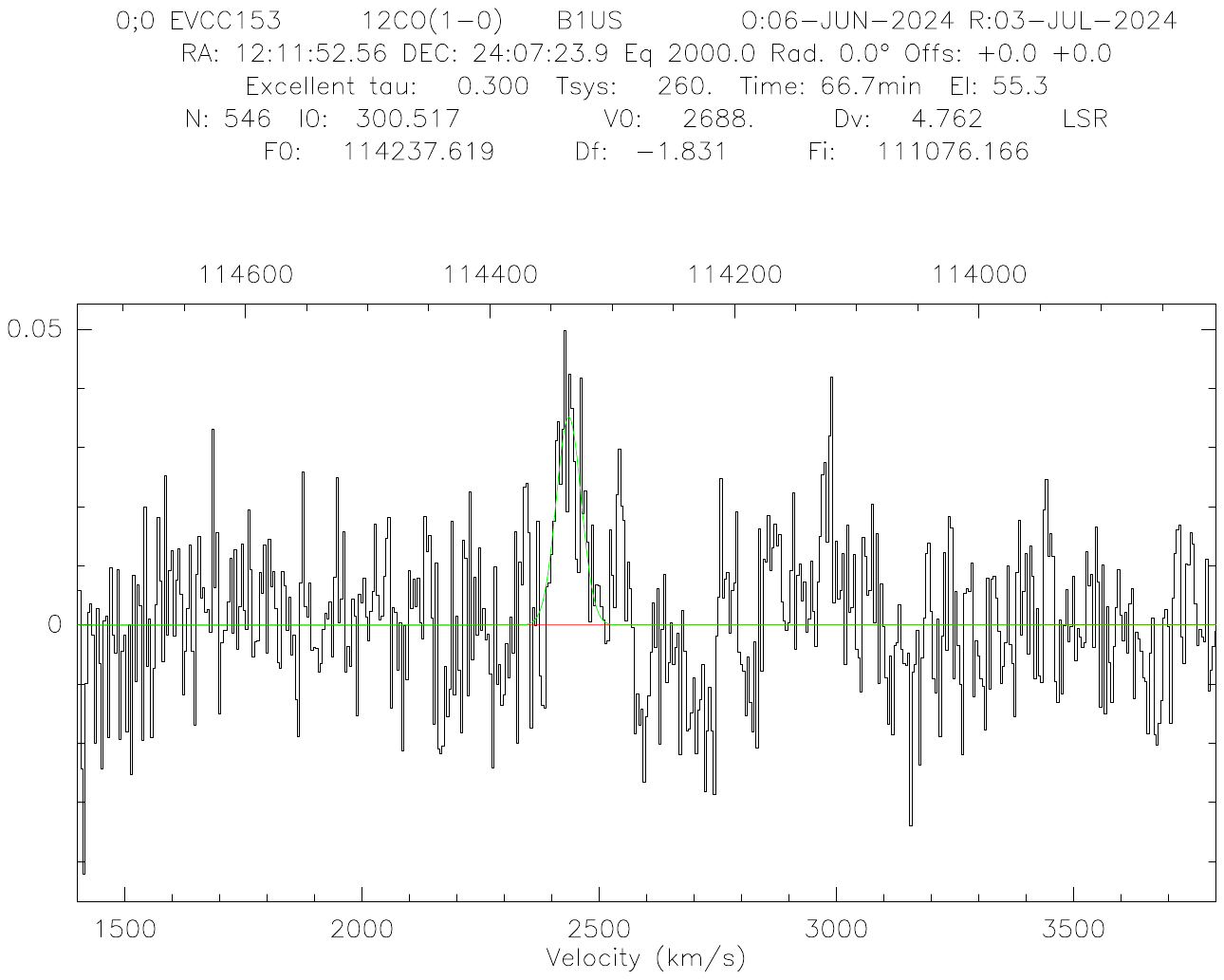}
     \addtocounter{figure}{-1} 
     \caption{(continued)} 
\end{figure}

\begin{figure}
     \centering
     \includegraphics[width=0.8\textwidth, angle=0]{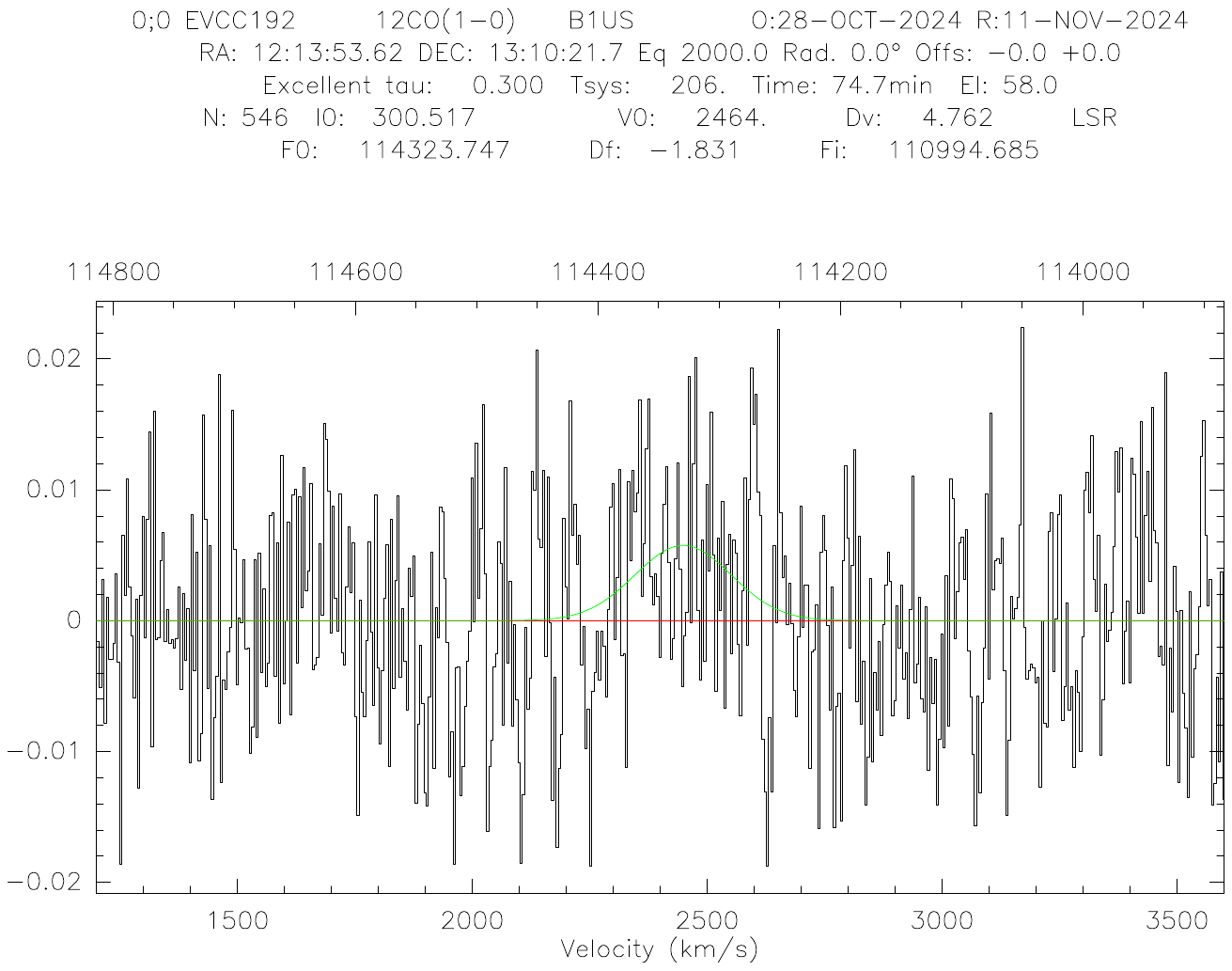}
     \includegraphics[width=0.8\textwidth, angle=0]{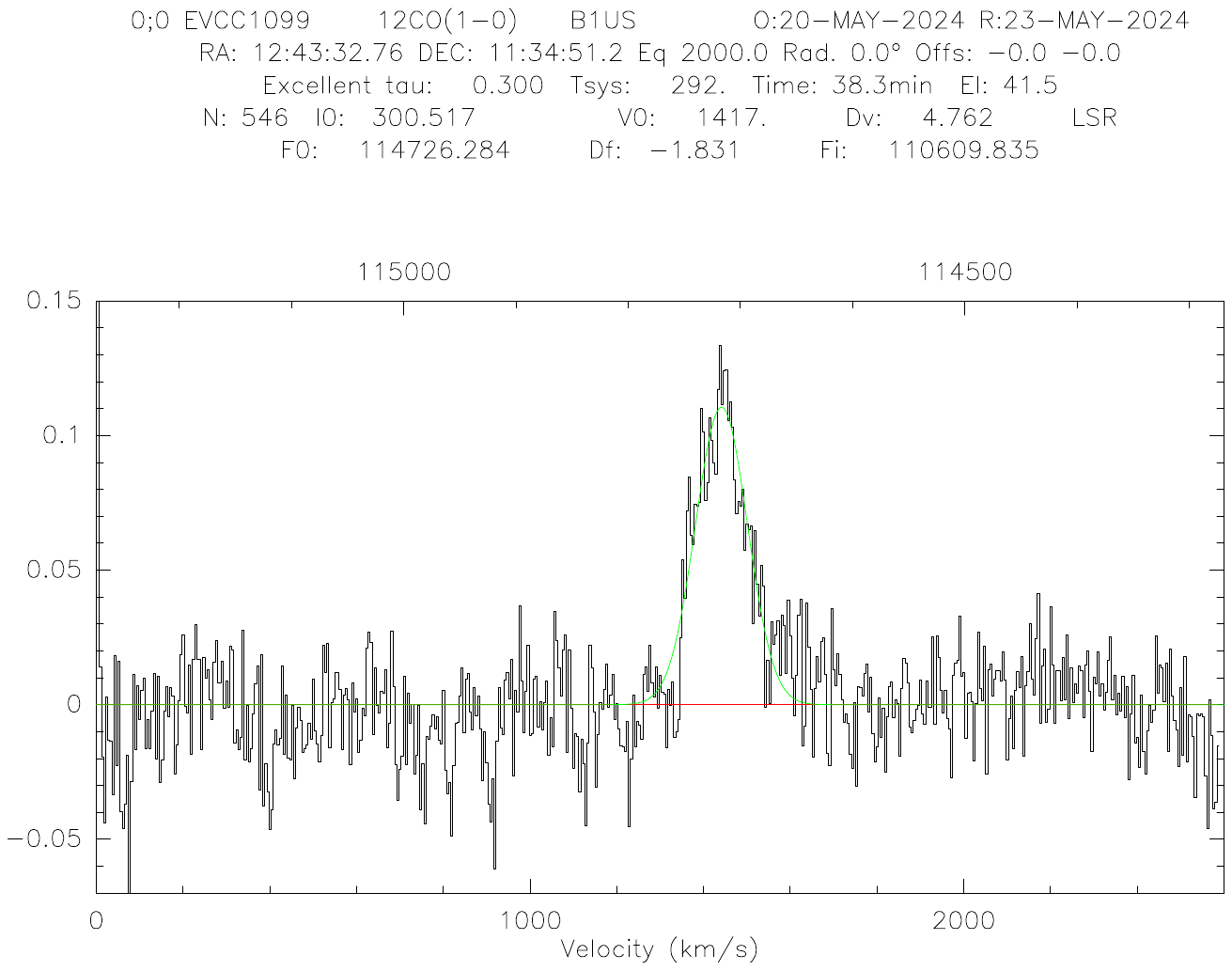} 
     \addtocounter{figure}{-1} 
     \caption{(continued)} 
\end{figure}

\begin{figure}
     \centering
     \includegraphics[width=0.8\textwidth, angle=0]{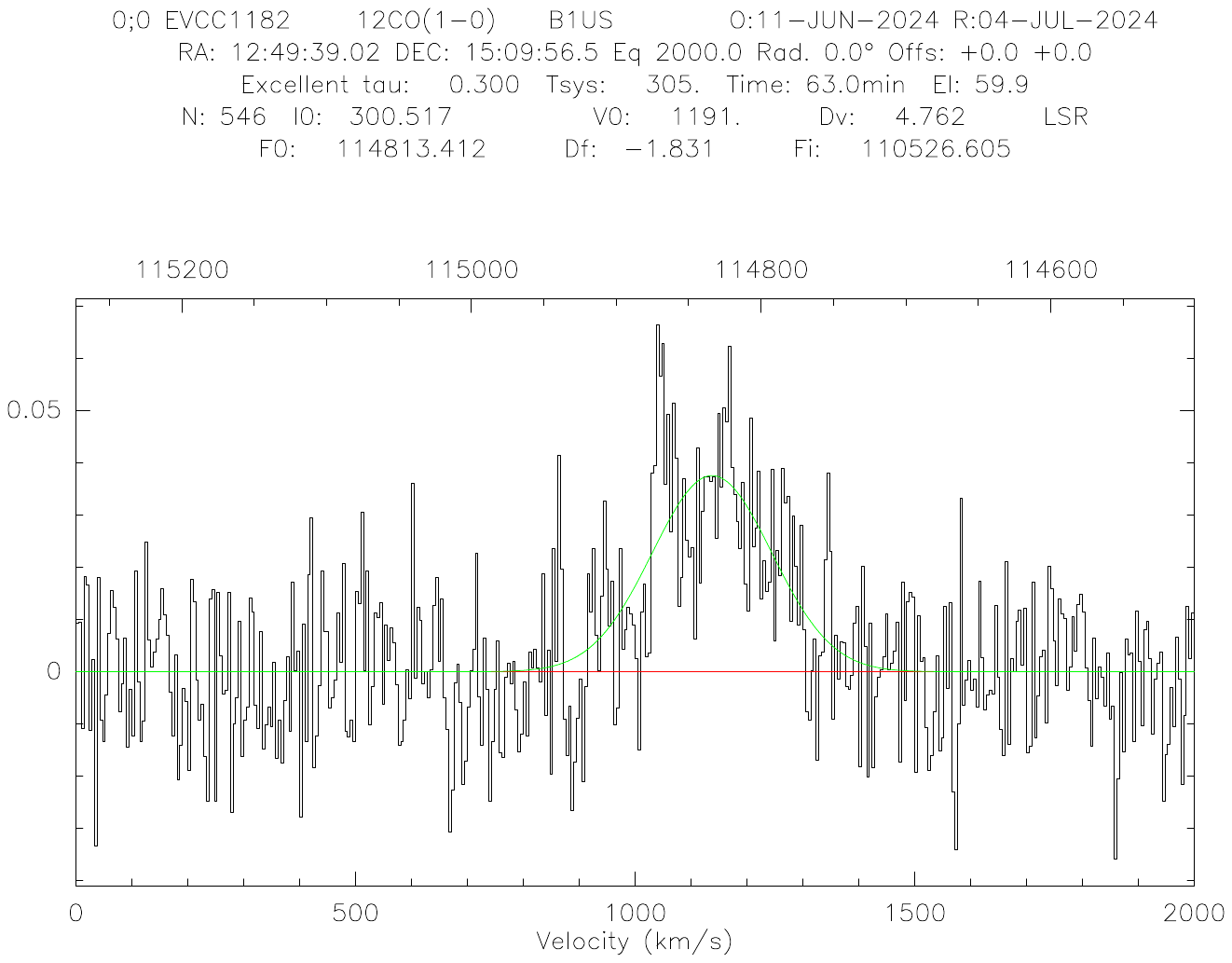}
     \includegraphics[width=0.8\textwidth, angle=0]{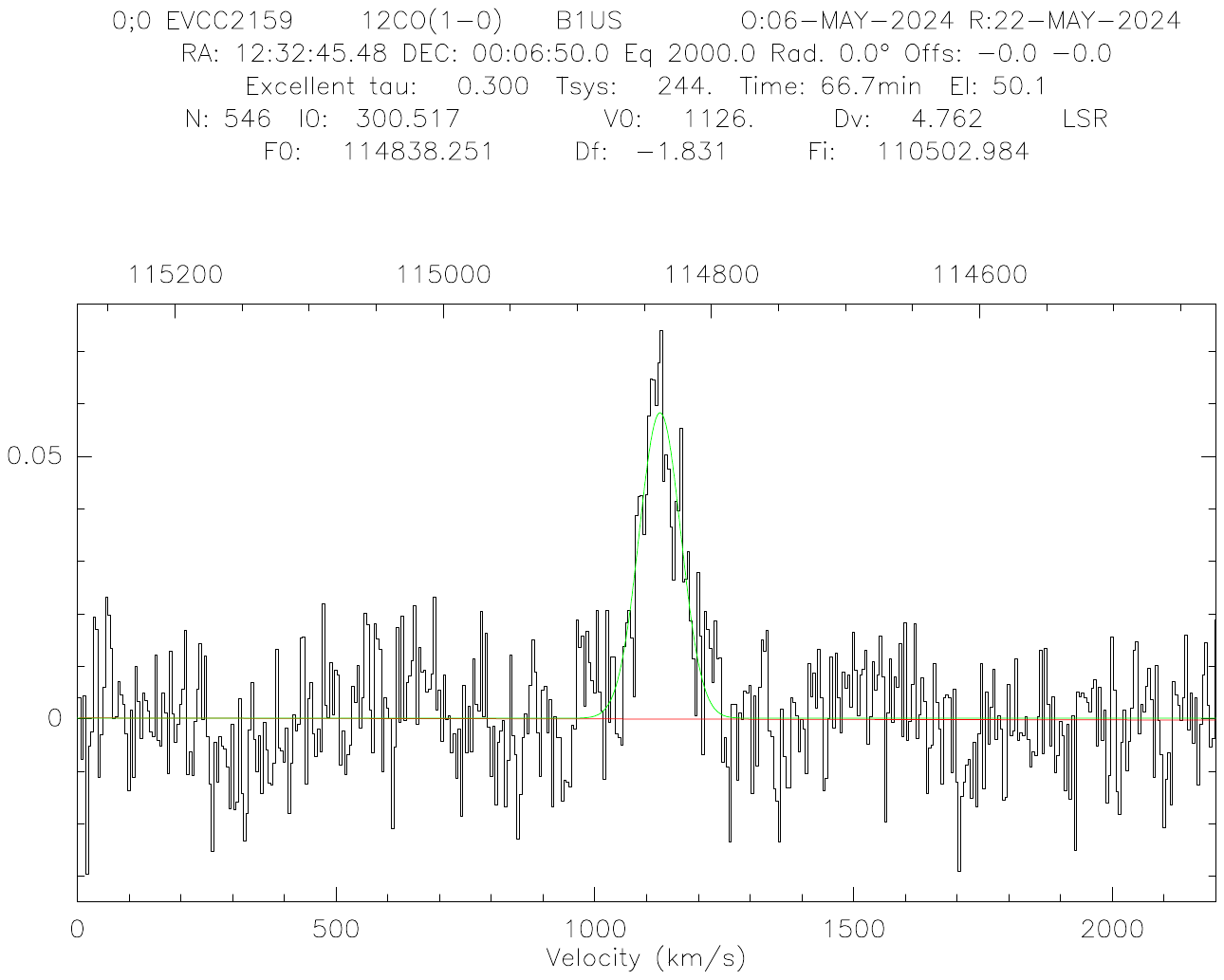}
     \addtocounter{figure}{-1} 
     \caption{(continued)} 
\end{figure}

\begin{figure}
     \centering
     \includegraphics[width=0.8\textwidth, angle=0]{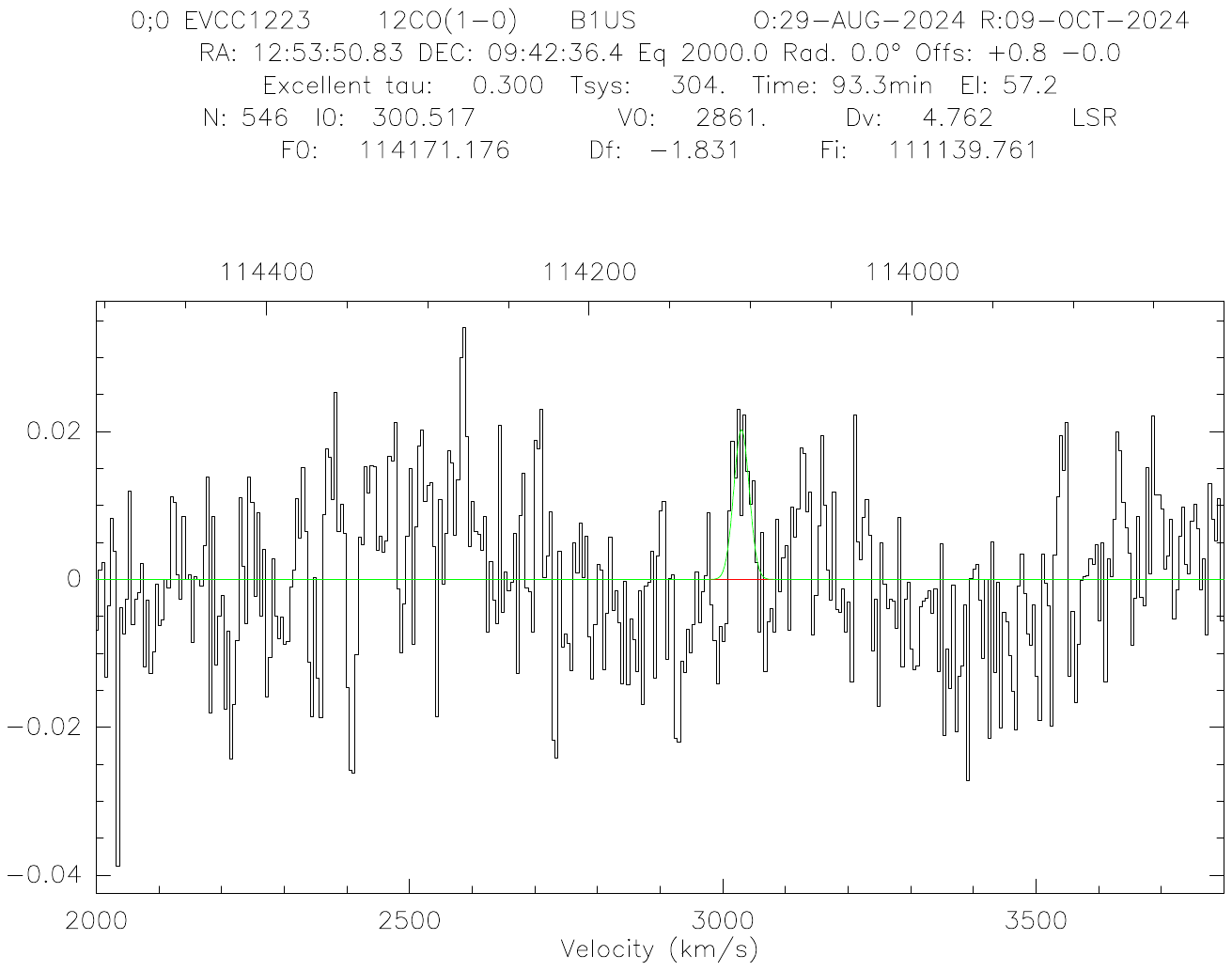}
     \includegraphics[width=0.8\textwidth, angle=0]{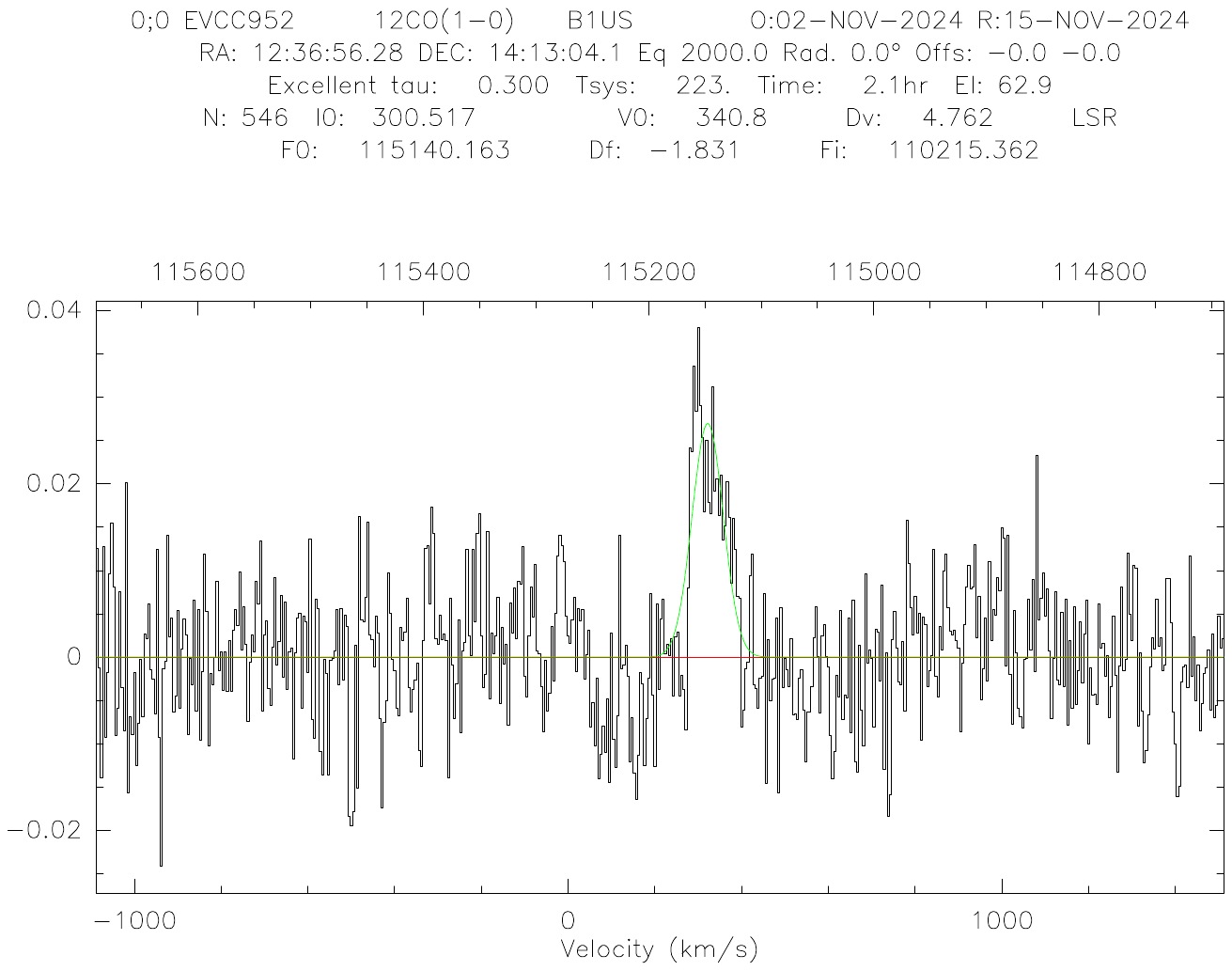}
     \addtocounter{figure}{-1} 
     \caption{(continued)} 
\end{figure}

\begin{figure}
     \centering
     \includegraphics[width=0.8\textwidth, angle=0]{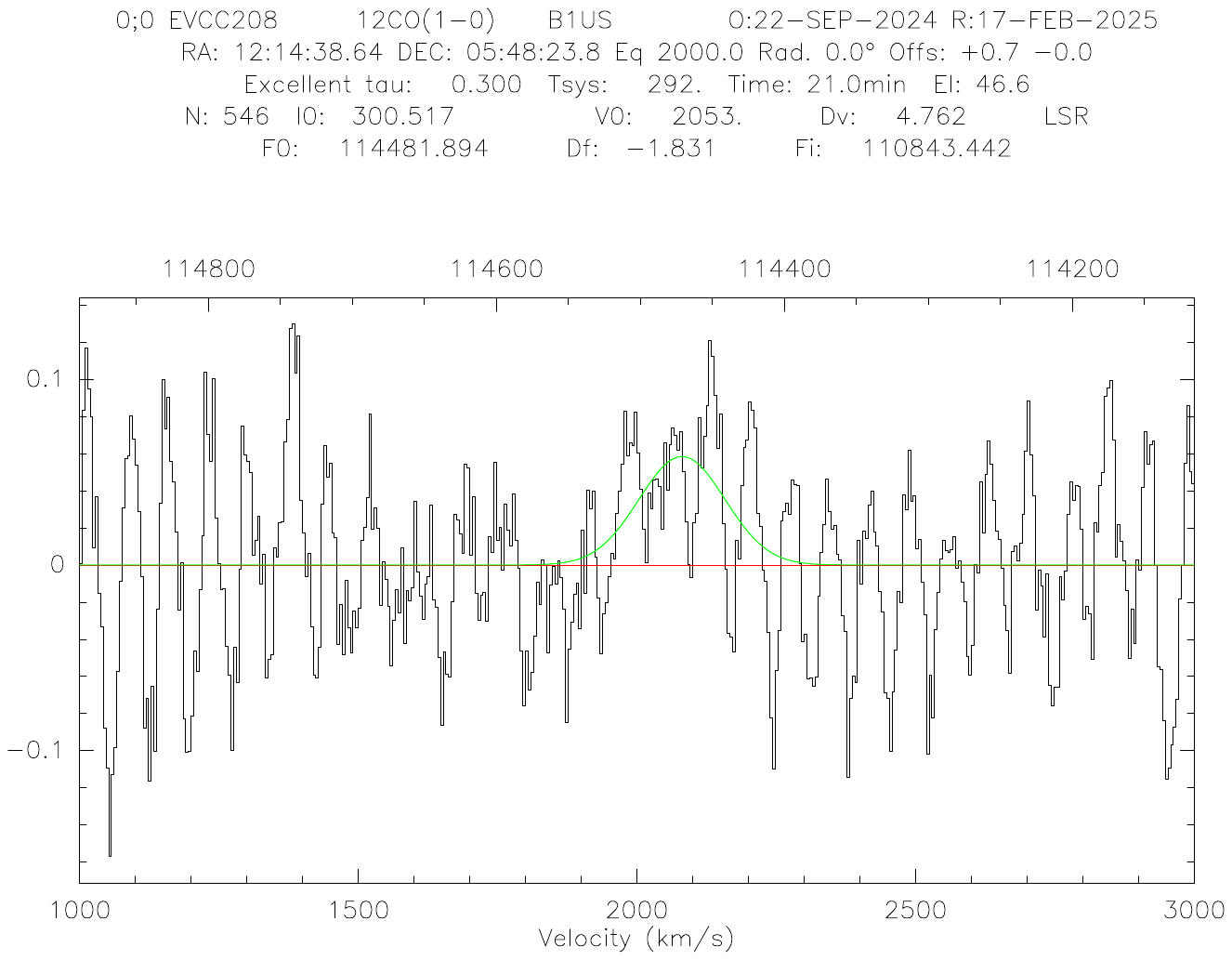}
     \includegraphics[width=0.8\textwidth, angle=0]{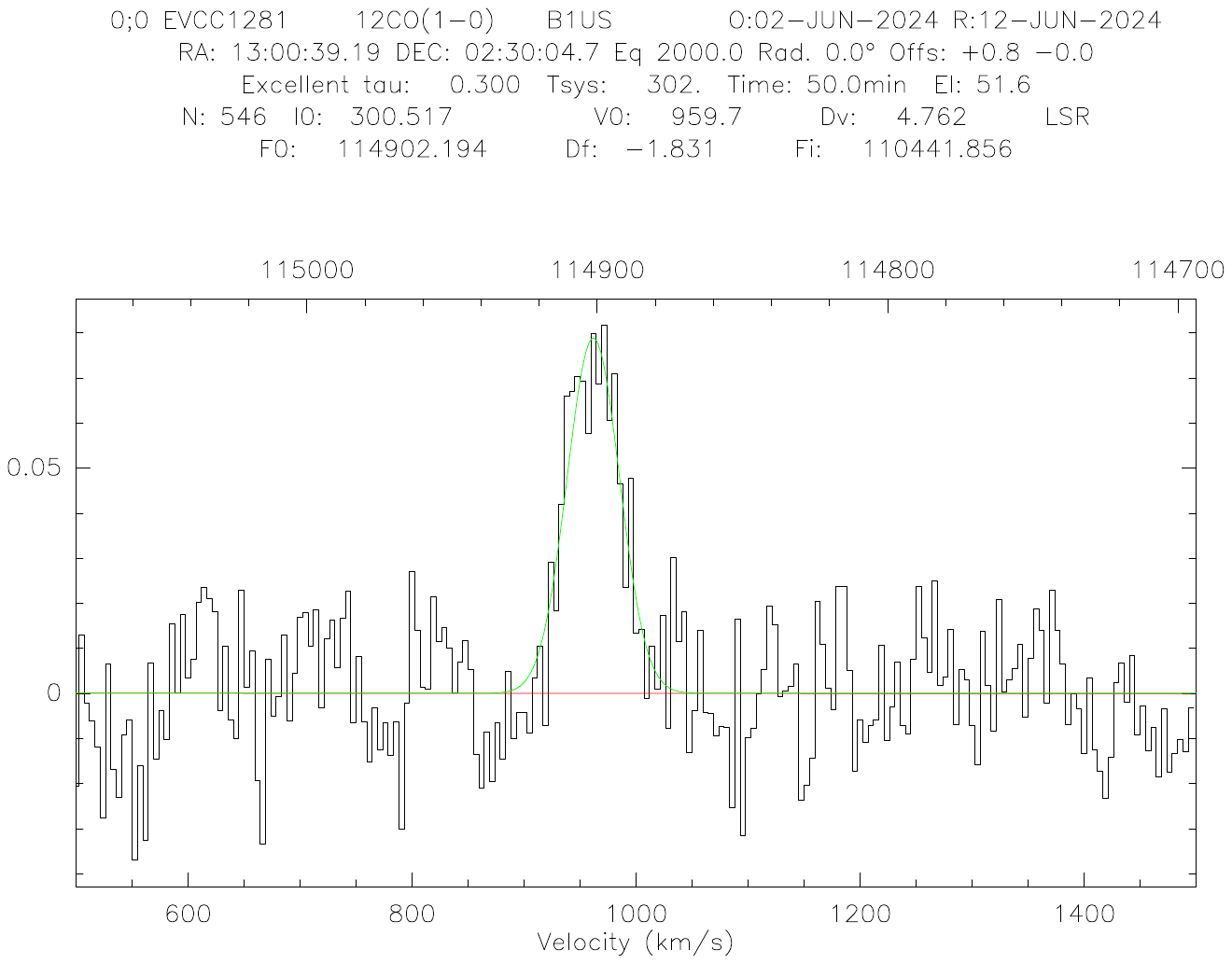}
     \addtocounter{figure}{-1} 
     \caption{(continued)} 
\end{figure}

\begin{figure}
     \centering
     \includegraphics[width=0.8\textwidth, angle=0]{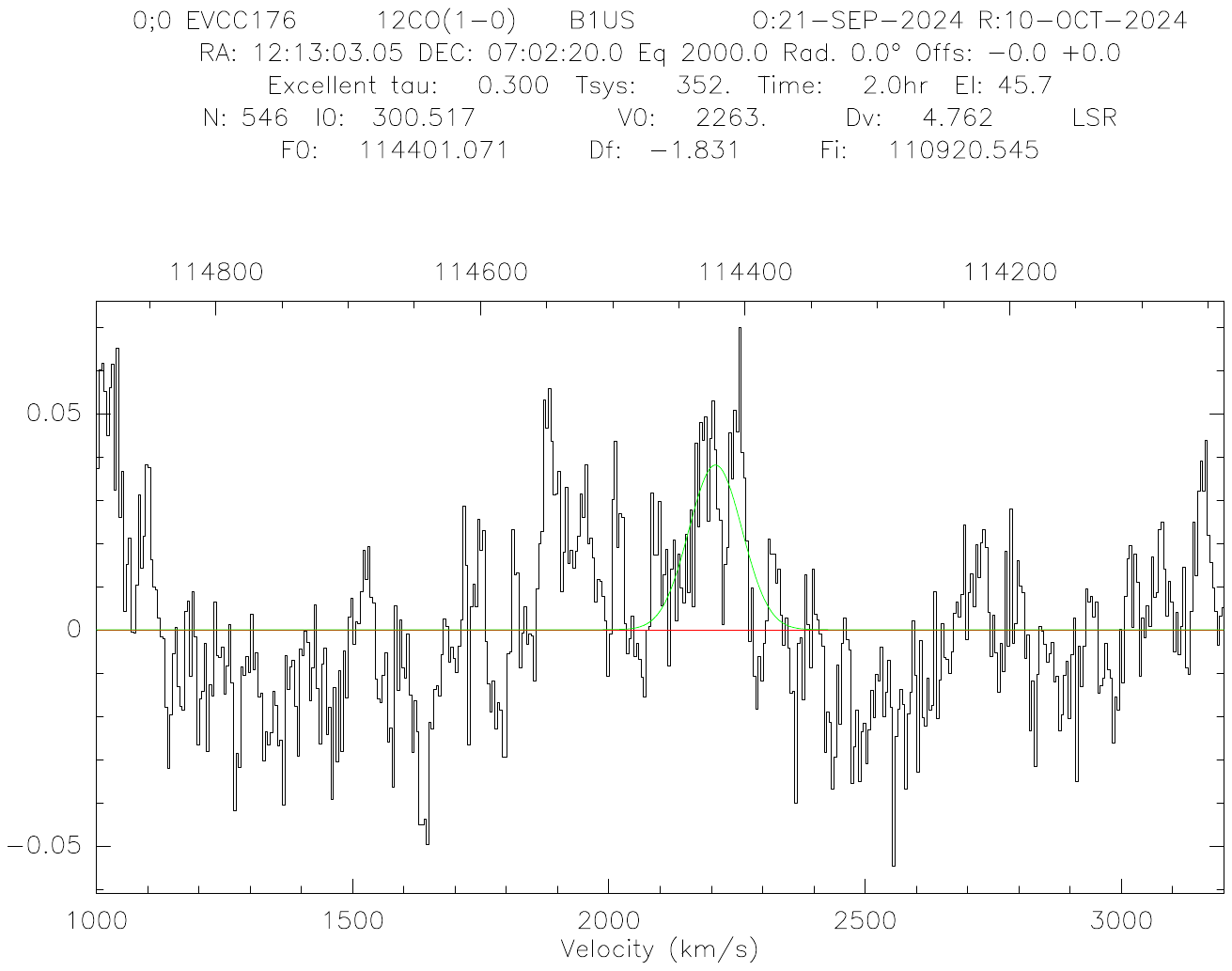}
     \includegraphics[width=0.8\textwidth, angle=0]{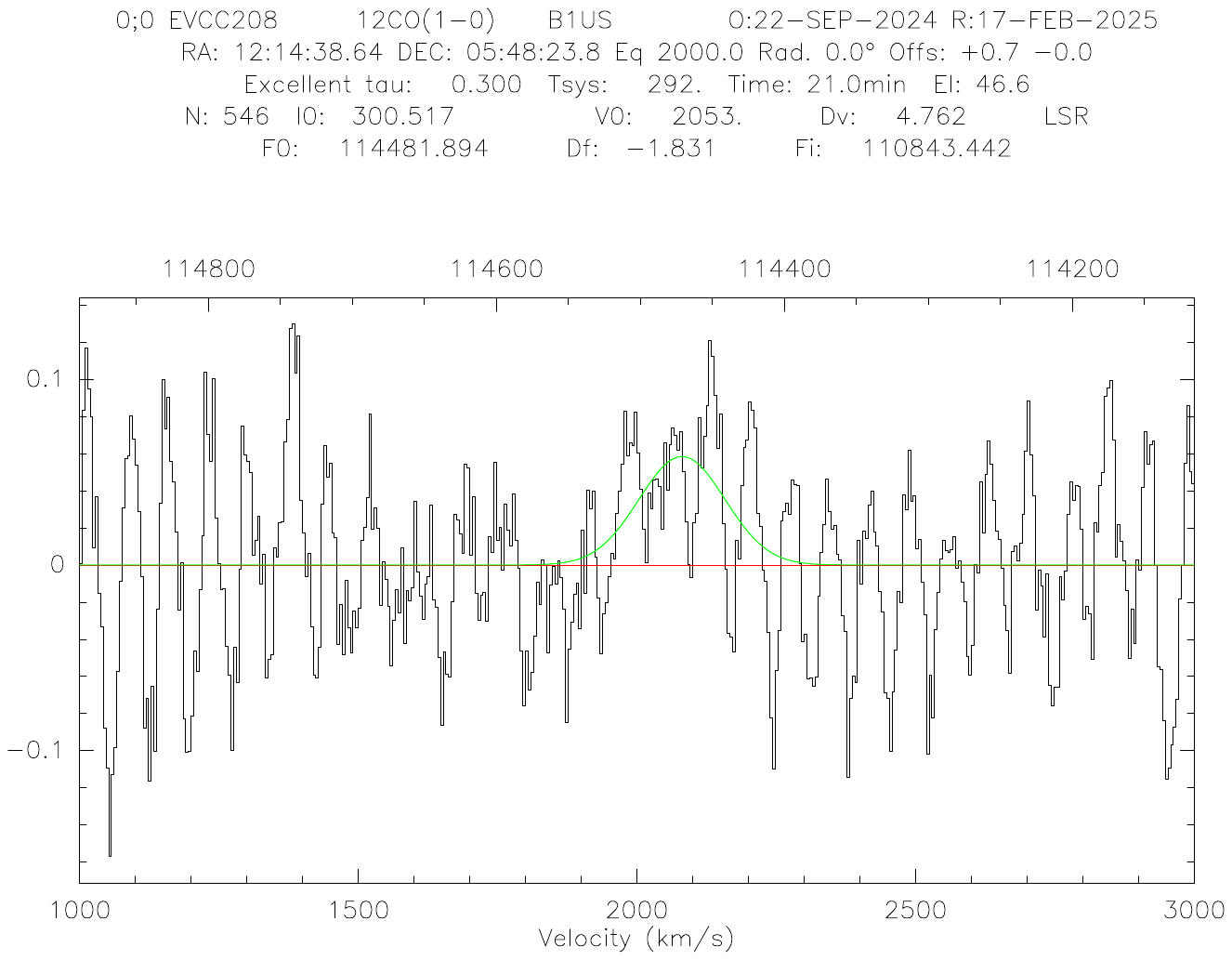}
     \addtocounter{figure}{-1} 
     \caption{(continued)} 
\end{figure}

\begin{figure}
     \centering
     \includegraphics[width=0.8\textwidth, angle=0]{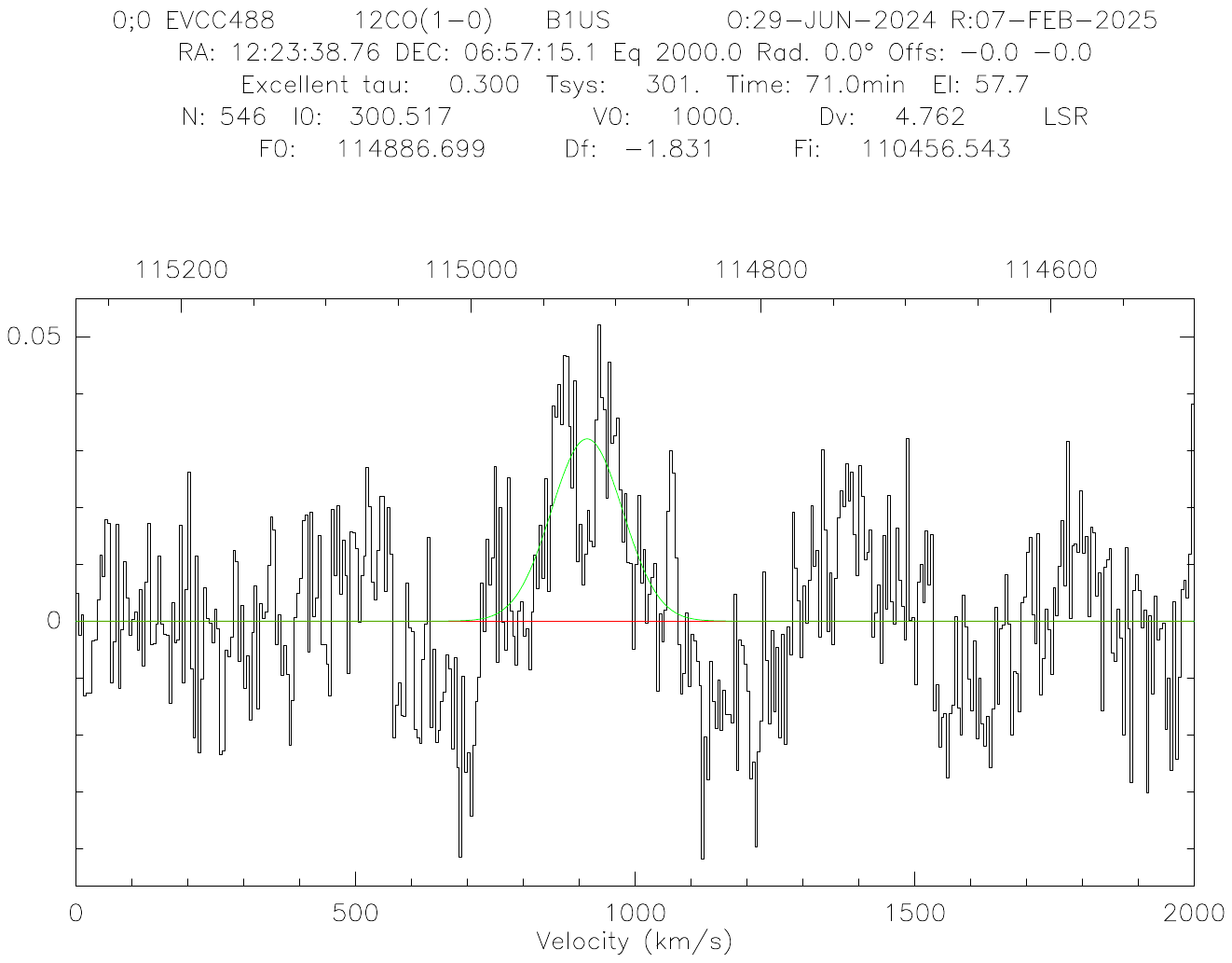}
     \includegraphics[width=0.8\textwidth, angle=0]{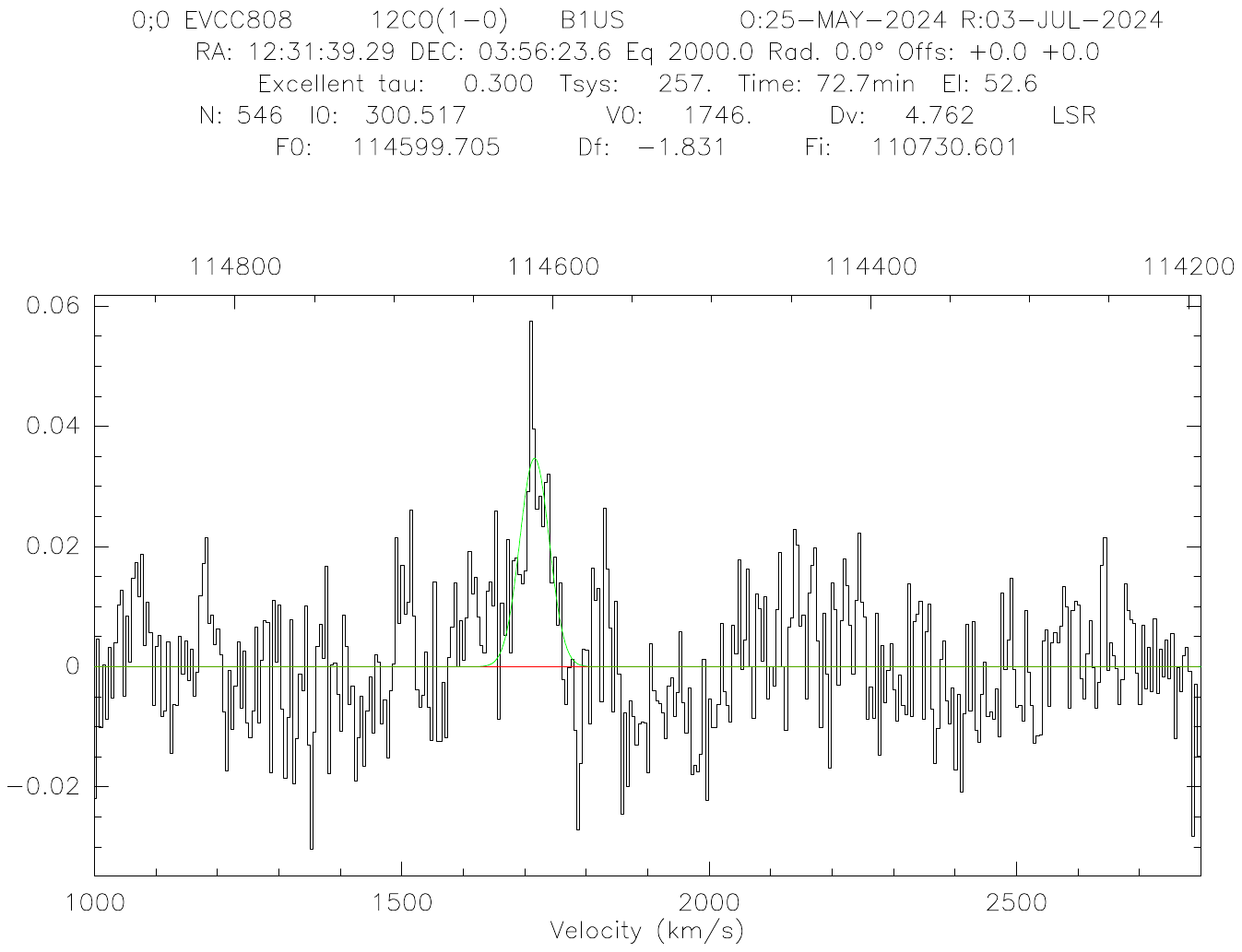}
     \addtocounter{figure}{-1}
     \caption{(continued)}
\end{figure}

\begin{figure}
     \centering
     \includegraphics[width=0.8\textwidth, angle=0]{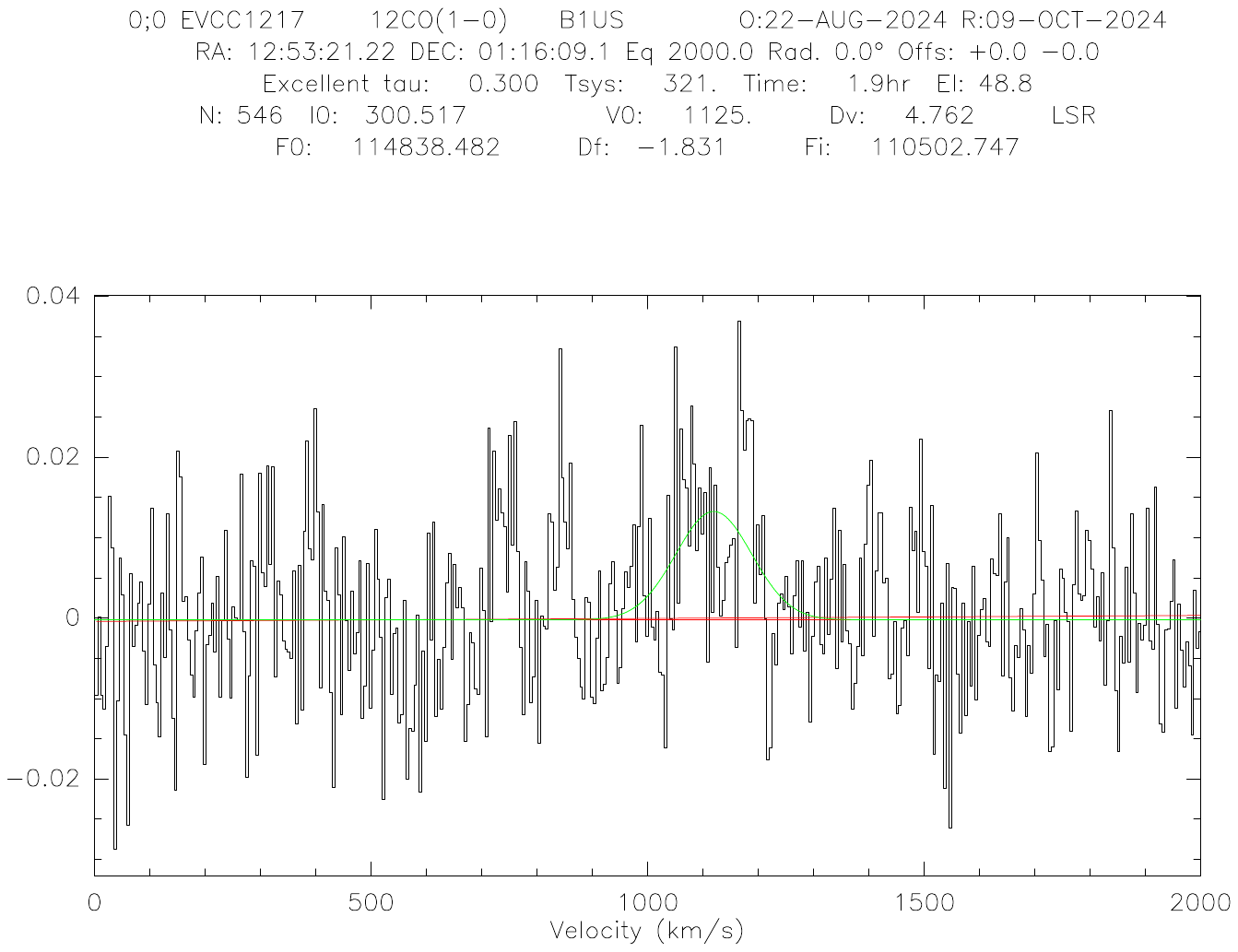}
     \includegraphics[width=0.8\textwidth, angle=0]{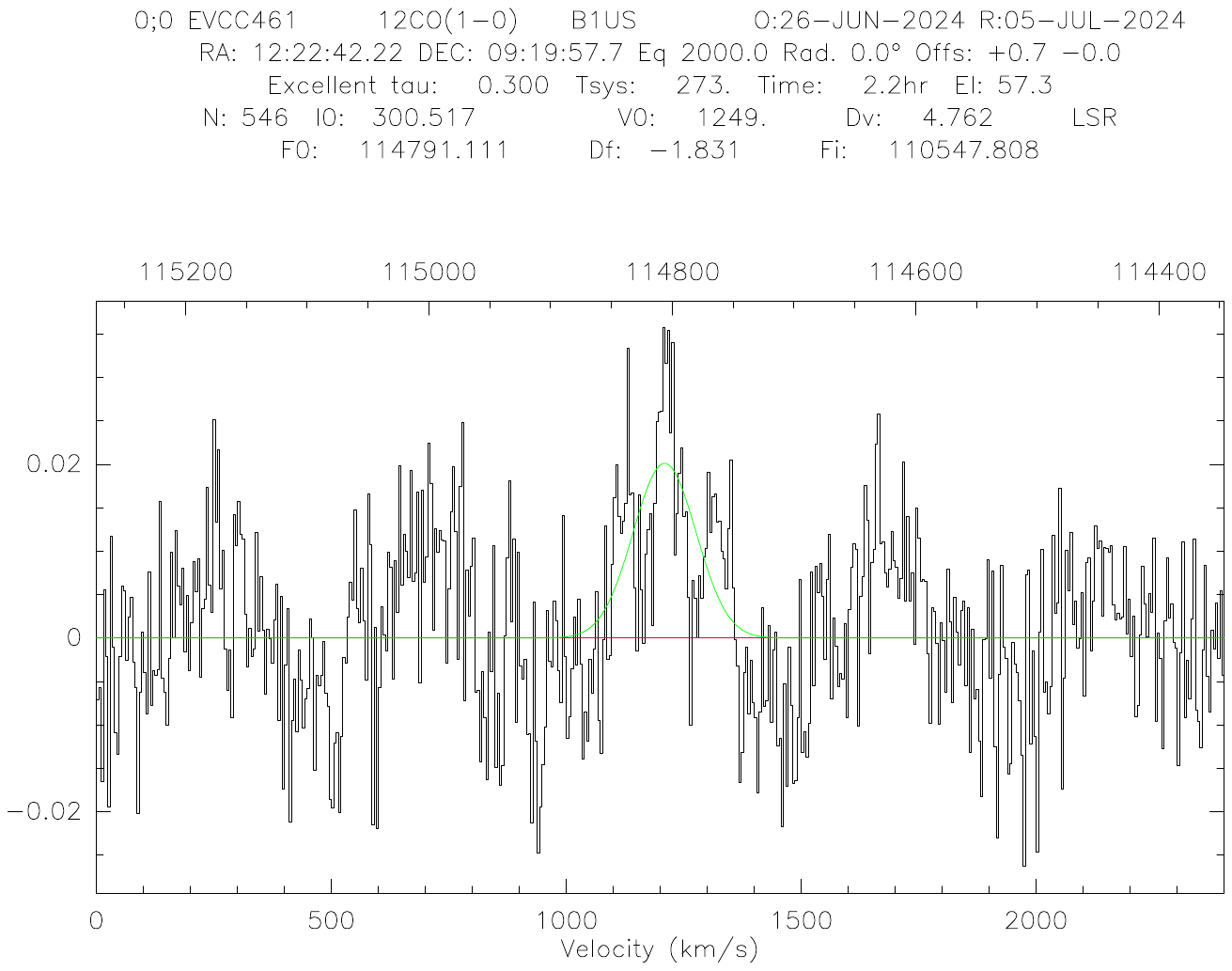}
     \addtocounter{figure}{-1} 
     \caption{(continued)}
\end{figure}

\subsection{sed fitting}
 
According to the Herschel and planck flux, the Spectral Energy Distribution (SED) of galaxies in the far infrared to submillimeter bands can be fitted.The sed fitting tool used in this paper is cigale.Code Investigating GALaxy Emission (CIGALE) is a Python code for the fitting of spectral energy distribution (SED) of galaxies\footnote{https://cigale.lam.fr/}. It has been developed for more than 1.5 decades (e.g.\citealt{Burgarella+etal+2005}; \citealt{Noll+etal+2009}; \citealt{Boquien+etal+2019}).

cigale can use different dust models and empirical templates to trace dust emission and thus synthesize a galaxy's sed. The emission of stars and dust is related to dust decay, which is estimated by taking into account the energy balance between the energy emitted by the different populations of stars absorbed by the dust and the energy re-emitted in the infrared.The model spectra are constructed assuming different star formation histories, metallicity, and decay laws.They are then convolved with transmissivity profiles of different photometric bands commonly used for space missions and compared with multi-frequency observations.The code determines the best fit model by $\chi^2$ minimization and performs a probability distribution function (PDF) analysis to determine the weighted mean and standard deviation of different physical parameters such as star mass and star formation rate.

Our template for using the cigale model is the cigale model of galaxy clusters in \citealt{Boselli+etal+2016}, which, in addition to the Herschel and Planck data, Four bands irsa(\citealt{Wheelock+etal+2019}) 16, 25, 100, 160 were also added in order to have more data points to make the fitting of the model more accurate. Table~\ref{Tab4} shows the data points for seed fitting with cigale and the simulated total IR luminosity and SFH.

\begin{longtable}{l l l l l l l l l } 

\caption{Properties of the Virgo} \label{Tab4} \\
% 表头
\toprule
EVCC & \(\log(M_{H_2})\) & \(M_{H_2}\) & \(M_{HI}\) & \(\text{def}_{HI}\) & \(\text{HIflux}\) & \(t_{\text{gas}}\) & \(t_{M_{H_2}}\) & \(t_{M_{HI}}\) \\
     & \(M_\odot\) &  &  &  & \(Jy\ km\ s^{-1}\) &  &  &  \\
\midrule
\endfirsthead
% 每页重复的表头
\toprule
EVCC & \(\log(M_{H_2})\) & \(M_{H_2}\) & \(M_{HI}\) & \(\text{def}_{HI}\) & \(\text{HIflux}\) & \(t_{\text{gas}}\) & \(t_{M_{H_2}}\) & \(t_{M_{HI}}\) \\
     & \(M_\odot\) &  &  &  & \(Jy\ km\ s^{-1}\) &  &  &  \\
\midrule
\endhead
% 每页重复的表尾
\midrule
\multicolumn{5}{r}{\textit{}} \\
\endfoot
% 最后一页的表尾
\bottomrule
\endlastfoot

EVCC429	&	8.75E+00  &	3.34E+08 &	1.17E+10 &	-0.503  &	84.71	  & 1.38E+09 & 3.81E+07 & 1.34E+09 \\
EVCC2159&	3.80E+00  &	1.12E+08 &	2.06E+09 &	0.411   &	126.55	  & 5.72E+08 & 2.96E+07 & 5.42E+08 \\
EVCC2184&	3.38E+00  &	2.00E+09 &	7.34E+08 &	0.974   &	11.15	  & 8.07E+08 & 5.90E+08 & 2.17E+08 \\
EVCC1099&	2.28E+00  &	5.63E+08 &	5.17E+08 &	0.723   &	7.86	  & 4.73E+08 & 2.47E+08 & 2.27E+08 \\
EVCC171	&	1.11E+00  &	1.65E+08 &	3.76E+09 &	0.173   &	62.20	  & 3.52E+09 & 1.48E+08 & 3.37E+09 \\
EVCC2174&	8.20E-01  &	1.77E+08 &	5.41E+08 &	0.935   &	10.18     & 8.74E+08 & 2.15E+08 & 6.59E+08 \\
EVCC233 &	2.46E+00  &	5.11E+08 &	9.16E+08 &	0.212   &	13.92	  & 5.81E+08 & 2.08E+08 & 3.73E+08 \\
EVCC808 &	1.22E+00  &	1.21E+08 &	2.15E+09 &	-0.132  &	43.41	  & 1.87E+09 & 9.94E+07 & 1.77E+09 \\
EVCC1314&	7.81E-01  &	1.22E+09 &	6.25E+09 &	-0.131  &	14.67     & 9.56E+09 & 1.56E+09 & 8.00E+09 \\
EVCC2209&	1.20E+00  &	1.64E+08 &	2.14E+09 &	-0.227  &	43.68	  & 1.91E+09 & 1.37E+08 & 1.78E+09 \\
EVCC952 &	8.79E-01  &	1.87E+08 &	7.00E+08 &	0.526   &	10.63	  & 1.01E+09 & 2.13E+08 & 7.96E+08 \\
EVCC84  &	1.96E+00  &	4.38E+09 &	5.75E+09 &	-0.236  &	21.45	  & 5.18E+09 & 2.24E+09 & 2.94E+09 \\
EVCC884 &	1.58E+00  &	7.04E+08 &	1.41E+07 &	2.579   &	0.20      & 4.54E+08 & 4.45E+08 & 8.94E+06 \\
EVCC1281&	3.05E+00  &	3.59E+08 &	1.31E+09 &	0.191   &	19.71	  & 5.48E+08 & 1.18E+08 & 4.30E+08 \\
EVCC1074&	1.69E+00  &	1.02E+09 &	5.00E+08 &	0.595   &	7.59	  & 9.00E+08 & 6.04E+08 & 2.96E+08 \\
EVCC107	&	2.25E+00  &	1.64E+09 &	6.64E+09 &	-0.427  &	53.69	  & 3.69E+09 & 7.30E+08 & 2.96E+09 \\
EVCC1202&	2.97E+00  &	2.53E+09 &	4.72E+09 &	0.289   &	17.6	  & 2.44E+09 & 8.52E+08 & 1.59E+09 \\
EVCC1182&	3.67E+00  &	6.80E+08 &	5.53E+07 &	1.893   &	0.84	  & 2.00E+08 & 1.85E+08 & 1.51E+07 \\
EVCC673 &	2.53E-01  &	2.66E+08 &	5.27E+08 &	0.955   &	8.01	  & 3.13E+09 & 1.05E+09 & 2.08E+09 \\
EVCC59	&	2.02E+00  &	1.54E+09 &	1.50E+10 &	-0.432  &	47.02     & 8.20E+09 & 7.63E+08 & 7.44E+09 \\
EVCC104	&	1.72E+00  &	1.51E+08 &	4.90E+09 &	-0.390  &	39.92	  & 2.94E+09 & 8.79E+07 & 2.85E+09 \\
EVCC574	&	1.87E+00  &	6.61E+08 &	3.39E+09 &	0.001   &	27.42	  & 2.17E+09 & 3.54E+08 & 1.82E+09 \\
EVCC631	&	8.13E+00  &	1.38E+08 &			 &          &		   &          & 1.70E+07 &          \\
EVCC461	&	6.36E-01  &	4.90E+08 &	1.18E+09 &	0.413   &	9.55	  & 2.63E+09 & 7.70E+08 & 1.86E+09 \\
EVCC497 &	1.38E+00  &	4.38E+08 &			 &          &		   &          & 3.17E+08 &          \\
EVCC153	&	2.22E+00  &	1.24E+09 &	6.16E+09 &	-0.280  &	13.54	  & 3.34E+09 & 5.57E+08 & 2.78E+09 \\
EVCC488	&   3.31E-01  &          &	6.79E+08 &  0.494   &   5.49      &          &          & 2.06E+09 \\
EVCC439	&	8.28E-01  &	1.78E+08 &	8.38E+07 &	1.655   &	1.32	  & 3.16E+08 & 2.15E+08 & 1.01E+08 \\
EVCC1080&	6.68E-01  &	1.78E+08 &	1.98E+09 &	0.042   &	19.01     & 3.23E+09 & 2.66E+08 & 2.96E+09 \\
EVCC126	&	2.07E+00  &	1.30E+08 &	1.73E+09 &	0.132   &	26.25	  & 8.97E+08 & 6.29E+07 & 8.34E+08 \\
EVCC1178&	1.25E+00  &	2.60E+08 &	3.10E+09 &	-0.077  &	47.61	  & 2.69E+09 & 2.09E+08 & 2.48E+09 \\
EVCC1282&   1.47E+00  &	7.62E+07 &	1.09E+09 &	0.023   &	14.63	  & 7.98E+08 & 5.19E+07 & 7.46E+08 \\
EVCC1217&	8.46E-01  &	2.60E+08 &	1.13E+09 &	0.438   &	11.69	  & 1.64E+09 & 3.08E+08 & 1.33E+09 \\
EVCC996	&   1.67E+00  &			 &			 &		    &             &          &          &          \\
EVCC587	&	1.43E-02  &	1.15E+08 &	7.34E+08 &	0.965   &	5.88	  & 5.93E+10 & 8.04E+09 & 5.12E+10 \\
EVCC660	&   1.26E+00  &	7.95E+07 &	4.19E+08 &	0.701   &	6.61      & 3.95E+08 & 6.30E+07 & 3.32E+08 \\
EVCC1223&	8.07E-01  &	2.55E+08 &	3.15E+09 &	0.076   &	8.81	  & 4.22E+09 & 3.16E+08 & 3.90E+09 \\
EVCC1069&   1.14E+00  & 4.90E+08 &	         &  0.264   &   7.53	  &          &          & 4.30E+08 \\
EVCC1104&   2.27E-01  &			 &			 &          &             &          &          &          \\
EVCC635	&   1.10E+00  & 8.02E+08 &	         &  -0.070  &   14.91	  &          &          & 7.27E+08 \\
EVCC176	&	1.26E+00  &	8.01E+08 &	4.37E+08 &	0.784   &   3.16	  & 9.85E+08 & 6.38E+08 & 3.48E+08 \\
EVCC208	&   8.15E-01  & 1.65E+09 &	         &  0.399   &   25.71	  &          &          & 2.03E+09 \\
EVCC231	&	1.06E+00  &	1.43E+08 &	4.32E+08 &	0.213   &   6.57	  & 5.45E+08 & 1.36E+08 & 4.09E+08 \\
EVCC1243&   1.84E-01  & 1.08E+09 &	         &  0.119   &   22.00     &          &          & 5.86E+09 \\
EVCC192	&	8.93E-01  &	3.68E+08 &	1.80E+09 &	0.376   &   8.41 	  & 2.42E+09 & 4.12E+08 & 2.01E+09 \\
EVCC776	&	3.29E-01  &	3.21E+09 &	5.05E+09 &	-0.093  &   11.31	  & 2.51E+10 & 9.77E+09 & 1.54E+10 \\
EVCC1209&   1.01E+00  &	3.55E+07 &	5.21E+08 &	0.370   &   13.07	  & 5.51E+08 & 3.51E+07 & 5.16E+08 \\
EVCC451	&	7.45E-01  &	3.68E+08 &	1.27E+08 &	1.116   &   1.93      & 6.64E+08 & 4.93E+08 & 1.70E+08 \\

\end{longtable}

\section{Data analysis}
\label{sect:analysis}

\subsection{CO Properties in Different Environments}
\subsubsection{CO Properties as a Function of the Optical Parameters}
We first investigated the relationship between the molecular gas and the optical properties of the galaxy, which is essential to minimize the possibility of confusing the properties of the inherent molecular gas with environmental influences.

\begin{figure}{H}
   \centering
   \includegraphics[width=\textwidth, angle=0]{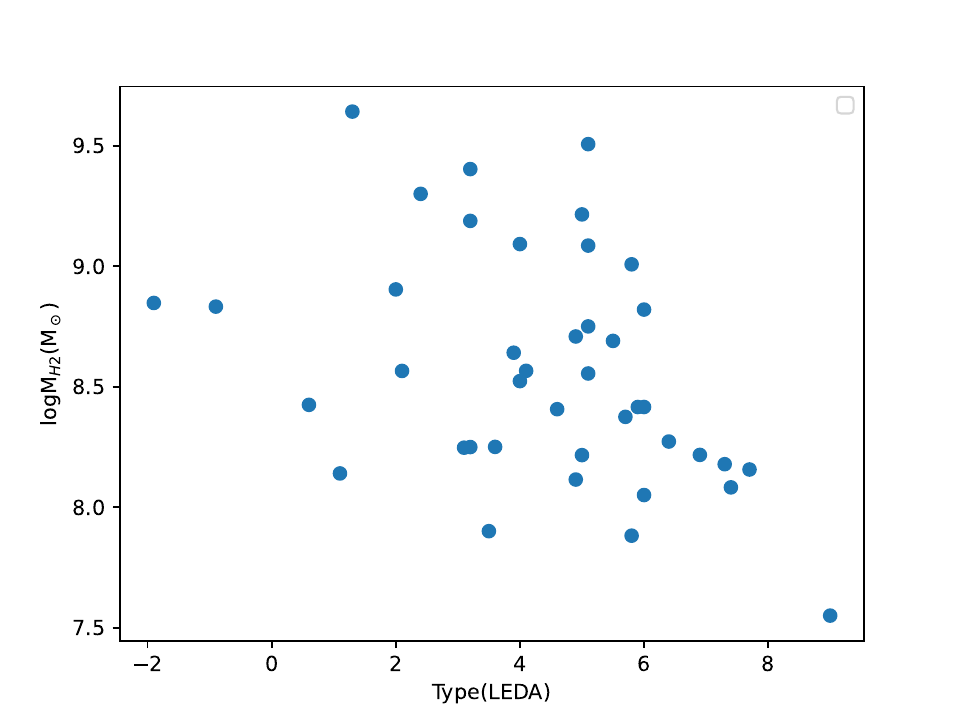}
   \caption{Molecular gas mass as a function of the morphological type}
   \label{Fig2}
   \end{figure}

\begin{figure}{H}
   \centering
   \includegraphics[width=\textwidth, angle=0]{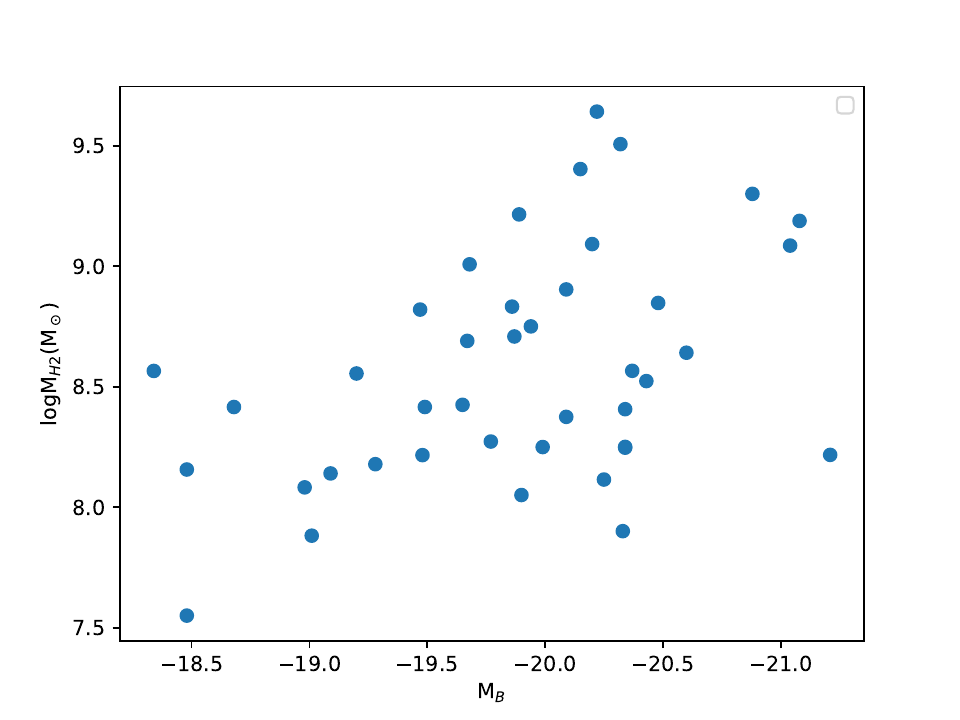}
   \caption{The corrected optical B-magnitude}
   \label{Fig3}
   \end{figure}

\begin{figure}{H}
   \centering
   \includegraphics[width=\textwidth, angle=0]{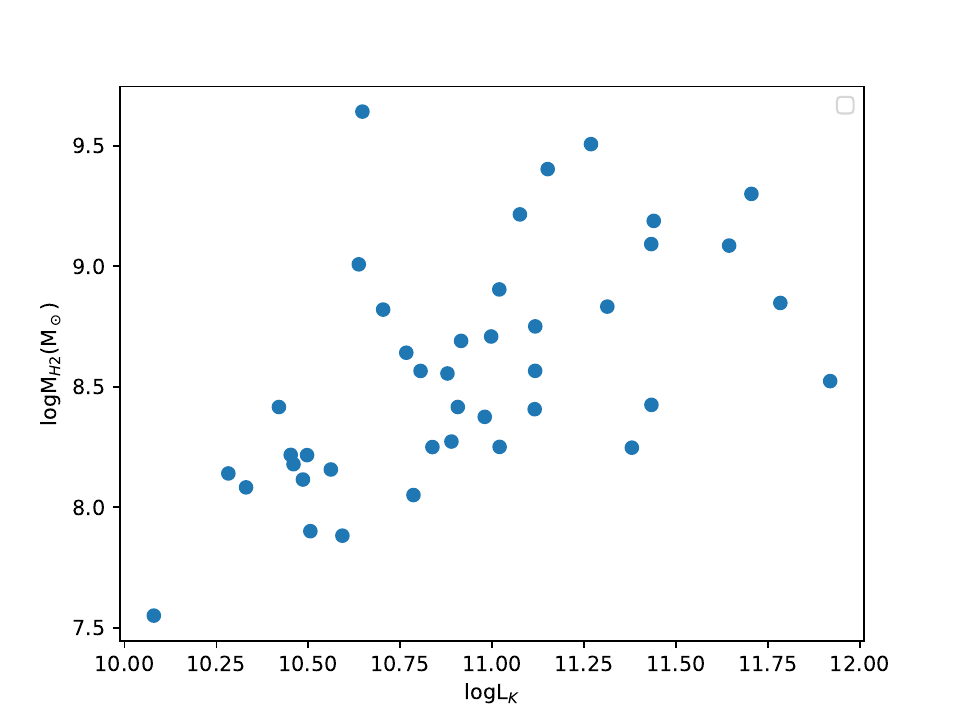}
   \caption{The K-band luminosity}
   \label{Fig4}
   \end{figure}

First of all, Figure~\ref{Fig2}, Figure~\ref{Fig3} and Figure~\ref{Fig4} show the change of general properties of molecular gas with morphological type(LEDA)\footnote{http://atlas.obs-hp.fr/hyperleda/ledacat.cgi?o=EVCC\%203}, B-band absolute magnitude(LEDA) and K-band luminosity respectively(2MASS; \citealt{Skrutskie+etal+2006}; \citealt{Jarrett+2000}; \citealt{Jarrett+etal+2003}).

As can be seen from Figure~\ref{Fig2}, galaxies with polymer gas converge to the sb-sc type, but generally speaking, the content of molecular gas has no significant relationship with the form type of galaxies, and the relationship between the mass of molecular gas and the form type is consistent with previous studies(e.g.,\citealt{Young+1991}; \citealt{Chung+etal+2017}).

In Figure~\ref{Fig3}, we can see that the mass of the molecular gas increases with the increase of the absolute magnitude of the B-band,but the scattering is larger.In Figure~\ref{Fig3}, we can see that the mass of the molecular gas is roughly linear with the luminosity of the K band, which is consistent with previous studies by others.(\citealt{Chung+etal+2017};\citealt{Lisenfeld+etal+2011})

%%%%%%%%%%%%%%%%%%%%%%%%%%%%%%%%%%%%%%%%%%%%%%%%%%%%%%%%%%%%%%
%%     Examples for figures using graphicx for LaTeX 2e
%%               -- our recommended way for embodying graphics
%%%%%%%%%%%%%%%%%%%%%%%%%%%%%%%%%%%%%%%%%%%%%%%%%%%%%%%%%%%%%%
%
%      A figure as large as the width of the column
%-------------------------------------------------------------
%
%      One column rotated figure
%-------------------------------------------------------------
  
%----------------------------------------------------- Figs 3 & 4:
%\begin{figure}{H}[h]
 % \begin{minipage}[t]{0.495\linewidth}
  %\centering
   %\includegraphics[width=60mm]{f1.eps}
	%  \caption{\label{Fig3}{\small Amplitudes evolution of . . .} }
  %\end{minipage}%
  %\begin{minipage}[t]{0.495\textwidth}
  %\centering
   %\includegraphics[width=60mm]{f2.eps}
	%  \caption{\label{Fig4}{\small Amplitude variation of AN Lyn.}}
  %\end{minipage}%
%\end{figure}

\subsection{CO and HI}
H I gas is typically the most vulnerable component in a galaxy, since it extends far beyond the optical size and is thus weakly bound by gravity.(e.g., \citealt{Hewitt+etal+1983}).As such, it is vulnerable to any external force, and can often serve as a useful tool for studying how galaxies interact with their surroundings (e.g., \citealt{Hibbard+etal+2001}; \citealt{Chung+etal+2009a}). By comparing molecular gas content and HI defects, we can better understand the gas composition of galaxies, the star formation process, and the physical mechanisms of galactic evolution.In this section, the relationship between CO luminosity normalized by K band luminosity and HI defect def$_{HI}$ is studied.

\begin{figure}{H}
   \centering
   \includegraphics[width=\textwidth, angle=0]{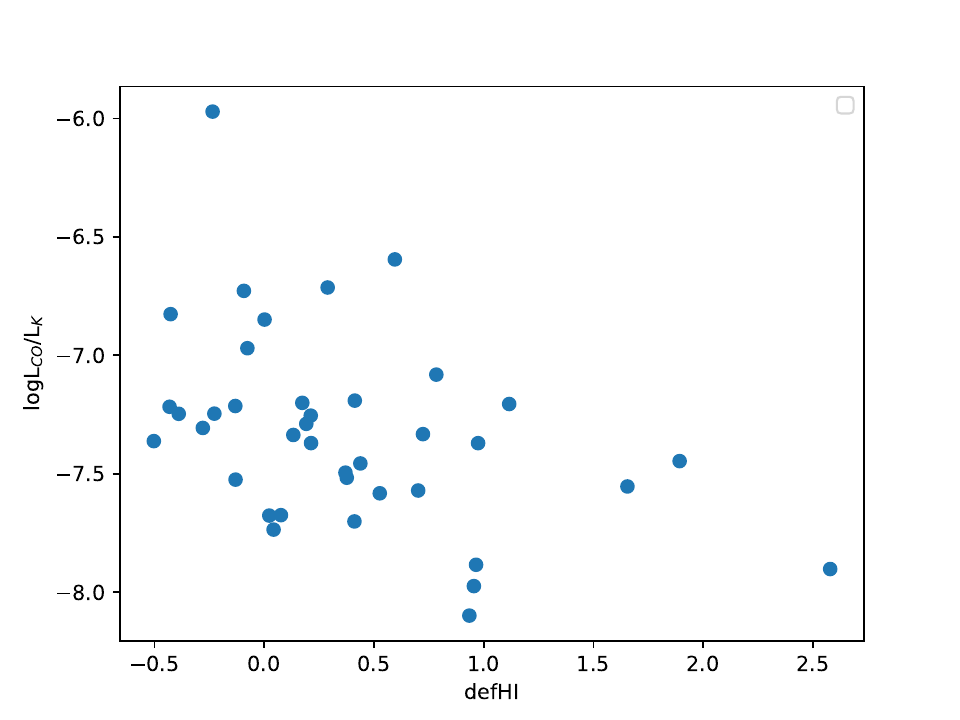}
   \caption{co luminosity normalized with K-band luminosity and def$_{HI}$}
   \label{Fig5}
   \end{figure}

In Figure ~\ref{Fig5}, we plot the relationship between co luminosity normalized with K-band luminosity and def$_{HI}$. As can be seen from the figure, the ratio of L$_{CO}/L_{K}$ decreases with increasing def$_{HI}$, indicating that galaxies lacking HI are also lacking molecular gas, which is consistent with previous findings (\citealt{Chung+etal+2017}).

\subsection{Star Formation Rates and Gas}

\begin{figure}{H}
   \centering
   \includegraphics[width=\textwidth, angle=0]{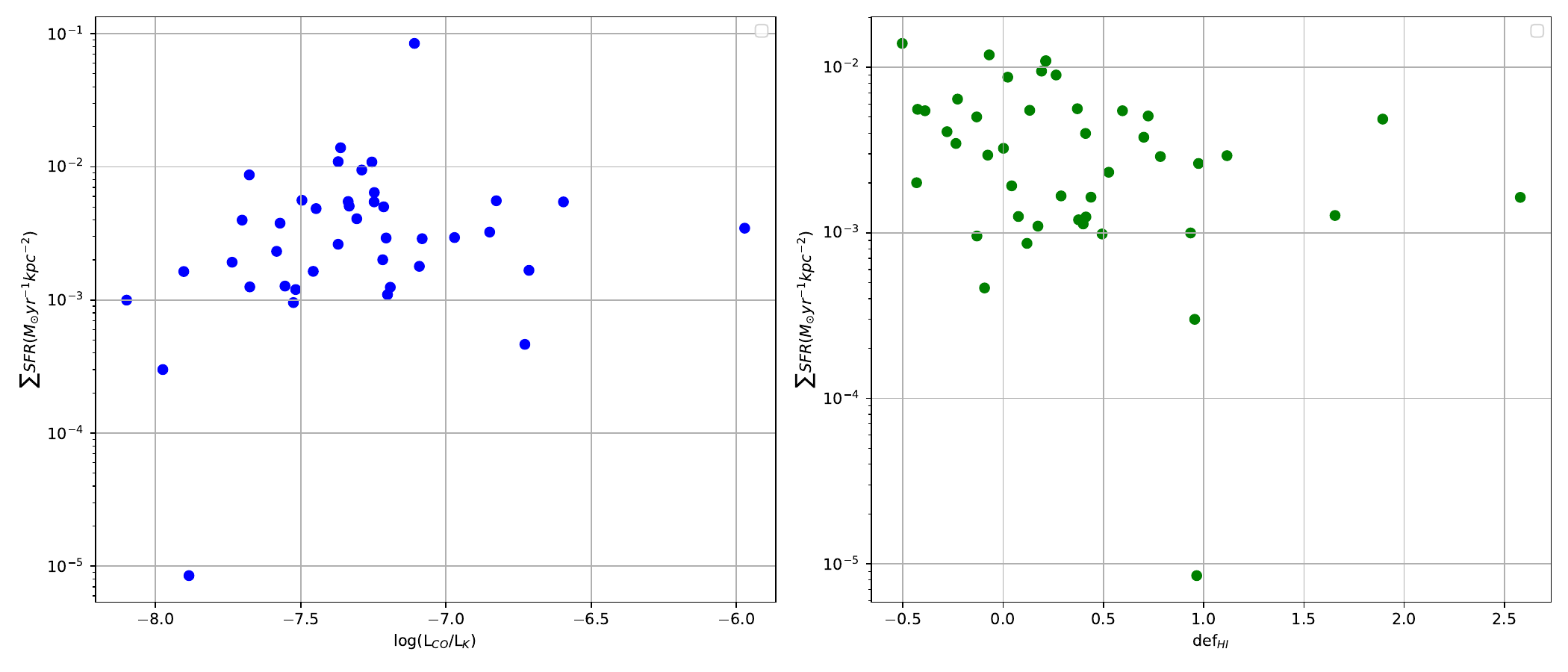}
   \caption{Galactic disk mean SFR ($\sum {SFR}$) compared to normalized CO luminosity (left) and H I defect (right).}
   \label{Fig6}
   \end{figure}

For star formation activity in galaxies in Virgo clusters, this study examined the total radiant heat star formation rate ( sed-fitting, section 3.2). The average disk star formation rate per unit area ($\sum{SFR} = SFR / disk area$) is expressed as a
function of defH I and log LCO/LK, as shown in Figure\ref{Fig6}. The Virgo clusters galaxy shows $\sum{SFR} \sim 10^{-3}-10^{-2} M_\odot yr^{-1} KPC^{-2}$.

The strong correlation between $\sum{SFR}$ and gas can be seen from the comparison of $\sum{SFR}$ and gas in Figure\ref{Fig6}, and in the left figure we can see that galaxies with higher CO content in Virgo clusters have higher $\sum{SFR}$. In the comparison of defHI and $\sum{SFR}$ in the graph on the right, galaxies with high $\sum{SFR}$ are mostly in normal HI-rich galaxies, while galaxies with poor HI have low $\sum{SFR}$ activity. This is consistent with our theory and previous research.

\subsection{Gas dissipation time}

   \begin{figure}{H}
   \centering
   \includegraphics[width=\textwidth, angle=0]{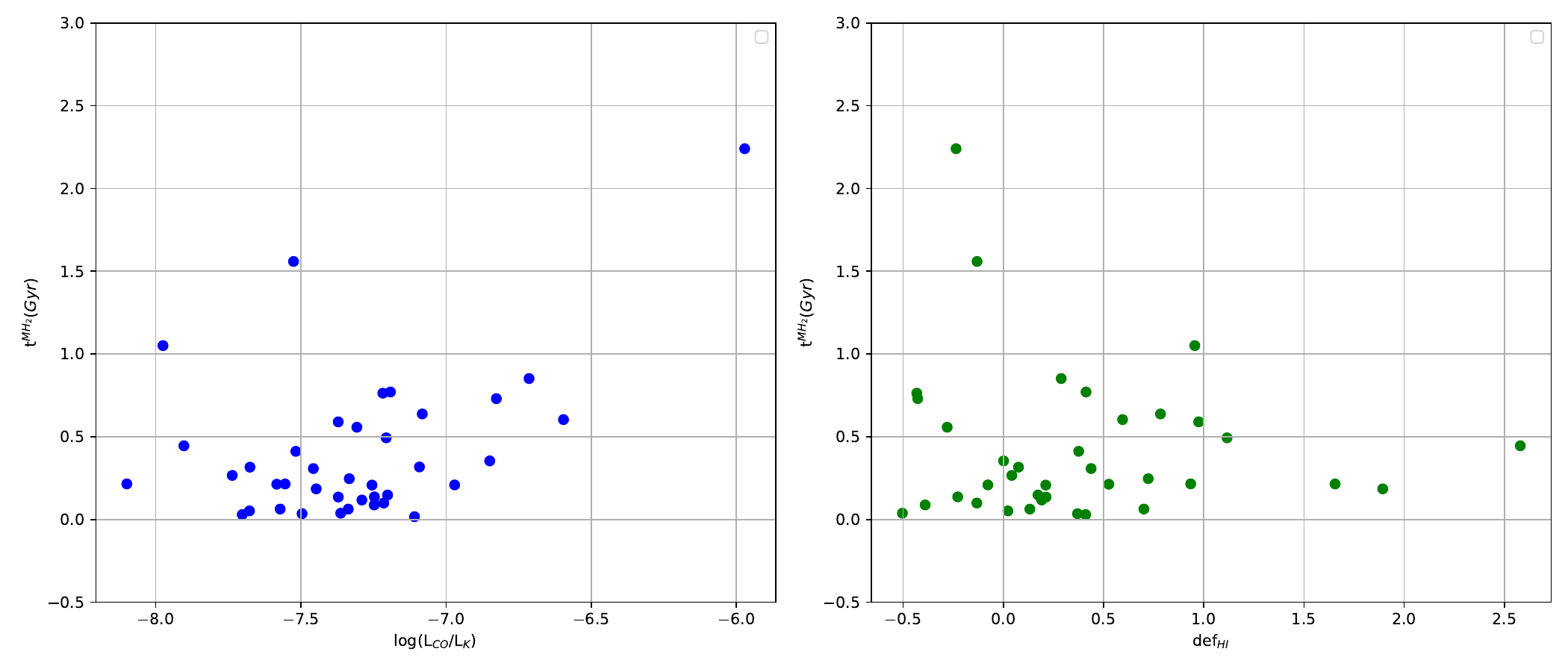}
   \caption{Gas consumption times for the H$_2$ gas mass t$_{H_2}$ and def$_{HI}$}
   \label{Fig7}
   \end{figure}

   \begin{figure}{H}
   \centering
   \includegraphics[width=\textwidth, angle=0]{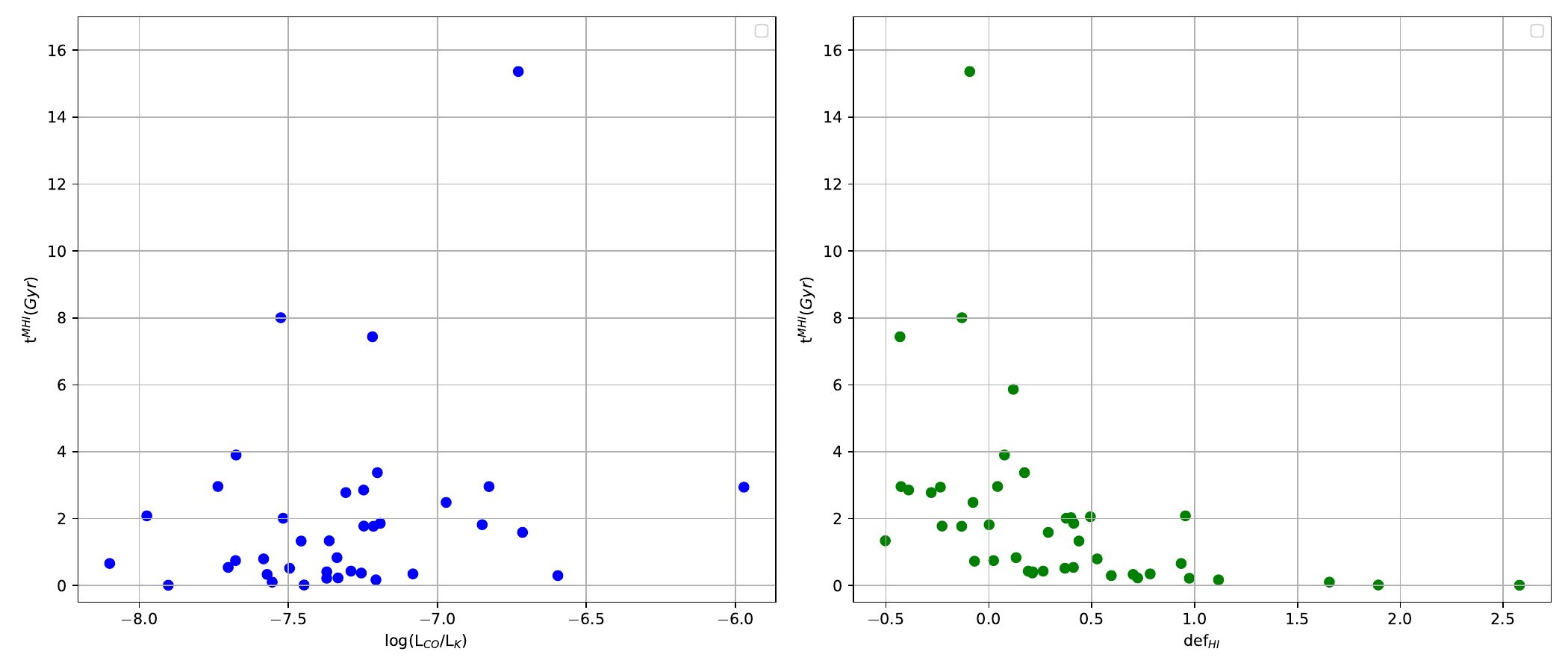}
   \caption{Gas consumption times for the H I gas mass t$_{HI}$ and def$_{HI}$}
   \label{Fig8}
   \end{figure}

   \begin{figure}{H}
   \centering
   \includegraphics[width=\textwidth, angle=0]{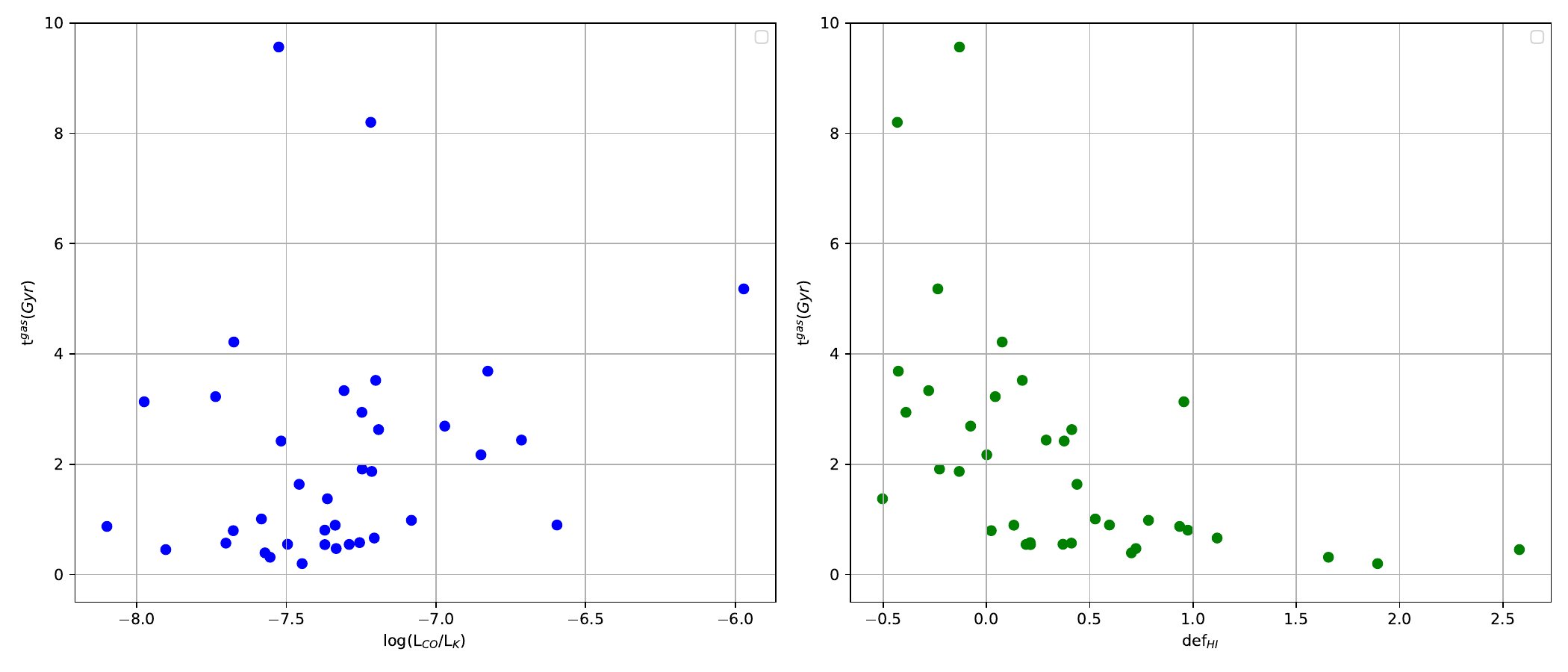}
   \caption{Gas consumption times for the total gas mass t$_{gas}$ and def$_{HI}$}
   \label{Fig9}
   \end{figure}

The kennicut-Schmidt relationship(\citealt{Kennicutt+etal+1998,Kennicutt+etal+1989}) shows that molecular gas and star formation are extremely closely linked across a wide range of galaxy types, both in the local universe and in high redshift galaxy types.The ratio of molecular mass to the rate of star formation, known as gas dissipation time, measures the time scale of molecular gas depletion, that is, the amount of time a galaxy can sustain star formation at its current rate without accreting gas.The gas dissipation mass used in this paper is: the time required for a galaxy to consume all of its cold gas through star formation. The gas mass and star formation rate(It is obtained in the 2.1 by cigale by sesing the galaxy) used are calculated as follows:

$$
t^{gas}(yr) = \frac{SFR(M_{\odot}yr^{-1})}{M_{gas}}
$$

where SFR is the star formation rate, $M_{gas}$ is the H I, H$_{2}$ and the total gas mass. Figure ~\ref{Fig7}, Figure ~\ref{Fig8} and Figure ~\ref{Fig9} show the relationship between atomic, molecular and total gas dissipation time and def$_{H I}$ and $L_{CO}/L_K$ respectively.The atomic and molecular gas consumption time ranges from 0.1-16 Gyr and 0.2-2.5Gyr,respectively.

The ratio of H$_{2}$ to def$_{HI}$ in Figure ~\ref{Fig7} shows that t$^{H_2}$ seems to increase with the increase of def$_{HI}$ when there is no shortage of HI gas, but t$^{H_2}$ seems to decrease with the increase of def${_HI}$ when there is the greatest shortage of HI.This can also prove that the efficiency of star formation is directly related to the dissipation time of molecular gas, and the lack of HI may be related to the rapid dissipation of gas during star formation.

In Figure~\ref{Fig8}, we see that the dissipation time of atomic gas is inversely related to def$_{HI}$,  and it can be seen that although the star formation rate and the atomic gas mass tend to be smaller in HI deficient galaxies, the lost hydrogen gas seems to have a stronger effect, resulting in an inverse correlation between the atomic gas consumption time and the HI deficient galaxies.

In Figure~\ref{Fig9}, the total cold air (HI + H$_{2}$) consumption time of the whole sample has a slightly decreasing trend with the increase of def$_{HI}$.The total cold gas consumption time is relatively insensitive compared to the distribution of total cold gas mass or SFR, which may reflect the highly local and short time scales associated with star formation activity.The more prominent trend in molecular gas consumption time compared to atomic or total gas consumption time suggests that internal processes are more important than external, ICM-driven processes and can explain the overall trend observed here.

For the ratio of L$_{CO}$ normalized with K-band luminosity to the dissipation time of the gas, whether it is the dissipation time of molecules, atoms and total cold gas. Both of them increase with the increase of $L_{CO}/L_K$, indicating that $L_{CO}/L_K$ can be a good standard to measure the dissipation time of cold gas.

\section{Conclusions}
\label{sect:conclusion}
The study measured CO (J=1-0) for a sample of dusty galaxies in the Virgo cluster, selected from the PLanck and Herschel surveys. The results show the molecular gas mass of 48 samples of galaxies. To study the statistical properties of molecular gas in these galaxies, combined with multi-band data (HI, optical, far infrared), to study the evolution of gas in the cluster environment and its relationship to star formation activity.  The following results were obtained:

H$_{2}$ masses is the CO data that we observed through the 13.7m telescope at Qinghai station, by using CO-to-H$_{2}$ conversion factor of $\chi =2.8 \times 10^{20}\ H_2\ cm^{-2}\ (K[T_R]\ km\  s^{-1})^{-1}$. This is the value found by \citeauthor{Bloemen+etal+1986}\citeyearpar{Bloemen+etal+1986}.in their analysis of the galactic gamma-ray emissivity.

We compared the molecular gas mass with different parameters($M_B$,$L_K$)to characterize the relationships followed by galaxies in Virgo clusters,In the limited number of cases used in this study, $M_{H_2}$ was strongly correlated with the $L_K$ of the field sample (Figure \ref{Fig3}), (Gavazzi et al.1996) showed the correlation of $M_{H_2}$ with a large sample. and the $L_{CO}/L_K$ or $M_{H_2}/L_K$ ratio can be a reasonable amount for studying the molecular gas content of galaxies in different environments, independent of the size effect.But the correlation between gas molecular mass and B-band optical size and luminosity is weak.

According to the data collected from irsa, planck and herschel, sed fitting of galaxies is performed to obtain the SFR of galaxies.  The corresponding gas dissipation time is derived from the gas mass and compared with the defHI ratio,The dissipation time of molecular gas increases with the HI deficiency, and star formation is less efficient in HI-deficient galaxies. This may be because HI is a precursor to molecular gas (H$_{2}$), and the lack of HI limits the formation of molecular gas, thus affecting the rate of star formation.

By comparing the functional relationship between SFR and def$_{HI}$, it is found that galaxies with high star formation rate mostly exist in the cluster of galaxies with normal/abundant HI, which is consistent with our expectation.  For SFR and $L_{CO}/L_K$, the SFR decreases with the decrease of $L_{CO}/L_K$, indicating that galaxies with lower co content have smaller SFR.  This also proves that the amount of molecular gas can affect the speed of SFR.The analysis of gas consumption time in this study shows that the internal processes of galaxies in Virgo clusters are more important in the sf process than the external, ICM-driven processes.

\section{Conclusions}
\label{sect:conclusion}
The study measured CO (J=1-0) for a sample of dusty galaxies in the Virgo cluster, selected from the PLanck and Herschel surveys. The results show the molecular gas mass of 48 samples of galaxies. To study the statistical properties of molecular gas in these galaxies, combined with multi-band data (HI, optical, far infrared), to study the evolution of gas in the cluster environment and its relationship to star formation activity.  The following results were obtained:

H$_{2}$ masses is the CO data that we observed through the 13.7m telescope at Qinghai station, by using CO-to-H$_{2}$ conversion factor of $\chi =2.8 \times 10^{20}\ H_2\ cm^{-2}\ (K[T_R]\ km\  s^{-1})^{-1}$. This is the value found by \citeauthor{Bloemen+etal+1986}\citeyearpar{Bloemen+etal+1986}.in their analysis of the galactic gamma-ray emissivity.

We compared the molecular gas mass with different parameters($M_B$,$L_K$)to characterize the relationships followed by galaxies in Virgo clusters,In the limited number of cases used in this study, $M_{H_2}$ was strongly correlated with the $L_K$ of the field sample (Figure \ref{Fig3}), (Gavazzi et al.1996) showed the correlation of $M_{H_2}$ with a large sample. and the $L_{CO}/L_K$ or $M_{H_2}/L_K$ ratio can be a reasonable amount for studying the molecular gas content of galaxies in different environments, independent of the size effect.But the correlation between gas molecular mass and B-band optical size and luminosity is weak.

According to the data collected from irsa, planck and herschel, sed fitting of galaxies is performed to obtain the SFR of galaxies.  The corresponding gas dissipation time is derived from the gas mass and compared with the defHI ratio,The dissipation time of molecular gas increases with the HI deficiency, and star formation is less efficient in HI-deficient galaxies. This may be because HI is a precursor to molecular gas (H$_{2}$), and the lack of HI limits the formation of molecular gas, thus affecting the rate of star formation.

By comparing the functional relationship between SFR and def$_{HI}$, it is found that galaxies with high star formation rate mostly exist in the cluster of galaxies with normal/abundant HI, which is consistent with our expectation.  For SFR and $L_{CO}/L_K$, the SFR decreases with the decrease of $L_{CO}/L_K$, indicating that galaxies with lower co content have smaller SFR.  This also proves that the amount of molecular gas can affect the speed of SFR.The analysis of gas consumption time in this study shows that the internal processes of galaxies in Virgo clusters are more important in the sf process than the external, ICM-driven processes.

\label{lastpage}
  
\end{document}